\documentclass[preprint, twocolumn]{aastex62}

\usepackage{graphicx}
\usepackage{amsmath}
\usepackage[flushleft]{threeparttable}
\usepackage{comment}
\usepackage{array,multirow,color,tablefootnote}
\usepackage{xcolor}
\usepackage{subfigure}

\defcitealias{Fraine2014}{F14}
\defcitealias{Mansfield2018}{M18}

\shorttitle{Atmosphere of HAT-P-11b}
\shortauthors{Y. Chachan et al.}

\begin{document}

\title{A {\it Hubble} PanCET Study of HAT-P-11\lowercase{b}: A Cloudy Neptune with a Low Atmospheric Metallicity}
\correspondingauthor{Yayaati Chachan}
\email{ychachan@caltech.edu}

\author[0000-0003-1728-8269]{Yayaati Chachan}
\affil{California Institute of Technology, 1200 E California Blvd, Pasadena, CA, 91125, USA}

\author{Heather A. Knutson}
\affil{California Institute of Technology, 1200 E California Blvd, Pasadena, CA, 91125, USA}

\author{Peter Gao}\altaffiliation{51 Pegasi b Fellow}
\affiliation{Department of Astronomy, University of California, Berkeley, CA 94720, USA}

\author{Tiffany Kataria}
\affiliation{Jet Propulsion Laboratory, California Institute of Technology, 4800 Oak Grove Drive, Pasadena, CA, USA}

\author{Ian Wong}\altaffiliation{51 Pegasi b Fellow}
\affil{Department of Earth, Atmospheric, and Planetary Sciences, Massachusetts Institute of Technology, Cambridge, MA 02139, USA}

\author{Gregory W. Henry}
\affiliation{Center of Excellence in Information Systems, Tennessee State University, Nashville, TN 37209, USA}

\author{Bjorn Benneke}
\affiliation{Department of Physics, Universite de Montreal, Montreal, QC, Canada}

\author{Michael Zhang}
\affil{California Institute of Technology, 1200 E California Blvd, Pasadena, CA, 91125, USA}

\author{Joanna Barstow}
\affiliation{Physics and Astronomy, University College London, London, UK}

\author{Jacob L. Bean}
\affiliation{Department of Astronomy, University of Chicago, Chicago, USA}

\author{Thomas M. Evans} 
\affiliation{MIT Kavli Institute for Astrophysics and Space Research, Vassar St, Cambridge, MA, USA}

\author{Nikole K. Lewis}
\affiliation{Department of Astronomy and Carl Sagan Institute, Cornell University, 122 Sciences Drive, 14853, Ithaca, NY, USA}

\author{Megan Mansfield}
\affiliation{Department of Astronomy, University of Chicago, Chicago, USA}

\author[0000-0003-3204-8183]{Mercedes L\'opez-Morales}
\affiliation{Center for Astrophysics ${\rm \mid}$ Harvard {\rm \&} Smithsonian, 60 Garden Street, Cambridge, MA 01238, USA}

\author{Nikolay Nikolov}
\affiliation{Astrophysics Group, School of Physics, University of Exeter, Exeter, EX4 4QL, UK}

\author{David K. Sing}
\affiliation{Department of Earth and Planetary Sciences, Johns Hopkins University, Baltimore, MD, USA}

\author{Hannah Wakeford}
\affiliation{Space Telescope Science Institute, 3700 San Martin Drive, Baltimore, Maryland 21218, USA}

\begin{abstract}
We present the first comprehensive look at the $0.35-5$ $\mu$m transmission spectrum of the warm ($\sim 800$ K) Neptune HAT-P-11b derived from thirteen individual transits observed using the \emph{Hubble} and \emph{Spitzer Space Telescopes}.  Along with the previously published molecular absorption feature in the $1.1-1.7$~$\mu$m bandpass, we detect a distinct absorption feature at 1.15~$\mu$m and a weak feature at 0.95~$\mu$m, indicating the presence of water and/or methane with a combined significance of 4.4 $\sigma$.  We find that this planet's nearly flat optical transmission spectrum and attenuated near-infrared molecular absorption features are best-matched by models incorporating a high-altitude cloud layer.  Atmospheric retrievals using the combined $0.35-1.7$~$\mu$m \emph{HST} transmission spectrum yield strong constraints on atmospheric cloud-top pressure and metallicity, but we are unable to match the relatively shallow \emph{Spitzer} transit depths without under-predicting the strength of the near-infrared molecular absorption bands.  HAT-P-11b's \emph{HST} transmission spectrum is well-matched by predictions from our microphysical cloud models.  Both forward models and retrievals indicate that HAT-P-11b most likely has a relatively low atmospheric metallicity ($<4.6 \; Z_{\odot}$ and $<86 \; Z_{\odot}$ at the $2 \sigma$ and $3 \sigma$ levels respectively), in contrast to the expected trend based on the solar system planets.  Our work also demonstrates that the wide wavelength coverage provided by the addition of the \emph{HST} STIS data is critical for making these inferences.
\end{abstract}

\section{Introduction}
The atmospheric compositions of extrasolar gas giant planets are expected to vary depending on their formation locations and accretion histories.  Variation in composition of disk gas and solids as well as the availability of polluting solids at different locations leaves an imprint on a planet's atmosphere \citep[e.g.][]{Oberg2011, Lambrechts2014, Venturini2016, Pudritz2018}.  By measuring the wavelength-dependent transit depth when one of these planets passes in front of its host star (the planet's ``transmission spectrum"), we can detect atmospheric absorption features that directly constrain the mean molecular weight and relative abundances of molecules including water, methane, carbon monoxide, and carbon dioxide.  Although some planets with strong and clear absorption features have been thus characterised (e.g. WASP 96b, \citealp{Nikolov2018a}; WASP 39b, \citealp{Wakeford2018}; WASP 107b, \citealp{Kreidberg2018}), large observing campaigns using the \emph{Hubble Space Telescope} (\emph{HST}) have revealed the presence of high-altitude clouds that attenuate the expected absorption signal in a majority of the close-in gas giant planets observed to date \citep[e.g.][]{Sing2016, Fu2017, Tsiaras2018, Wakeford2019}.  This problem is even more acute for the current sample of Neptune-sized planets, whose relatively small radii, high surface gravities, and low temperatures all combine to reduce the expected amplitude of atmospheric absorption as compared to their better-studied Jovian counterparts \citep[e.g.][]{Crossfield2017}.  This limits our ability to search for trends in atmospheric properties with other parameters of the system, e.g. planet mass, radius, and temperature --- all of which are crucial for improving our understanding of planet formation and evolution.

Although the current body of observed transmission spectra clearly require the presence of high altitude scattering particles, there is considerable debate about the nature and origin of these particles.  At high temperatures, we expect refractory species such as metal oxides, silicates, and sulphides to condense in exoplanetary atmospheres \citep[e.g.][]{Helling2018, Powell2018, Morley2012}.  However, cloud formation is a complex process that depends on both microphysical processes, such as sedimentation, nucleation, and growth, and the material properties of the condensing species, many of which are highly uncertain or unknown \citep{Helling2018}.  Consequently, the use of different underlying assumptions can lead to significantly different cloud properties, severely limiting the predictive power of these models.

While some of these questions may be resolved by ongoing laboratory experiments \citep{Johnson2017, Johnson2018, Horst2018, He2018, He2018a}, observational constraints on the properties of clouds in exoplanetary atmospheres provide complementary leverage to further refine and develop microphysical cloud models.  The nature of these constraints varies depending on the wavelength of the observations: optical and near-infrared transmission spectroscopy can be used to investigate the sizes, number density, and vertical distribution of cloud particles, while vibrational modes in the mid-infrared can be used to directly determine the compositions of cloud particles  \citep[e.g.][]{Wakeford2015, Pinhas2017, Kitzmann2018}.

Although clouds represent a substantial challenge for compositional inferences from transmission spectroscopy, previous \emph{HST} studies have demonstrated that we can nonetheless obtain reasonable constraints on atmospheric composition for planets with detectable near-infrared water features by utilizing information at optical wavelengths to break degeneracies between cloud-top pressure and atmospheric metallicity (\citealp[e.g. HAT-P-26b, ][]{Wakeford2017b}; \citealp[WASP-39b, ][]{Wakeford2018}).  Spectroscopic observations in the near infrared have been instrumental in the detection of molecular absorption in exoplanetary atmospheres but they are usually unable to put tight constraints on the composition, i.e. the absolute mixing ratios, of these molecules.  This is because the transmission spectra of an atmosphere with a deep cloud and low mixing ratios is statistically indistinguishable (with currently available precision) from an atmosphere with a high cloud and high mixing ratios.  These distinct scenarios can be distinguished by their differing spectral behavior in the optical.  In this spirit, the Panchromatic Comparative Exoplanet Treasury (PanCET) survey is a multi-cycle \emph{HST} treasury program whose primary goal is to characterize the atmospheres of a sample of 20 transiting gas giant planets at wavelengths ranging from the ultraviolet to the near-infrared \citep[e.g.][] {Evans2017, Wakeford2017, Evans2018, Nikolov2018, Alam2018, Bourrier2018}.  In this study we present new optical \emph{HST} STIS PanCET observations of HAT-P-11b, a warm Neptune sized planet with a radius of 4.4 Earth radii and mass of 23 Earth masses on a 4.88 days orbit around a 0.81 $M_{\odot}$, 0.68 $R_{\odot}$ K4 star ($ T_{\mathrm{eff}} = 4780 \pm 50$K).  This planet has a significantly eccentric orbit ($e = 0.218$) and as a result its predicted equilibrium temperature varies between $\sim 600 - 900$ K \citep{Bakos2010, Deming2011, Yee2018}. The planet therefore crosses multiple condensation lines, which enhances its potential for cloud formation.

HAT-P-11b has been previously observed with both ground- \citep[e.g.][]{Bakos2010, Sanchis-Ojeda2011} and space-based telescopes \citep[e.g.][]{Deming2011, Fraine2014, Huber2017} and is one of the most favorable Neptune-sized planets for atmospheric characterization due to its large atmospheric scale height and host star brightness ($V \sim 9$).  It is one of the smallest planets with a published detection of water absorption in its 1.1--1.7 $\mu$m \textit{HST} WFC3 transmission spectrum \citep[][hereafter \citetalias{Fraine2014}]{Fraine2014}.  Although there is an optical detection of the planet's secondary eclipse using \emph{Kepler} photometry \citep{Huber2017}, no corresponding infrared detection has been reported to date.  Measurements of absorption in the He metastable 10830 \AA~line during transit provide complementary constraints on the size of the planet's exosphere and corresponding mass loss rate \citep{Allart2018, Mansfield2018}.  Although the relatively high activity level of HAT-P-11b's K dwarf primary can bias the shape of the planet's measured transmission spectrum \citep[e.g.][]{Sing2011, Rackham2018, Rackham2019}, the planet's nearly polar orbit \citep{Winn2010, Hirano2011} has enabled exquisitely detailed studies of the starspot distribution and active latitudes \citep[e.g.][] {Sanchis-Ojeda2011, Deming2011, Morris2017, Morris2017a} that can be used to effectively correct for these effects.

Here, we combine previously published transit observations from \textit{HST} WFC3 ($0.8-1.7$~$\mu$m) and \textit{Spitzer} (3.6, 4.5 $\mu$m) \citep{Fraine2014, Mansfield2018} with new optical \textit{HST} STIS observations to obtain the first comprehensive look at HAT-P-11b's transmission spectrum between $0.35-5$~$\mu$m.  We compare the resulting transmission spectrum to predictions from forward models for cloud condensation and use retrievals to independently constrain the planet's atmopsheric composition and cloud properties.  Sections~\ref{sec:observations} and \ref{sec:extraction} describe our spectral and photometric extraction methods, while Section~\ref{sec:systematics} discusses instrumental and astrophysical noise sources in our data.  Section~\ref{sec:analysis} details our fits to these data, and Section~\ref{sec:forward_models} discusses predictions from forward models for HAT-P-11b's atmosphere.  Adopting some material properties and tools from this section, we then use simple models to directly fit the observed transmission spectrum in order to derive statistical constraints on atmospheric parameters in Section~\ref{sec:retrieval}, which we compare to the forward models in Section~\ref{sec:discussion}.

\section{Observations}
\label{sec:observations}
A summary of the observations used in our analysis is given in Table~\ref{table:observations}. We analyze 13 transits in total and describe each of them below.

We observed three transits of HAT-P-11b with the Space Telescope Imaging Spectrograph (STIS) on the \textit{Hubble Space Telescope} (\textit{HST}) (PI Sing \& L\'opez-Morales, GO 14767).  Two observations were conducted using the G430L grism (0.29-0.57 $\mu$m) on UT 2017 Feb 22 and UT 2017 May 26, while a third visit on UT 2017 April 12 used the G750L grism (0.524-1.027 $\mu$m).  All of our observations were obtained using the $52''\times2''$ slit.  This was done to minimize slit losses and the effect of telescope breathing.  Each visit consists of 5 \textit{HST} orbits.  Short (1~s) exposures were taken before each orbit to mitigate the severity of the exponential ramp at the beginning of each orbital light curve, but this step did not appear to be effective for these observations.  The wavelength calibration and flat field exposures were taken during the occultation of \textit{HST} by Earth during the last orbit.  Along with the \emph{HST} STIS data, we independently re-reduce and fit all of the prior data collected with \emph{HST} and \emph{Spitzer} as part of our updated global analysis, which we discuss below in chronological order.

HAT-P-11b was observed with \textit{HST}'s Wide Field Camera 3 (WFC3) instrument in 2012 (PI Deming, GO 12449) using the G141 grism in the 256$\times$256 sub-array mode, which provides a low resolution spectrum in the 1.1 - 1.7 $\mu$m wavelength range.  The data were collected over 4 \textit{HST} orbits using only forward scans \citep{Mccullough2012} with a scan rate of $0.3891''$ s$^{-1}$.  The second orbit covers part of ingress.  A buffer dump occurred during the third orbit, which partially resets the ramp that is used to model the instrumental behaviour \citep{Deming2013, Knutson2014a, Kreidberg2014}; see \S~\ref{sec:systematics} for more details.  These data were originally published in \citetalias{Fraine2014}.

Four transits of HAT-P-11b were observed in 2011 using the Infrared Array Camera (IRAC) mounted on \emph{Spitzer Space Telescope}, with two transits in each of the two warm-\emph{Spitzer} channels (3.6 and 4.5 $\mu$m).  The observations were taken in the sub-array mode, which yielded 32$\times$32 pixel images with an integration time of 0.4 s.  The \emph{Spitzer} data were published along with the WFC3 G141 data in \citetalias{Fraine2014}.

Finally, five transits of HAT-P-11b were also observed using the WFC3 G102 grism ($0.8-1.15$ $\mu$m) in the 256$\times$256 subarray mode (PI Bean, GO 14793) on UT 2016 Sep 14, 2016 Oct 13, 2016 Nov 7, 2016 Nov 26, and 2016 Dec 26.  This grism is complementary to the G141 observations, as both grisms together span a series of adjacent and overlapping water and methane bands.  During each visit, the planet was observed in scan mode over 4 orbits.  The use of forward and backward scans and longer exposure times for G102 observations yielded a higher observational efficiency ($\sim 75 $\%) than the 2012 G141 observations ($\sim 50$\%).  These data were published in \cite{Mansfield2018} (hereafter \citetalias{Mansfield2018}), which reported a strong helium absorption from escaping gas in the planet's outer atmosphere but did not see the expected molecular (water) absorption features in this bandpass.

\begin{table*}
	\centering
	\caption{Observations}
	\begin{tabular}{cccccccc}
		\hline \hline
		Date (UT) & Start Time & Duration & Observatory & Band pass & Integration & Exposures & Reference \\
		 & & & & ($\mu$m) & Time (s) & &  \\
		\hline
		2011 Jul 11 & 23:11:41   & 7.43 h & Spitzer & 3.05 - 3.95   & 0.4 & 62592 & \cite{Fraine2014}  \\
		2011 Aug 5 & 07:02:48   & 7.43 h & Spitzer & 4.05 - 4.95  & 0.4 & 58112 & \cite{Fraine2014}  \\
		2011 Aug 15 & 01:49:20   & 7.43 h  & Spitzer & 3.05 - 3.95 & 0.4 & 52633 & \cite{Fraine2014}  \\
		2011 Aug 29 & 17:37:18   & 7.43 h  & Spitzer & 4.05 - 4.95 & 0.4 & 62592 & \cite{Fraine2014}  \\
		2012 Oct 18 & 04:58:38   & 6.87 h & Hubble & 1.1 - 1.70   & 44.4 & 113 & \cite{Fraine2014}   \\
		2016 Sep 14 & 10:36:07   & 5.65 h & Hubble  & 0.8 - 1.15   & 81.1 & 99 & \cite{Mansfield2018}   \\
		2016 Oct 13 & 18:44:21   & 5.83 h & Hubble  & 0.8 - 1.15   & 81.1 & 99 & \cite{Mansfield2018}    \\
		2016 Nov 7 &  05:12:38   & 5.88 h & Hubble   & 0.8 - 1.15   & 81.1 & 99 & \cite{Mansfield2018}    \\
		2016 Nov 26 & 18:22:55   & 5.83 h & Hubble  & 0.8 - 1.15   & 81.1 & 99 & \cite{Mansfield2018}    \\
		2016 Dec 26 & 02:17:30   & 5.77 h & Hubble  & 0.8 - 1.15   & 81.1 & 99 & \cite{Mansfield2018}   \\
		2017 Feb 22 & 17:04:39   & 7.17 h & Hubble  & 0.29 - 0.57    & 140 & 82 & This work  \\
		2017 Apr 12 & 14:15:13  & 7.67 h & Hubble  & 0.524 - 1.027  & 140 & 81  & This work  \\
		2017 May 26 & 13:50:02  & 7.28 h & Hubble  & 0.29 - 0.57    & 140   & 81 & This work \\ \hline
	\end{tabular}
	\label{table:observations}
\end{table*}

\section{Spectral \& Photometric Extraction}
\label{sec:extraction}
We use the ExoTEP framework \citep{Benneke2019} for the extraction and fitting of all datasets.  The extraction process for each of the instruments is described below.

\subsection{HST STIS Spectroscopy}
We correct for cosmic ray hits and other transient phenomena by stacking all of the images from a given visit and examining flux as a function of time at each pixel position.  Because these data have relatively sparse time sampling ($<100$ images per visit) and time-correlated instrumental effects, we find that we obtain optimal results when we flag $4\sigma$ outliers in each pixel's time series and replace them with the median pixel value.  We then estimate the background in each image by taking the median pixel value in two rectangular regions located far enough from the spectral trace to avoid contamination.  We optimize the aperture width (in the cross-dispersion direction) for extraction of 1-dimension (1D) spectra and decide whether or not to remove the background by minimizing the scatter in the white-light residuals after subtracting the best-fit transit and instrumental noise model for each visit \citep[e.g.][]{Deming2013}.  For each visit, we consider aperture sizes of 7, 9, 11, and 13 pixels. In the G750L visit we obtain optimal results when we use a 9 pixel wide aperture centered on the peak of the point spread function and do not subtract the background.  For the G430L observations, we prefer to subtract the background and utilize 13 and 11 pixel wide apertures for the first and second visits, respectively.  We find that in all visits the white-light transit depths and transmission spectral shapes are relatively insensitive to our choice of aperture width.

Data taken with the G750L grism exhibit a fringing effect due to internal reflection within individual pixels.  We correct for this effect using a fringe flat field obtained contemporaneously with our data following the methods outlined in \cite{Nikolov2014, Nikolov2015} and \cite{Sing2016}.  Using the first frame as a template, we then fit for the shift in position in the dispersion direction and relative amplitude of all subsequent frames in order to align the frames in wavelength.  These best-fit relative amplitudes give us the normalised white light curve for each visit.  For the wavelength-dependent light curves, we sum the flux within a 200 pixel wide bin for the G750L grism and a 100 pixel wide bin for the G430L grism.  We also check for the presence of sodium and potassium absorption in the G750L bandpass by extracting the flux in two narrow bandpasses centered on the corresponding absorption lines (588.7--591.2 and 770.3--772.3 nm respectively).

\subsection{HST WFC3 Spectroscopy}
We reduce data from both the G102 and G141 grisms following the method outlined in \cite{Tsiaras2016}.  Unlike that study, we begin with the bias- and dark-corrected \texttt{ima} images produced by the standard \textit{calwfc3} pipeline rather than calibrating the \texttt{raw} images ourselves.  Each of the exposures consists of 5 non-destructive reads.  We create difference sub-exposures by subtracting consecutive reads \citep[e.g.][]{Deming2013, Kreidberg2014, Evans2016}.  We determine the extent of the sub-exposure in the scan direction by finding the rows where the median flux profile in the spatial scan direction falls to 20\% of the peak flux and add an additional buffer of 15 pixels above and below these rows.  The extraction is not very sensitive to the number of pixels used for this buffer and any value between 10 and 20 suffices.  We then mask out the rows exterior to this $y$ pixel range and estimate the background using a 20 column wide rectangular region within the sub-exposure spanning columns between the end of the spectral trace and the edge of the array, taking care to avoid any secondary sources in the image.  We remove any bad pixels by discarding $3\sigma$ outliers from this background region and then subtract the median of the remaining pixels from the unmasked part of the image.  We then create a combined full frame image by co-adding all of the background subtracted sub-exposures.

Although the pointing of WFC3 is generally very stable, our scanned observations nonetheless exhibit small image to image variations in the position of the spectral trace in the $x$ (dispersion) direction.  
By default, we estimate the magnitude of these shifts relative to the first frame by summing each image in the $y$ direction and using this rough 1D-extracted spectrum to calculate the corresponding $x$ offset.  We find that the magnitude of this shift is less than 0.1 pixel over the entire duration of the WFC3 G102 visits.  The WFC3 G141 data were taken shortly after the spatial scanning mode was first implemented on \emph{HST} and exhibit a larger shift of approximately one pixel over the visit, most likely due to the sub-optimal scanning strategy utilized in these older observations. We find that using the centroid of each exposure and determining the horizontal offset relative to the centroid of the first exposure significantly decreases the scatter in the best-fit residuals for the G141 visit.  
We then use the wavelength and trace calibration functions provided by STScI \citep{Kuntschner2009, Kuntschner2009a} for each grism to calculate the full 2D wavelength solution for each image.

We flat-field all frames using the calibration files provided by STScI \citep{Kuntschner2011} following the method outlined in \cite{Wilkins2014} and identify bad pixels in each individual image using a 6$\sigma$ moving median filter in both the $x$ and $y$ directions.  Although we also consider lower filter thresholds, we find that these result in overly aggressive outlier correction.  We replace these outliers with the mean value within the moving filter and repeat the same filtering a second time to ensure that we have identified and removed all outliers.

The width of the spectral trace in the dispersion direction varies with the $y$-position of the star on the detector.  As a result, lines of constant wavelength are slanted relative to the columns of the detector.  For the wavelength dependent light curve extraction, we follow the method outlined in \cite{Tsiaras2016} and use the wavelength solution to determine the boundaries of each slanted wavelength bin and sum the flux within each bin.  When the bins intersect with pixels, we use a second-order 2D polynomial to interpolate and integrate the flux over each partial-pixel region.  This procedure ensures flux conservation and leads to a small reduction in the photometric scatter relative to other commonly employed methods, which usually smooth the data in the dispersion direction before light curve extraction \citep[e.g.][]{Deming2013, Fraine2014, Knutson2014a}.

For the wavelength dependent light curves obtained with the G141 grism, we use 30 nm wide bins spanning the wavelength range 1.1-1.7~$\mu$m.  \citetalias{Fraine2014} utilized narrower wavelength bins, but also convolved their 1D spectra with a 4 pixel wide Gaussian filter prior to binning.  Since we do not smooth our data, we adopt a lower wavelength resolution.  For the G102 data, we utilize bins with a width of $\sim 23.3$ nm spanning the wavelength range 0.8-1.15$ \mu$m, identical to those adopted by \citetalias{Mansfield2018}.  The white light curve is simply obtained by summing the flux from all the spectroscopic light curves.

\subsection{Spitzer 3.6 and 4.5 $\mu$m Photometry}
We extract the photometric light curve for each \textit{Spitzer} visit following the method described in \cite{Knutson2012}, \cite{Wong2016}, and \cite{Zhang2018}.  We determine the star's position in each $32\times32$ pixel \textit{Spitzer} subarray image by iteratively calculating the flux-weighted centroid within a circular aperture with a radius of 3 pixels.  To estimate the sky background, we first mask pixels located within a 12 pixel radius of the star's position and then iteratively trim 3$\sigma$ outliers \citep[e.g.][]{Knutson2012}.  We calculate the mean value of the remaining background pixels using the biweight location method \citep{Collaboration2013, Collaboration2018} and subtract it from each image.  We then use the \texttt{photutils} package \citep{Bradley2018} to extract the photometry using circular apertures with radii ranging from 1.5 to 3 pixels in 0.1 pixel increments and $3-5$ pixels in 0.5 pixel increments.  We select the optimal aperture for each visit by minimizing the scatter in the best-fit residuals, which are binned in 60 s intervals (see \S\ref{spitz_instr} for more information).  This procedure gives extraction apertures of 2.8 and 2.3 pixels for first and second transit in the 3.6 $\mu$m channel respectively, and 2.3 and 2.6 pixels for the first and second transit in the 4.5 $\mu$m channel respectively.

We iteratively trim outliers in the resulting timeseries using a 50 point moving median filter and discarding photometric points that lie more than 3$\sigma$ away.  We also fit 3rd order polynomials to the star's $x$ and $y$ positions and discard any photometric points more than $3 \sigma$ away from the polynomial model position during the observation, as these points are not well-corrected by our instrumental noise model.  The number of points removed in each of these steps ranges between $0.09-0.97\%$ for each individual visit and is commensurate with expectations for normally distributed data.

\section{Systematics and Astrophysical Models}
\label{sec:systematics}
\subsection{HST/STIS Instrumental Model}
We remove the first orbit in each of the STIS datasets as the instrumental systematics are notably worse than they are in subsequent orbits.  This difference is attributed to the thermal relaxation of \textit{HST} following target acquisition due to the change in incidence angle of solar radiation.  In addition, we remove the first exposure within each orbit as it has a much lower flux that is not well-matched by our parametric model.  Both of these steps are standard practice for STIS datasets \citep[e.g.][]{Sing2011, Nikolov2015, Wakeford2017b}.  For the instrumental systematics model, we use a fourth order polynomial in orbital phase and a linear trend in time \citep{Sing2008}.  We also fit for a linear trend in the $x$ (dispersion) position of the star on the array for the G750L visit and the first G430L visit as it significantly reduces the Bayesian Information Criterion (BIC: change of 35 and 8, respectively) and lowers the residual scatter in our light curve fits from $1.51$ and $1.6$ times the photon noise limit to $1.26$ and $1.53$ times, respectively.  

As discussed in \cite{Sing2019}, we find that the scatter in our white-light residuals is further reduced if we decorrelate against additional parameters related to variations in telescope pointing.  We find that the white-light residuals from our initial fit exhibit a strong correlation with the recorded RA and Dec, V2 and V3 roll, and latitude and longitude from the image file headers.  However, these parameters are highly correlated with each other and we therefore use Principal Component Analysis (PCA) to reduce the number of independent fit parameters.  We start with 6 principal components and retain those that capture $\geq 95 \%$ of the systematic variation in the light curves. Using this criterion, we retain 4 and 3 parameters for the first and second visit in the G430L bandpass respectively and 3 parameters in the G750L bandpass.  We include linear contributions from these PCA parameters as part of our final systematics model.  The addition of these linear jitter parameters has a negligible effect on the BIC ($\mid$BIC$\mid <2$) for all three visits but it reduces the scatter in our residuals by $5 - 8$\%. The full systematics model $S(t)$ is given as

\begin{multline}
S (t) = c + vt_v + mx + \sum_i j_i \; p_{\mathrm{jitter}} + \sum_{k = 1}^{4} p_k  t_{orb}^k,
\end{multline}
where $t_v$ is the time from the beginning of the visit, $t_{orb}$ is the time from the beginning of an orbit, $p_{\mathrm{jitter}}$ are the PCA vectors that describe the telescope pointing jitter, and $c$, $v$, $m$, $j_i$ and $p_i$ are free parameters in the fit.

\subsection{HST/WFC3 Instrumental Model}
\subsubsection{G141 grism}
\citetalias{Fraine2014} used the spectral template fitting method to derive wavelength-dependent transit depths for the WFC3 data.  Here, we fit the timeseries for each individual spectroscopic light curve independently following the method described in \cite{Tsiaras2016}.  As with the STIS data, we trim the first orbit and the first exposure of each orbit, as they are not well-matched by our instrumental noise model.

Although there is an alternative physically motivated model that would in theory allow us to fit these data \citep{Zhou2017}, we do not expect that this would improve the precision of our transit depth measurement as we already have an out-of-transit baseline that is comparable in duration to our in-transit data.  We fit the remaining orbits using a linear function of time and an exponential function of orbital phase, which is needed in order to correct for charge-trapping in the array \citep[e.g.][]{Deming2013, Zhou2017}.

Our WFC3 systematics model $S (t)$ is 
\begin{equation}
S (t) = (c + vt_v) +  \left( 1-e^{-at_{orb}-b-D_t} \right),
\label{eq:wfc3_systematics}
\end{equation}
where $c$, $v$, $a$, and $b$ are free parameters in the fit, $t_v$ is the time from the beginning of the visit, $t_{orb}$ is the time from the beginning of an orbit, and $D_t$ is a vector (same length as $t_v$) that is used to add duration-specific non-zero phase offsets.  We use it to model the partial reset of the exponential ramp after a mid-orbit buffer dump in the third orbit (free parameter $e$) and to account for the slightly different ramp amplitude of the first fitted orbit \citep[free parameter $d$, see][]{Kreidberg2015a}.  $c$ and $v$ characterize the linear dependence of systematic noise on time.  For the exponential dependence, $a$ controls the dependence on $t_{orb}$, and $b$ sets the overall time-independent amplitude of the exponential term.

\subsubsection{G102 grism}
Unlike the G141 data, which only scanned in a single direction, the G102 observations were taken with an alternating scan direction.  The behavior of the ramp is slightly different for each scan direction, likely due to small offsets in the relative position of the scanned spectrum on the array.  We carry out an initial fit in which we allow the full exponential ramp model to vary independently for each of the scan directions and find that all parameters except the constant $c$ in Equation~\ref{eq:wfc3_systematics} are consistent with a single common value.  We therefore carry out our final fits assuming the same slope $v$ and exponential ramp coefficients $a, b,$ and $D_t$ for both scan directions, but assign the forward and backward directions separate constant terms $c_f$ and $c_b$ that are allowed to vary independently.

\subsection{Spitzer Instrumental Model}\label{spitz_instr}
\textit{Spitzer} 3.6 and 4.5 $\mu$m photometry exhibits a ramp-like behavior \citep[e.g.][]{Lewis2013, Wong2016, Zhang2018} at the start of each new observation.  Rather than fitting this ramp with a model, we simply trim the first $0.5-2$ hours of data and find that the optimum trim duration for each visit that minimizes the scatter in our binned best-fit residuals is $1$ hour.  Even after truncation, we find that the second visit in the 3.6 $\mu$m bandpass possesses a significant ramp.  Fitting this visit with the standard systematics model we adopt (see Equation~\ref{eq:Spitzer_systematics} below) yields a much shallower transit depth and larger BIC ($\Delta$ BIC $\sim$ 20) compared to the values we obtain when we fit for this ramp.  We do not use the ramp model for the other Spitzer visits because it changes the transit depths by $\lesssim 1 \sigma$ and increases the BIC.  Prior to fitting we bin the data in 60~s intervals.  This binning results in a lower level of time-correlated noise in our best-fit residuals while still resolving the transit ingress and egress (for a discussion of binning practices with \textit{Spitzer} data see \citealp{Deming2015} and \citealp{Kammer2015}).  

The primary instrumental noise source in the 3.6 and 4.5 $\mu$m \textit{Spitzer} arrays is intra-pixel sensitivity variations combined with telescope pointing jitter. We model this behavior using Pixel-Level Decorrelation (PLD) following \cite{Deming2015}:

\begin{equation}
S (t) = 1 + vt_v + \sum_{i = 1}^{9} w_i P_i (t),
\label{eq:Spitzer_systematics}
\end{equation}
where $t_v$ is the time from the beginning of the visit, $P_i$ is the normalized pixel count in the 3$\times$3 array around the source, and $w_i$ are the weights assigned to each of these arrays, which are determined using linear regression after dividing out the transit light curve at each step in the fit.  The slope parameter $v$ is left to vary as a free parameter.  For the second visit in the 3.6 $\mu$m bandpass, we have an additional ramp term in the model with an amplitude $A$ and decay timescale $\tau$: $A e^{-t_v/\tau}$.

\subsection{Transit Model}

\begin{figure} 
	\centering
	\includegraphics[width=\linewidth]{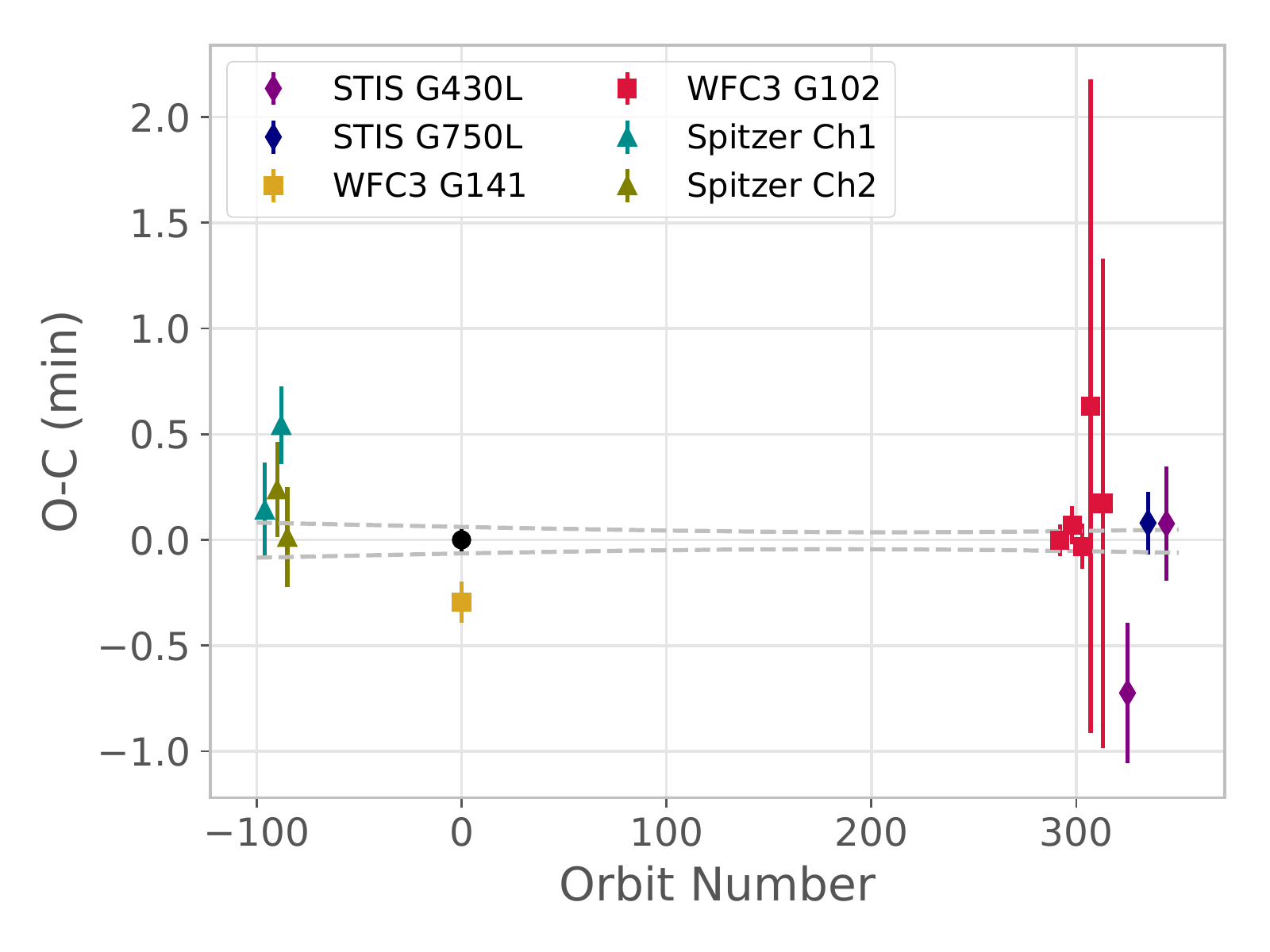}
	\caption{Observed minus calculated mid-transit times from fits to individual visits, where the color indicates the instrument.  Predicted transit times are calculated using the best-fit ephemeris from the global fit, with $1\sigma$ uncertainties indicated by the dashed grey lines. Visits with minimal data during ingress or egress have significantly larger uncertainties.}
	\label{fig:O_C}
\end{figure}

We use the \texttt{BATMAN} package \citep{Kreidberg2015} to model the transit light curve.  The astrophysical model depends on the planet-star radius ratio $ R_p / R_*$, planet semi-major axis to stellar radius ratio $a / R_*$, impact parameter $b$, period $P$, and transit center time $T_c$.  We fit for all of these parameters in our global fit, but use fixed values for $P$, $a / R_*$, and $b$ when fitting individual transits.  We fix the orbital eccentricity $e = 0.218$ and longitude of periastron $w = 199^o$ to the best-fit values from \cite{Yee2018}.  We validate our assumption of a linear ephemeris by comparing the best-fit mid-transit times from individual visits with the best-fit ephemeris from our global fit in Figure~\ref{fig:O_C}.  The best-fit mid-transit times for all visits are consistent with a linear ephemeris at the $2\sigma$ level or better.  

Our updated ephemeris is consistent with the values reported in \cite{Deming2011} and \cite{Southworth2011} to within $0.2\sigma$.  However, there is only moderate agreement with the values reported in \cite{Sanchis-Ojeda2011} and \cite{Huber2017}.  Curiously enough, the values reported by \cite{Huber2017}, \cite{Sanchis-Ojeda2011}, and \cite{Southworth2011} are for the same epoch and they disagree at the 10 $\sigma$ level.  We suspect this is due to errors in reporting of the mid-transit time in the stated time convention.  For example, \cite{Southworth2011} and \cite{Sanchis-Ojeda2011} report almost identical values for the mid transit time but the former report it in BJD UTC while the latter do so in BJD TDB.  These two time conventions differ by 66.184 s (an additional leap second was added in the first month of \emph{Kepler}'s quarter 14).  Similarly, the value reported by \cite{Huber2017}, supposedly in BJD UTC, match that of \cite{Southworth2011} converted to BJD TDB.  Careful accounting of these errors might resolve the paradoxes posed by these differing mid-transit times.

\begin{figure*}
	\centering
	\includegraphics[width=\linewidth]{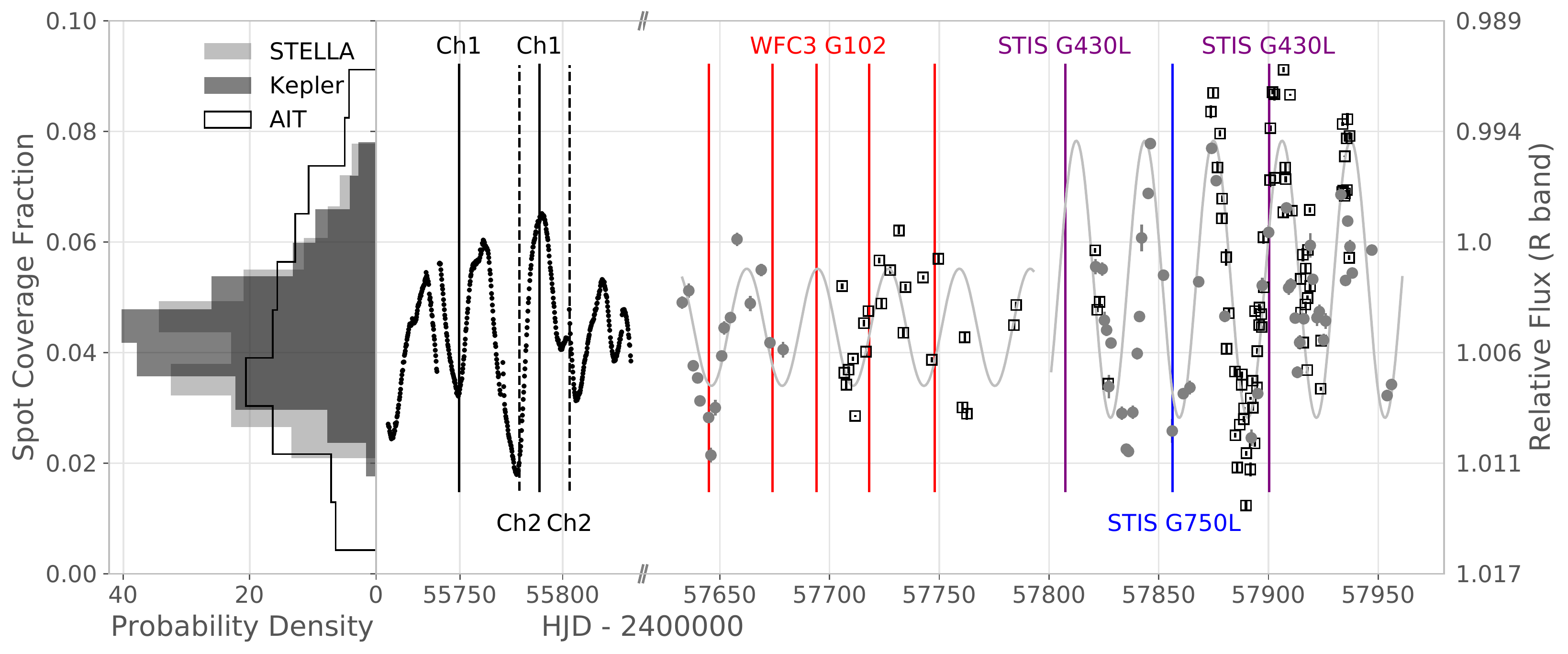}
	\caption{Change in HAT-P-11's $R$ band flux and corresponding spot coverage fraction in 2011 and from late 2015 to early 2017.  Points are calculated using photometric monitoring data obtained in the Cousins $R$ band pass using the AIT telescope at Fairborn Observatory, in the Johnson $B$ and $V$ bands using the STELLA telescope at Iza\~na Observatory, and from the \emph{Kepler} telescope.  We assume that the relative flux baseline for all three telescopes corresponds to a spot coverage fraction of 4.4\%, and use a photospheric temperature of 4780 K and spot temperature of 4500 K to convert these observations to the equivalent $R$ band fluxes.  Visit times for \emph{Spitzer} 3.6 $\mu$m (black) and 4.5 $\mu$m channels (black dashed), \emph{HST} WFC3 G102 (red), STIS G430L (purple), and STIS G750L (blue) observations are indicated by vertical lines.  The grey curves are sinusoidal functions that best match the observed variability at different epochs and are used to infer spot coverage fractions for \emph{HST} visits that do not have contemporaneous ground-based monitoring.}
	\label{fig:stellar_var}
\end{figure*}

As part of ExoTEP, we employ the Python package \texttt{LDTk} \citep{Parviainen2015} to calculate limb darkening coefficients for all of our observations except the \textit{Spitzer} transits.  \texttt{LDTk} queries spectral intensity profiles from the \texttt{PHOENIX} library \citep{Husser2013} and computes a mean limb darkening profile for a star given its effective temperature, surface gravity, and metallicity (and associated uncertainties).  We then fit this profile with a 4-parameter non-linear limb darkening model, and we fix the limb darkening coefficients to the model values in our light curve fits.  \texttt{PHOENIX} profiles extend from 50 nm to 2600 nm in wavelength space and can therefore only supply limb darkening coefficients for the \textit{HST} bandpasses.  For the \textit{Spitzer} bandpasses, we use the (4-parameter non-linear) limb darkening coefficients tabulated by (\citealp{Sing2010}, assuming $T_{\mathrm{eff}} =$ 4750 K, log g = 4.5, [Fe/H] = 0.3), which are calculated from \texttt{ATLAS} models.  We investigate the importance of our choice of limb-darkening models in the \textit{Spitzer} bands by re-fitting the \textit{Spitzer} light curves with quadratic limb darkening coefficients as free parameters.  We obtain transit depths that agree to within $0.5\sigma$ with those obtained with \texttt{ATLAS} limb darkening coefficients.  We therefore conclude that our use of \texttt{ATLAS} models instead of \texttt{PHOENIX} models at 3.6 and 4.5 $\mu$m has a negligible effect on our results.

\subsection{Stellar Activity}
\label{sec:stellar_activity}

HAT-P-11 is a relatively active K dwarf with a Ca II H \& K emission line strength of log($R_{\mathrm{HK}}^{'}$) $=-4.57$ \citep{Knutson2010}, and it is therefore important to address the impact of its activity on the transmission spectrum \citep{Mccullough2014, Morris2017, Morris2017a}.  Both occulted and unocculted spots introduce wavelength dependent biases in the transmission spectrum \citep[e.g.][]{Pont2008, Sing2011, Rackham2018}.  These biases must be corrected to combine transit depth measurements from different epochs and different wavelength bandpasses.

We find no evidence for any spot crossings during the \textit{HST} observations included in this analysis, with the exception of two G102 visits.  Following \citetalias{Mansfield2018}, we simply trim the data associated with the spot occultation rather than including this effect in our models.  While two of the four \textit{Spitzer} transits with contemporaneous \emph{Kepler} transit photometry included a spot occultation, this occultation was evident only in the \emph{Kepler} light curve.  Given the relatively small chromatic effect of spot crossing at infrared wavelengths, \citetalias{Fraine2014} concluded that these spots would have had a negligible effect on the measured \textit{Spitzer} transit depths.

\begin{figure*}
    \centering
    \includegraphics[width=0.45\linewidth]{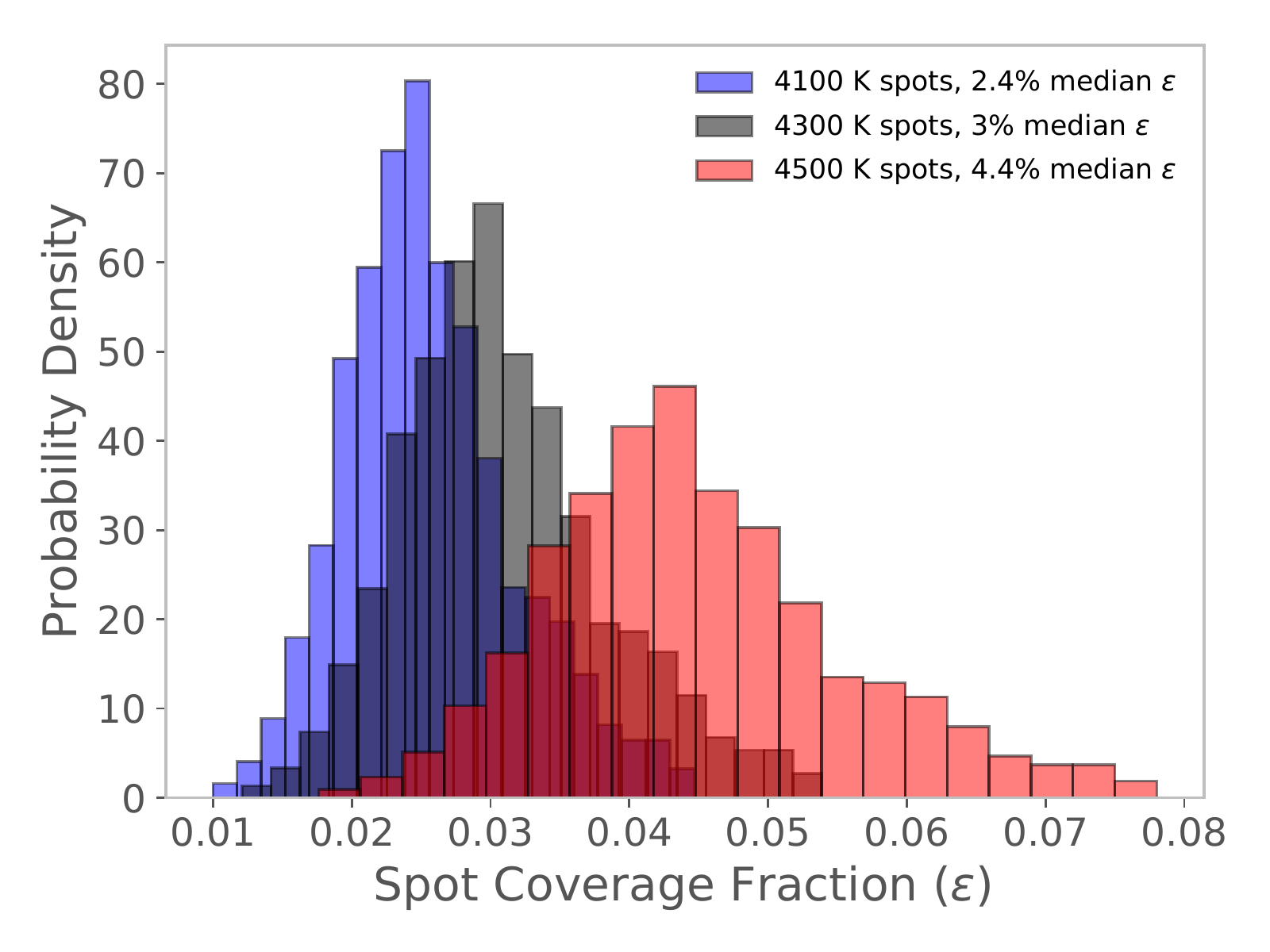}
    \includegraphics[width=0.45\linewidth]{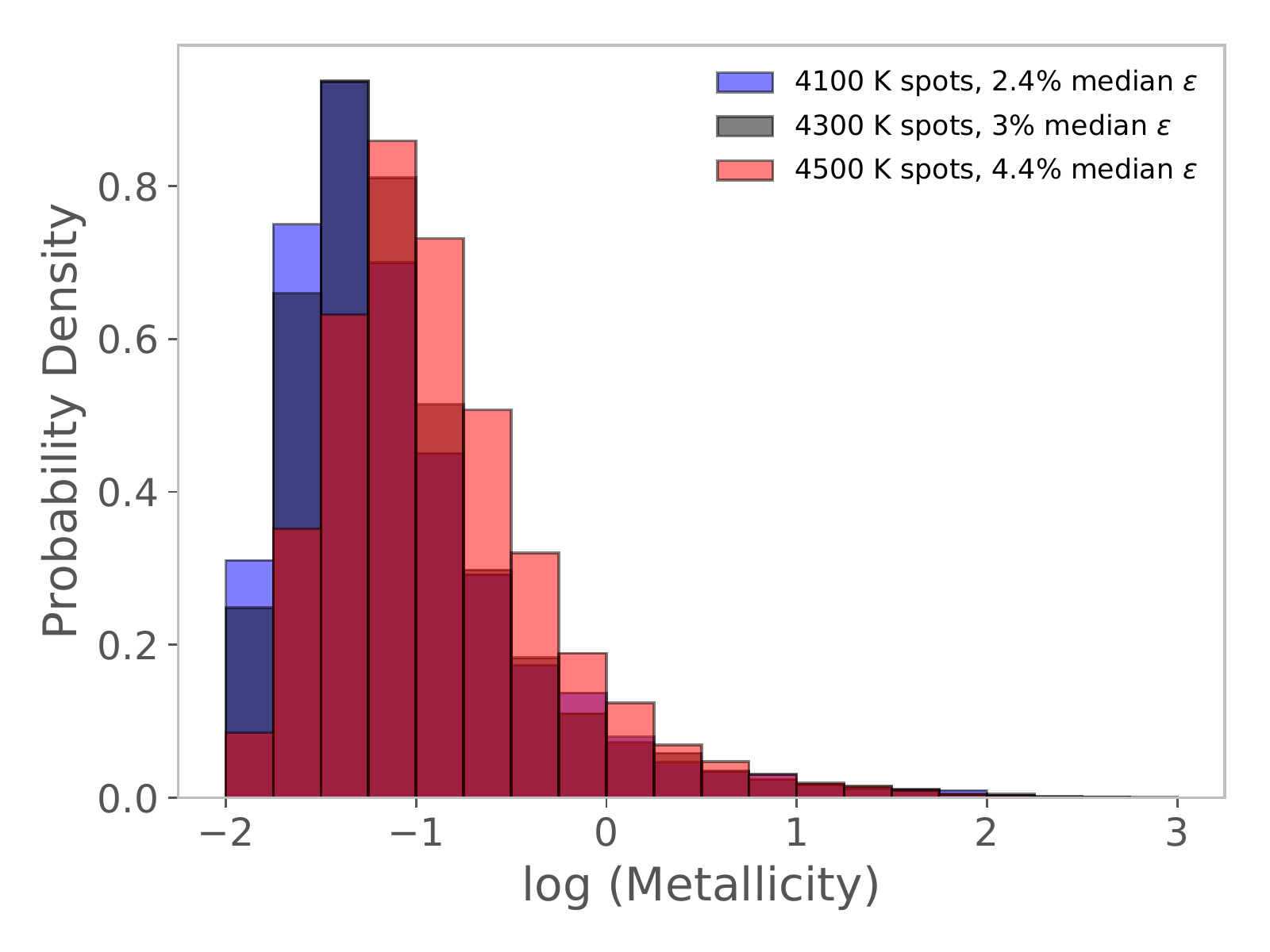}
    \caption{We vary star spot temperatures and spot coverage fractions such that they produce the same absolute correction in the \emph{Kepler} bandpass.  The spot coverage fractions in the left panel are deduced from \emph{Kepler} long cadence photometry.  We fit light curves for these different stellar spot properties and quantify their effect on the retrieved atmospheric metallicity.  In the right panel, we show that the posterior for metallicity is relatively insensitive to our choice of spot temperature.  We adopt a value of 4500 K in the rest of this study following \cite{Morris2017}.}
    \label{fig:spot_prop_variation}
\end{figure*}

Unocculted spots are usually much harder to correct for as accounting for their effect requires knowledge of the fractional surface area of the star that is covered by the spots as well as the average spot temperature.  Fortunately, HAT-P-11 has some of the best constraints on spot properties amongst all stars that host transiting planets.  This is because HAT-P-11b orbits its star from pole to pole \citep[its orbit is misaligned with the stellar spin axis by 106 degrees;][]{Sanchis-Ojeda2011, Deming2011} and the star was monitored by \textit{Kepler} in a broad optical bandpass from $2009-2012$, allowing us to observe more than 200 transits of the planet.  This essentially provides us with a latitude-longitude map of the entire stellar surface and constrains the spot covering fraction of the stellar surface to be $3^{+6}_{-1} \%$ \citep{Morris2017, Morris2017a}.

The \emph{Kepler} data span the epoch of the \textit{Spitzer} transit observations and although the G141 observations were taken in 2012, they unfortunately coincided with a gap in the \emph{Kepler} coverage \citepalias{Fraine2014}.  We also obtained photometric monitoring data in the Cousins R band pass with the Celestron 14-inch (C14) Automated Imaging Telescope (AIT) at Fairborn Observatory \citep{Sing2015} and in the Johnson B and V filters from the 1.2 m robotic STELLA telescope at Iza\~na Observatory (\citealp[]{Strassmeier2004}; data taken from \citetalias{Mansfield2018}).  These data were obtained between $2015-2017$, covering the epochs of the WFC3 G102 and STIS observations but not the 2012 WFC3 G141 observations.  This introduces a source of uncertainty, as there is no uniform source of monitoring data spanning the epochs of all of the datasets included in our global analysis.

We use the \emph{Kepler} and ground-based photometric monitoring data to estimate the spot coverage fraction during the \emph{Spitzer}, \emph{HST} WFC3 G102, and \emph{HST} STIS observation epochs.  We assume that the baseline of the relative flux from each telescope corresponds to a median spot coverage fraction $\bar{\epsilon}$ and calculate the absolute values of $\epsilon$ for all the other relative flux values.  We account for the difference in the telescope bandpasses while calculating the spot coverage fraction.  In Figure~\ref{fig:stellar_var}, we show the photometric data, relative flux in R band, and the corresponding spot coverage fraction from \emph{Kepler}, STELLA, and the AIT for a median spot coverage fraction $\bar{\epsilon}$ of 4.4\% and average spot temperature of 4500 K.  Histograms for the inferred spot coverage fraction from the photometric data are consistent with each other and with the $3^{+6}_{-1} \%$ estimate obtained by \cite{Morris2017}.  We find that during the STIS observations, the stellar variability is best matched by a sine curve with a period of 30 days and peak-to-peak relative flux of about 1.5\%.  The star appears to have been somewhat less active and variable during the epoch of the WFC3 G102 observations with peak-to-peak relative flux of 0.7\% and a period of 33 days.  These observations imply that there is almost a $1-2$\% difference in the relative transit depth between epochs due to changes in stellar brightness.  These periods and variability are also in good agreement with inferences from \emph{Kepler}. 

The spot coverage fraction $\epsilon$, stellar photospheric temperature, and spot temperature determine the ratio of the observed ($D_{\lambda, obs}$) to true ($D_{\lambda}$) transit depths \citep{Rackham2018}:

\begin{equation}
D_{\lambda, obs} = \frac{D_{\lambda}}{1 - \epsilon \left(1 - F_{\lambda,spots} / F_{\lambda,star} \right) },
\label{eq:stellar_activity}
\end{equation}
where $F_{\lambda,spots}$ and $F_{\lambda,star}$ are the stellar intensity profiles corresponding to the temperature of the spots and the unspotted stellar photosphere respectively.  We apply this correction by re-scaling the model transit light curves at each step in our fits by the denominator in Equation~\ref{eq:stellar_activity}.  We do not include faculae in our model because they produce a distinct spectral signature in the optical region of the transmission spectrum \citep[e.g.][]{Zhang2018a}, and we observe no such effect in our three \textit{HST} STIS visits (see \S~\ref{sec:analysis}).

To model the star spots and the surface fluxes, we use BT-NextGen (AGSS2009) stellar models \citep{Allard2012} and fix the photospheric temperature to 4780 K.  The brightness contrasts estimated from spot crossings in the \emph{Kepler} light curves give a range for spot temperatures.  We explore the effect of changing median spot coverage fraction $\bar{\epsilon}$ and spot temperature on the retrieved atmospheric metallicity of the planet.  We choose combinations of spot temperatures and $\bar{\epsilon}$ such that the absolute corrections to the transit depths in the \emph{Kepler} bandpass are identical.  Spot temperatures of 4100 K, 4300 K, and 4500 K are thus combined with $\bar{\epsilon}$ of 2.4\%, 3\%, and 4.4\% respectively.  Figure~\ref{fig:spot_prop_variation} shows histograms for spot coverage fractions for the range of variability observed in the \emph{Kepler} light curves and the corresponding atmospheric metallicity constraints for HAT-P-11b obtained from retrievals.  We find that the metallicity posterior is relatively insensitive to our choice of spot temperature.  Following the more detailed stellar activity study of HAT-P-11 conducted by \cite{Morris2017} and spot temperature characterisation by \citetalias{Mansfield2018}, we choose to adopt a spot temperature of 4500 K in the rest of this study.

For the \emph{HST} WFC3 G141 data, we assume a fixed spot coverage fraction of 4.4\% as this visit is not covered by any photometric observation.  For the \emph{Spitzer}, WFC3 G102, and STIS visits, we apply a visit-specific correction.  We fit periodic curves to the spot coverage fraction to determine its value for the third G102 visit and the first G430L visit as ground based data at these epochs are scarce.  For the other WFC3 G102 and STIS visits, we use the closest observation to obtain an estimate of the spot coverage fraction, if the next closest observation is more than 0.5 days away (i.e., on a different night). Otherwise, we use the average of the two nearest observations.

\begin{table*}
	\centering
	\begin{threeparttable}
    	\caption{Global Broadband Light Curve Fit Results\textsuperscript{a}} \label{table:wlc_fits}
    	\begin{center}
    		\renewcommand{\arraystretch}{1.2}
            \begin{tabular}{ c c c c }
                \hline \hline
                Parameter & Instrument & Band pass ($\mu$m) & Value \\ \hline 
                Planet radius, $R_p/R_*$ & STIS G430L & 0.29 - 0.57 & $0.05806_{-0.00028}^{+0.00036}$     \\
                Planet radius, $R_p/R_*$ & STIS G750L & 0.524 - 1.027 & $0.05783_{-0.00035}^{+0.00034}$     \\
                Planet radius, $R_p/R_*$ & WFC3 G102 & 0.8 - 1.15 & $0.05788_{-0.00011}^{+0.00016}$   \\			
                Planet radius, $R_p/R_*$ & WFC3 G141 & 1.1 - 1.7 &   $0.05847_{-0.00015}^{+0.00016}$       \\		
                Planet radius, $R_p/R_*$ & IRAC Channel 1 & 3.16 - 3.93   & $0.05778_{-0.00026}^{+0.00024}$  \\
                Planet radius, $R_p/R_*$ & IRAC Channel 2 & 3.97 - 5.02  & $0.05811_{-0.00027}^{+0.00028}$  \\ 
                Transit center time $T_c$ (BJD$_{\mathrm{TDB}}$) & -- & -- & $2456218.866182_{-0.000044}^{+0.000042}$ \\
                Period $P$ (days) & -- & -- & $4.88780228_{-0.00000018}^{+0.00000016}$ \\
                Impact parameter $b$ & -- & -- & $0.135_{-0.078}^{+0.064}$ \\
                Relative semi-major axis $a/R_*$ & -- & -- & $17.46_{-0.20}^{+0.14}$ \\
                Inclination\textsuperscript{b} $i$ & -- & -- & $89.56_{-0.22}^{+0.26}$ \\
                \hline
            \end{tabular}  
            \begin{tablenotes}
             \small
			 \item {\bf Notes.} 
			 \item \textsuperscript{a}{The $R_p/R_*$ values reported here have been corrected for unocculted spots assuming a photosphere temperature of 4780 K, spot temperature of 4500 K, and spot covering fraction of 4.4\%.}
			 \item \textsuperscript{b}Calculated from posteriors for $b$ and $a/R_*$.
			\end{tablenotes}
    	\end{center}
	\end{threeparttable}
\end{table*}

\begin{figure*}
	\centering
	\includegraphics[width=0.8\textwidth]{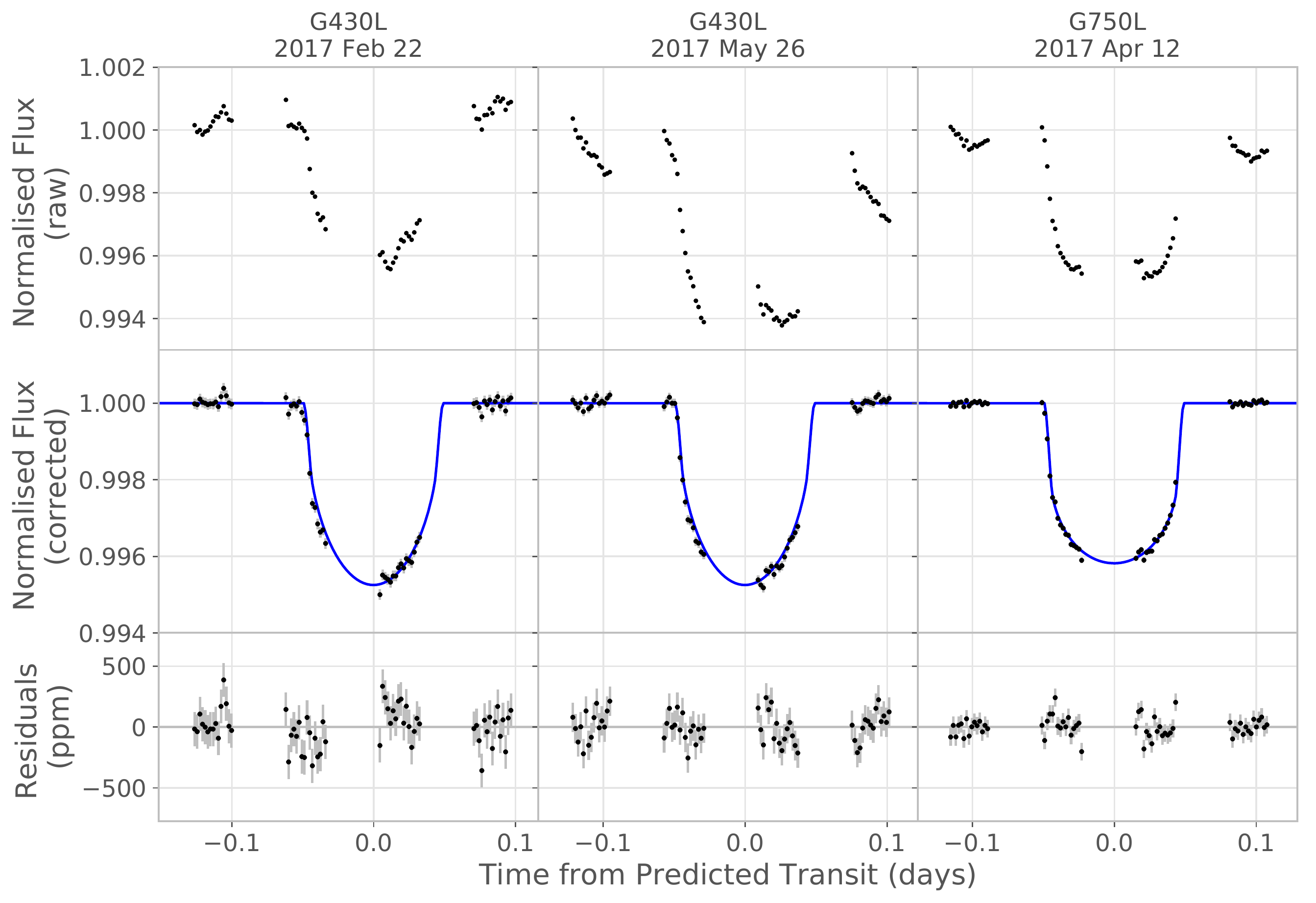}
	\caption{STIS white light transit light curves before (top) and after (middle) dividing out the best-fit instrumental systematics model.  The best-fit transit light curve is shown in blue for comparison, and the fit residuals are shown at the bottom.}
	\label{fig:STIS_WLC_fits}
\end{figure*}

\section{Analysis}
\label{sec:analysis}
The log-likelihood $\mathcal{L}$ (logarithm of the posterior probability) of our astrophysical transit model $M$ and systematics model $S$ given data $D$ with uncertainty $\sigma$ is

\begin{equation}
    \mathcal{L} = \sum_{i=1}^{n} \left [ \left( \frac{D_i - (M_i \times S_i)}{2 \sigma_i} \right)^2 + \mathrm{ln} (\sqrt{2 \pi} \sigma_i) \right].
\end{equation}

We use the Markov Chain Monte Carlo (MCMC) method to fit the white-light timeseries for each visit individually and then carry out a joint fit where the same transit shape and ephemeris parameters are used for all datasets, while the planet-star radius ratio is allowed to vary across different bandpasses.  In all cases we fit an independent instrumental systematics model for each individual transit.  We carry out our fits using the \texttt{emcee} package, which is an affine-invariant ensemble sampler \citep{Foreman-Mackey2012}.

\begin{figure*}
	\centering
	\includegraphics[width=0.8\textwidth]{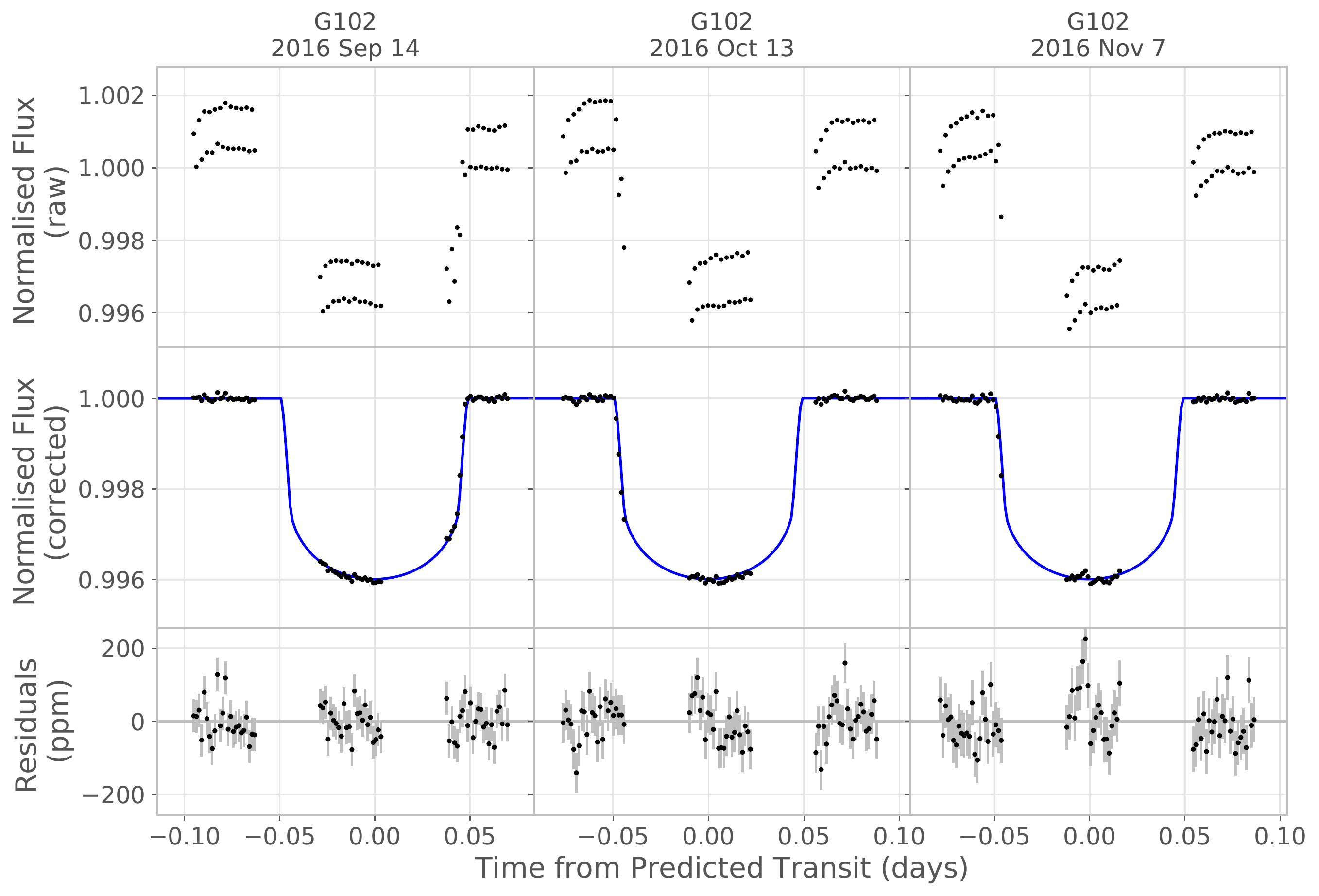}
	\includegraphics[width=0.8\textwidth]{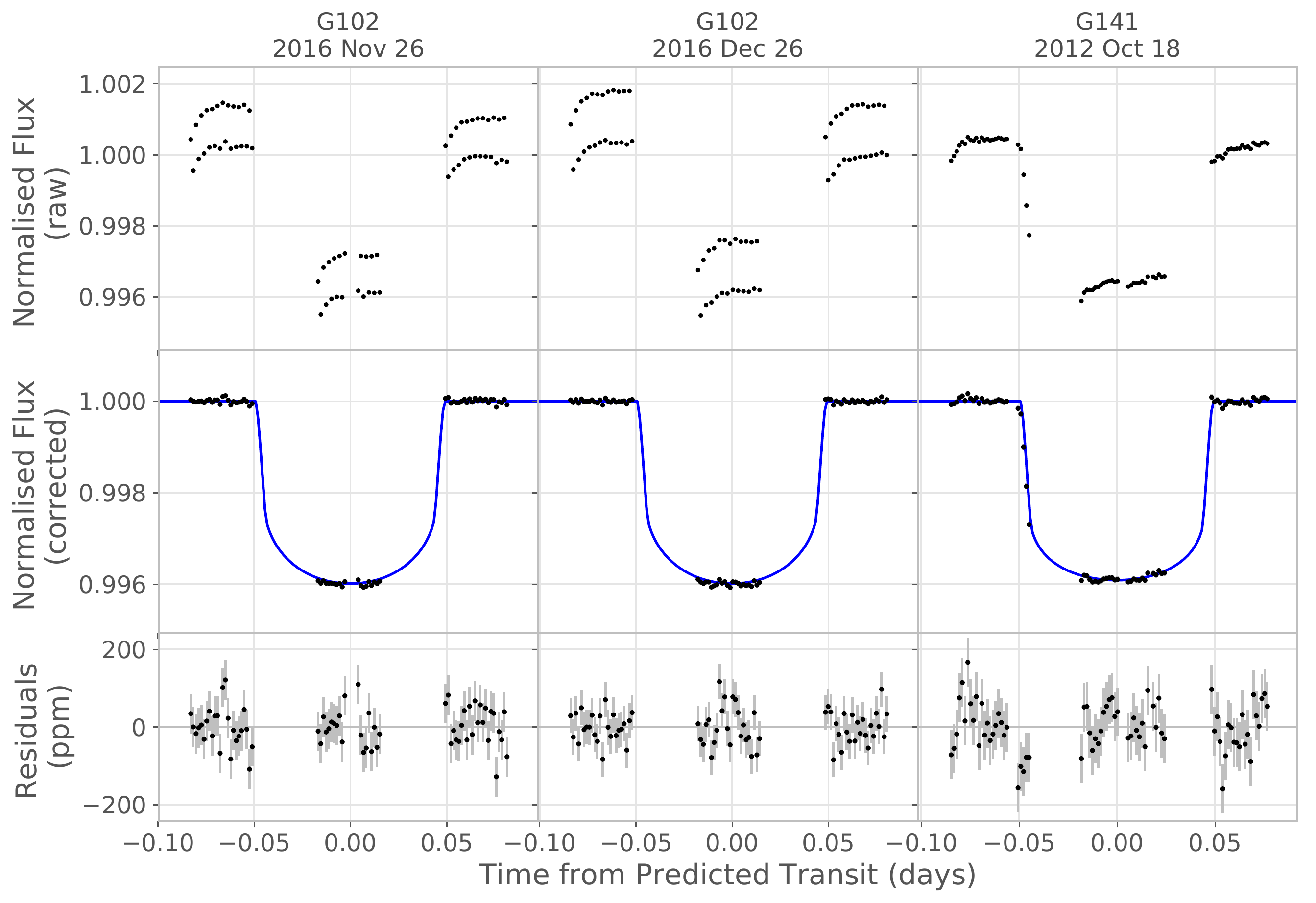}
	\caption{WFC3 G102 and G141 white light transit light curves before (top) and after (middle) dividing out the best-fit instrumental systematics model.  The best-fit transit light curve is shown in blue for comparison, and the fit residuals are shown at the bottom.}
	\label{fig:WFC3_WLC_fits}
\end{figure*}

We first fit each dataset individually to obtain an initial set of best-fit parameters.  For these individual fits, in addition to fitting for astrophysical and systematics model parameters, we allow the measurement error $\sigma$ to vary as a free parameter to ensure we obtain a reduced $\chi^2$ of unity and to accurately model uncertainties in the parameters due to the intrinsic scatter in the light curves.  We then use the results of these individual fits as initial guesses in the joint fit and fix the measurement error $\sigma$ for each visit to the best-fit value obtained from its corresponding individual fit.  We run an initial burn-in phase with 2000 steps for individual datasets and 40,000 steps for the global fit.  We identify and discard walkers that become trapped in local minima by removing any chain whose maximum likelihood value is lower than median likelihood value of any of the other chains.  We set the initial number of walkers to four times the number of free parameters and typically reject $\lesssim 10 \%$ of these walkers.  Whenever an odd number of walkers remain, we randomly remove a walker from the remaining set.  After burn-in, the fit is continued with the remaining walkers for another 3000 steps for individual fits and 60,000 steps for the global fit.  We assume flat priors within a suitable range for each parameter.  We check for convergence by inspecting the chain plots and running these fits with long chains three times.  We find that the parameter estimates are consistent at the $0.5\sigma$ level or better and the transmission spectrum is consistent to within $0.5\sigma$.

We fit a total of thirteen individual transits in our global analysis, each with their own instrumental systematics model.  This corresponds to a total of 93 free parameters, which is too large to reliably explore with MCMC.  We therefore utilize linear optimization to reduce the number of free parameters in our MCMC fit.  At each step in the fit, we calculate new best-fit values for all linear parameters in the global systematics model using linear regression while keeping all other model parameters fixed to their values at that step in the MCMC.  This reduces the number of free parameters in the MCMC fit to 48.  We additionally fix the $\sigma$ parameters for all visits in our global fit to the values obtained in our individual fits, which reduces the number of free parameters to 35.  This is small enough to ensure reliable convergence within a reasonable number of steps.  We acknowledge that in principle this approach might cause us to underestimate the uncertainties in our astrophysical model parameters, as we are optimizing rather than marginalizing over the linear instrumental model parameters \citep[see e.g.][]{Benneke2019}.  However, we find that in practice these linear instrumental model parameters contribute negligibly to the uncertainties in our astrophysical model parameters.  Optimizing the linear instrumental parameters in a global fit to the data excluding G102 light curves reduces the uncertainties in $R_p/R_*$ by less than $5$\%.

\subsection{White Light Curve Fits}
\label{sec:wlc_fits}

\begin{figure*}[t]
	\centering
	\includegraphics[width=\textwidth]{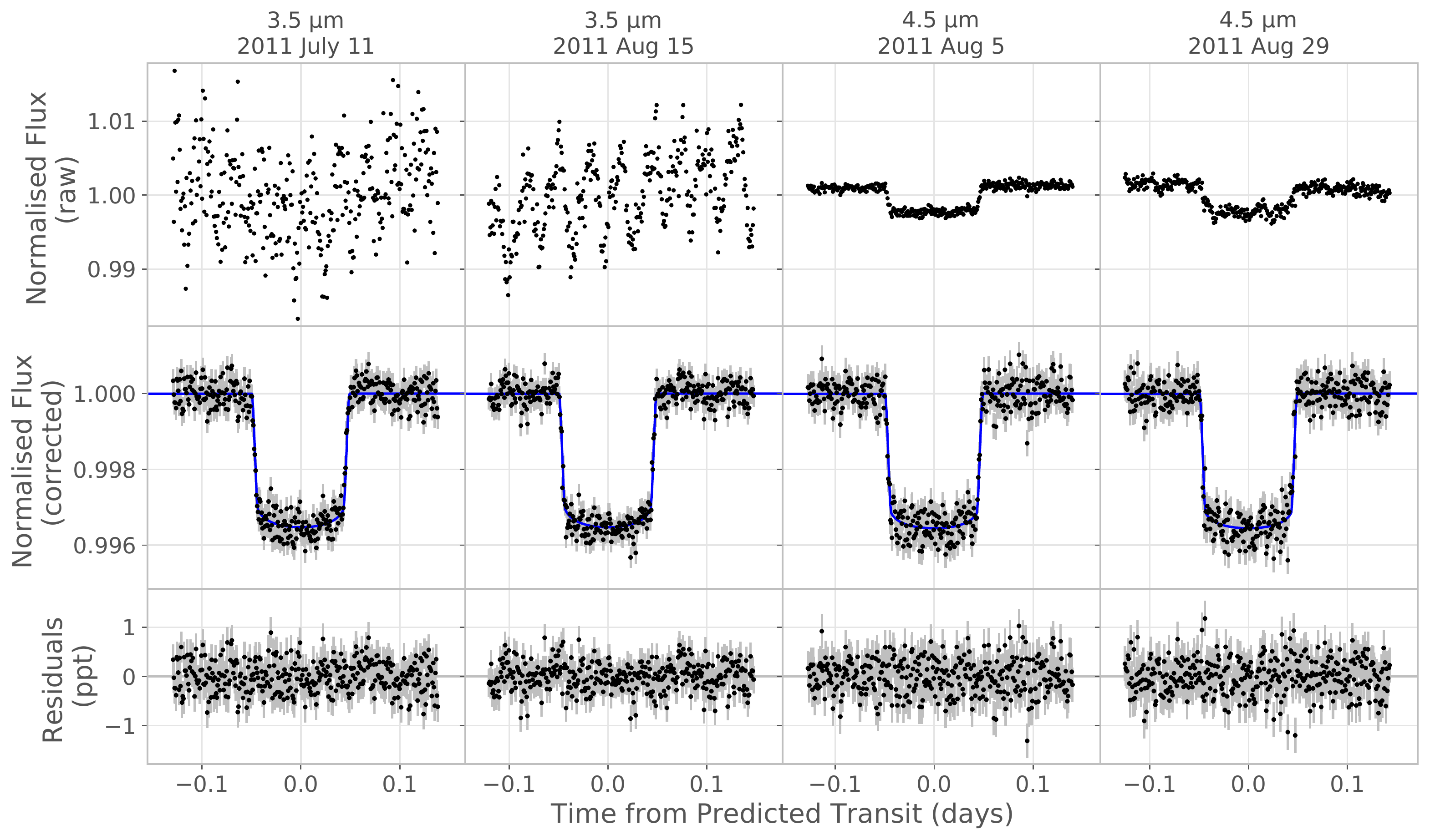}
	\caption{\emph{Spitzer} transit light curves before (top) and after (middle) dividing out the best-fit instrumental systematics model.  The best-fit transit light curve is shown in blue for comparison, and the fit residuals are shown at the bottom.}
	\label{fig:Spitzer_WLC_fits}
\end{figure*}

We confirm that the individual transit depths in bandpasses with multiple visits agree to within $2\sigma$ after correcting for the effects of unocculted star spots, as discussed in \S~\ref{sec:stellar_activity}.  We therefore report the global best-fit transit depths for each band in Table~\ref{table:wlc_fits}.  The best-fit transit light curves and their residuals are shown in Figures~\ref{fig:STIS_WLC_fits}-\ref{fig:Spitzer_WLC_fits}.  The white light curve depths for the WFC3 G141 visit and G102 visits agree with the values reported by \citetalias{Fraine2014} and \citetalias{Mansfield2018} at the $1.5 \sigma$ and $0.6 \sigma$ level respectively.  Our visit-averaged 3.6 and 4.5 $\mu$m \emph{Spitzer} transit depths are in good agreement ($1.5 \sigma$ lower and $0.5 \sigma$ higher respectively) with the values obtained by \citetalias{Fraine2014}. The residuals from our \emph{Spitzer} fits display the predicted root-\textit{n} scaling expected for Gaussian noise.

We find that both our 3.6 and 4.5 $\mu$m \textit{Spitzer} transit depths are somewhat lower than our WFC3 G141 white light transit depth.  The difference in white light curve depths between the WFC3 G141 observations and the \emph{Spitzer} observations is consistent with the results reported by \citetalias{Fraine2014}.  \citetalias{Fraine2014} attributed this difference to stellar activity and used an offset of 93 ppm for the WFC3 spectrum to obtain their best-fit model.  However, this difference cannot be explained by stellar activity for plausible star spot properties.  For the \emph{Spitzer} transit depths to be $\gtrsim 100$ ppm higher than the \emph{HST} measurements, HAT-P-11 would need to be 3\% brighter during the \emph{Spitzer} epochs than the \emph{HST} ones, which is larger than the observed peak-to-peak variability of the star.  For representative spot temperatures of 4500 K and 4300 K, the spot coverage fraction would need to be different by $>$10\% and $\sim$5\% respectively to obtain such a large relative correction to the transit depths.  Finally, such a large correction to the \emph{HST} measurements would strongly distort the transmission spectrum from 0.3 $\mu$m -- 1.7 $\mu$m and impart an almost unphysical upward slope (with increasing wavelength) to it.  We discuss this difference in the \emph{HST} and \emph{Spitzer} transit depths and our efforts to interpret it in \S~\ref{sec:retrieval}.

\subsection{Wavelength-Dependent Light Curves}

\begin{figure*}[t]
    \centering
    \includegraphics[width=\linewidth]{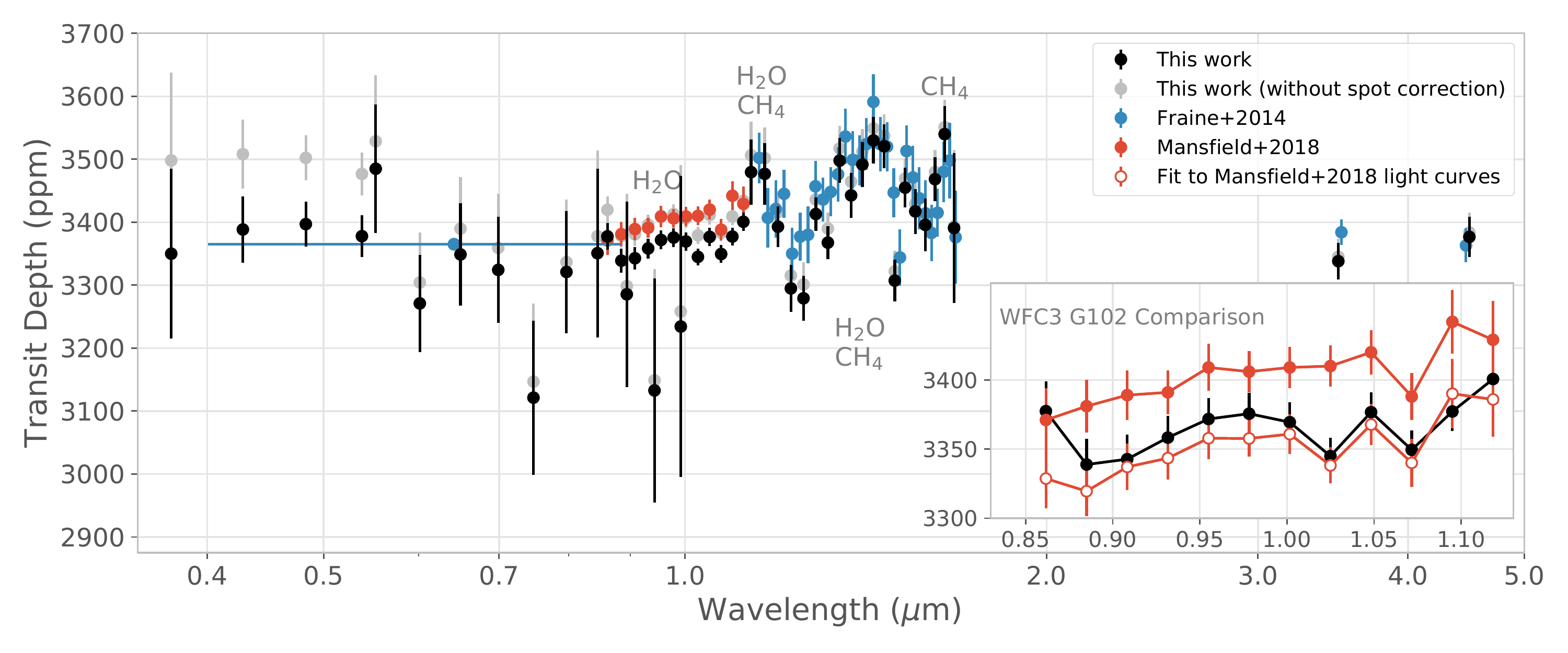}
    \caption{The transmission spectrum of HAT-P-11b both with and without stellar activity correction.  Our transmission spectrum is in good agreement with \citetalias{Fraine2014}'s published spectrum.  In the inset figure, we compare our WFC3 G102 spectrum with a fit to \citetalias{Mansfield2018}'s light curves, as well as \citetalias{Mansfield2018}'s published spectra.  Our G102 spectrum deviates most significantly from the published spectrum at 0.86 $\mu$m, 1.025 $\mu$m, and 1.095 $\mu$m, which has the effect of washing out the small absorption feature at 0.95 $\mu$m in the published version.}
    \label{fig:transmission_spectrum}
\end{figure*}

When fitting for the wavelength-dependent radius ratio $ R_p / R_*$ within each \textit{HST} STIS and WFC3 bandpass, we fix the orbital parameters $P$, $T_c$, $a/R_*$ and $b$ to the best-fit values from the global fit.  We re-fit the full systematics model in each individual bandpass without recourse to values obtained from the white-light fit.  We found that fitting the individual spectroscopic light curves with the full systematics model significantly improved the quality of the fit as compared to using the (scaled) systematics models from the global fit.  For the \emph{HST} STIS data, all parameters in the full systematics model are obtained by linear optimization and we simply use this model for the spectroscopic light curves as well.  We find that fitting the individual spectroscopic light curves with the full systematics model as compared to using the (scaled) systematics models from the global fit significantly improves the quality of the fit for the WFC3 G102 data ($\Delta$BIC $>10$ for 8 out of 12 wavelength bins) but not for the WFC3 G141 data.  Applying a common-mode correction to the spectroscopic light curves obtained by dividing the white light curve flux with the best-fit transit model \citep[e.g.][]{Deming2013} and employing a simpler model for the residual systematics in the spectroscopic light curves is strongly favored ($\Delta$BIC $> 10$ for 16 out of 19 wavelength bins).  Our simple model for the WFC3 G141 spectroscopic light curves is a linear function of the measured shift $(x - x_o)$ in the dispersion direction relative to the first exposure with offset $c$ and slope $v$ :

\begin{equation}
    S (t) = c + v (x - x_o) 
\end{equation}
We present $ R_p / R_*$ and associated errors for each bandpass in Table~\ref{table:spectroscopic_fits}, the transmission spectrum in Figure~\ref{fig:transmission_spectrum}, and show the corresponding wavelength-dependent light curves in Figures~\ref{fig:G430_WaveLC_fits}--\ref{fig:G141_WaveLC_fits} in the Appendix.

\begin{table}
\small
    \centering
    \caption{Spectroscopic Light Curve Fit Results}
    \begin{tabular}{lcc}
    \hline \hline
    Wavelength ($\mu$m) & $R_p / R_*$ & $\pm 1 \sigma$ \\ \hline 
    STIS G430L \\
    0.346-0.401 & 0.05788 & 0.00117 \\
    0.401-0.456 & 0.05821 & 0.00045 \\
    0.456-0.511 & 0.05828 & 0.00031 \\
    0.511-0.566 & 0.05812 & 0.00029 \\
    STIS G750L \\
    0.528-0.577 & 0.05903 & 0.00086 \\
    0.577-0.626 & 0.05719 & 0.00068 \\
    0.626-0.674 & 0.05787 & 0.00070 \\
    0.674-0.723 & 0.05766 & 0.00073 \\
    0.723-0.772 & 0.05587 & 0.00109 \\
    0.772-0.821 & 0.05763 & 0.00084 \\
    0.821-0.870 & 0.05789 & 0.00116 \\
    0.870-0.919 & 0.05732 & 0.00129 \\
    0.919-0.967 & 0.05597 & 0.00159 \\
    0.967-1.016 & 0.05687 & 0.00210 \\
    0.589-0.591$^*$ & 0.06244 & 0.00361 \\
    0.766-0.773$^*$ & 0.05689 & 0.00192 \\
    WFC3 G102 \\
    0.850-0.873 & 0.05812 & 0.00019 \\
    0.873-0.897 & 0.05778 & 0.00016 \\
    0.897-0.920 & 0.05782 & 0.00015 \\
    0.920-0.943 & 0.05795 & 0.00014 \\
    0.943-0.967 & 0.05807 & 0.00013 \\
    0.967-0.990 & 0.05810 & 0.00013 \\
    0.990-1.013 & 0.05805 & 0.00013 \\
    1.013-1.037 & 0.05784 & 0.00011 \\
    1.037-1.060 & 0.05811 & 0.00013 \\
    1.060-1.083 & 0.05787 & 0.00012 \\
    1.083-1.107 & 0.05811 & 0.00012 \\
    1.107-1.130 & 0.05831 & 0.00012 \\
    WFC3 G141 \\
    1.120-1.150 & 0.05899 & 0.00044 \\
    1.150-1.180 & 0.05896 & 0.00041 \\
    1.180-1.210 & 0.05825 & 0.00028 \\
    1.210-1.240 & 0.05740 & 0.00033 \\
    1.240-1.270 & 0.05726 & 0.00031 \\
    1.270-1.300 & 0.05842 & 0.00023 \\
    1.300-1.330 & 0.05803 & 0.00023 \\
    1.330-1.360 & 0.05914 & 0.00030 \\
    1.360-1.390 & 0.05867 & 0.00031 \\
    1.390-1.420 & 0.05909 & 0.00030 \\
    1.420-1.450 & 0.05941 & 0.00031 \\
    1.450-1.480 & 0.05933 & 0.00030 \\
    1.480-1.510 & 0.05751 & 0.00029 \\
    1.510-1.540 & 0.05878 & 0.00027 \\
    1.540-1.570 & 0.05846 & 0.00030 \\
    1.570-1.600 & 0.05827 & 0.00036 \\
    1.600-1.630 & 0.05889 & 0.00030 \\
    1.630-1.660 & 0.05950 & 0.00037 \\
    1.660-1.690 & 0.05823 & 0.00102 \\
    \hline
	\end{tabular}
	\label{table:spectroscopic_fits}
\end{table}

In Figure~\ref{fig:transmission_spectrum}, we show both stellar activity corrected and uncorrected transit depths.  We obtain the uncorrected depths by fixing the orbital parameters $b$ and $a/R_*$ to values obtained from the global white light curve fit (shown in Table~\ref{table:wlc_fits}) and fitting the light curves without any wavelength or epoch dependent correction.  This allows us to isolate the effect of activity correction on the transit depths.  We note that activity correction is crucial for obtaining correct inferences from the optical data.  The uncorrected upward slope in the STIS G430L bandpass would dramatically affect our interpretation of the planet's atmospheric properties.  In addition, the magnitude of the correction is commensurate with values necessary to produce a consistent and connected spectrum across multiple bandpasses.  For example, the uncorrected STIS G750L depths are fairly low compared to the STIS G430L measurements, but STIS G750L observations are taken at a time when spot coverage of the star is at a minimum and the STIS G430L measurements are obtained when the star is fairly spotted (see Figure~\ref{fig:stellar_var}).  This produces a small correction for the STIS G750L measurements and a large one for the STIS G430L depths, as one would expect.

We see evidence for molecular absorption in the WFC3 G141 bandpass, in good agreement with the results from \citetalias{Fraine2014}.  Our spectrum is not as smooth as that of \citetalias{Fraine2014}, but this is likely due to their use of a 4-pixel wide smoothing kernel (Figure~\ref{fig:transmission_spectrum}).  Our spectrum agrees within $\sim 1 \sigma$ with the previously published spectrum in almost all the wavelength bins.  Stellar activity correction introduces a slightly different slope than that of \citetalias{Fraine2014}, with shallower transit depths at short wavelengths and larger transit depths at longer wavelengths.  Notably, our updated spectrum (both with and without correction) possesses a steeper rise longward of 1.5 $\mu$m compared with \citetalias{Fraine2014}'s, suggesting the presence of methane in the planet's atmosphere (see \S~\ref{sec:retrieval_results}).

Our WFC3 G102 spectrum differs from the version published by \citetalias{Mansfield2018} in subtle but significant ways (see inset, Figure~\ref{fig:transmission_spectrum}).  We diagnose the reason for this discrepancy by carrying out an additional set of fits using our models applied to the light curves from \citetalias{Mansfield2018}.  We find that a majority of the observed vertical offset between the spectrum published in \citetalias{Mansfield2018} and our fit to \citetalias{Mansfield2018}'s light curves is due to differences in the assumed values for the orbital parameters.  We fit for period, while fixing impact parameter and $a/R_*$ to the best-fit values from our global fit, and eccentricity and argument of pericenter values to the values obtained from \cite{Yee2018}.  In contrast, \citetalias{Mansfield2018} fix the period and eccentricity to values from \cite{Huber2017} and use impact parameter and $a/R_*$ values from \citetalias{Fraine2014} with Gaussian priors.  Small differences in the stellar activity correction were found to be insignificant.  Our spectrum is not a perfect match for the one we derive using \citetalias{Mansfield2018}'s light curves.  The spectral shape of our fit to \citetalias{Mansfield2018}'s light curves is intermediate to that of our spectrum and the published spectrum.  This implies that although our choice of systematics model (especially the use of an additional ramp delay parameter $d$ for the first fitted orbit) and global fitting of orbital parameters improves the agreement between our spectra, some differences must partly arise due to choices made in the light curve extraction.  In particular, there are significant differences in our light curves for the first visit, which arise due to \citetalias{Mansfield2018}'s decision to exclude the last non-destructive read (for forward scan, first read for backward) of the scan.  These differences are important, as the absorption features at 1.15 $\mu$m and 0.95 $\mu$m are barely discernible in the spectrum published by \citetalias{Mansfield2018}.  In our updated spectrum, the combination of WFC3 G102 and G141 data reveals three molecular absorption features: two strong features centered at 1.15 $\mu$m and 1.4$\mu$m and a weak feature at 0.95 $\mu$m (Figure~\ref{fig:transmission_spectrum}).  This allows us to infer the presence of water and/or methane with a combined significance of 4.4 $\sigma$ (see \S~\ref{sec:retrieval_results}).

Our new STIS observations indicate that HAT-P-11b has a relatively featureless transmission spectrum at optical wavelengths with a hint of increasing transit depth with decreasing wavelength (scattering slope).  This is in agreement with recently reported measurements obtained from ground-based observations \citep{Murgas2019}.  As mentioned above, a careful accounting for the effects of unocculted spots produces a much flatter optical transmission spectrum than the uncorrected version.  This plays an important role in constraining atmospheric metallicity and places constraints on the effective size and number density of the particles responsible for scattering in the atmosphere.  We see no evidence for narrow-band sodium or potassium absorption, although these features are expected to form at relatively low pressures where cloud opacity should be less important. This is not surprising, as HAT-P-11b's atmosphere is predicted to be too cold for these elements to remain in vapour form \citep[e.g.][]{Lodders1999}.  Additionally, we do not see the jump in transit depth at 0.8 $\mu$m that \cite{Lothringer2018} report for GJ 436b and note for HAT-P-26b.

\section{Comparison to Forward Models}
\label{sec:forward_models}

We next compare HAT-P-11b's observed transmission spectrum to predictions from a 1D microphysical cloud model originally developed for use with solar system planets \citep[e.g.][]{Toon1979, Toon1992, James1997, Colaprete1999, Gao2017}.  These cloud models require a temperature-pressure profile and a prescription for the vertical mixing in the atmosphere as inputs.  We draw both of these profiles from results of a 3D general circulation model (GCM) for HAT-P-11b.  We discuss the details of both models in the following two sub-sections.

\subsection{General Circulation Model}
We use a GCM to put constraints on the extent of (1D) mixing in the atmosphere.  This allows us to take into account the effect of three-dimensional (3D) dynamics on the 1D atmospheric profiles used in transmission spectroscopy studies.  This is particularly important for eccentric short-period planets like HAT-P-11b, which are presumed to be tidally locked and therefore may have a pressure and temperature structure that varies significantly with longitude.  The appreciable eccentricity of HAT-P-11b also leads to the convolution of latitudinal structure and orbital phase of the planet.  We take the planet's eccentricity into account in our GCM and use atmospheric profiles (for temperature, pressure, eddy diffusion coefficient) from the planet's transit.  In this case, we utilize the Substellar and Planetary Radiation and Circulation (SPARC) model \citep{Showman2009, Kataria2016}, which couples the MITgcm dynamical core \citep{Adcroft2004} with a plane-parallel, two-stream version of the multi-stream radiation code developed by \cite{Marley1999}.  As we will discuss in \S\ref{sec:retrieval}, our retrievals using the \emph{HST} data prefer relatively low metallicity values, so we choose models with atmospheric metallicities of 1$\times$ and 50$\times$ solar (we multiply relative abundances of elements heavier than hydrogen and helium by this metallicity value and renormalize the sum of relative abundances to 1); this range is therefore a good match for the posterior probability distribution for this parameter.

\begin{figure*}
	\centering
	\includegraphics[width=0.48\linewidth]{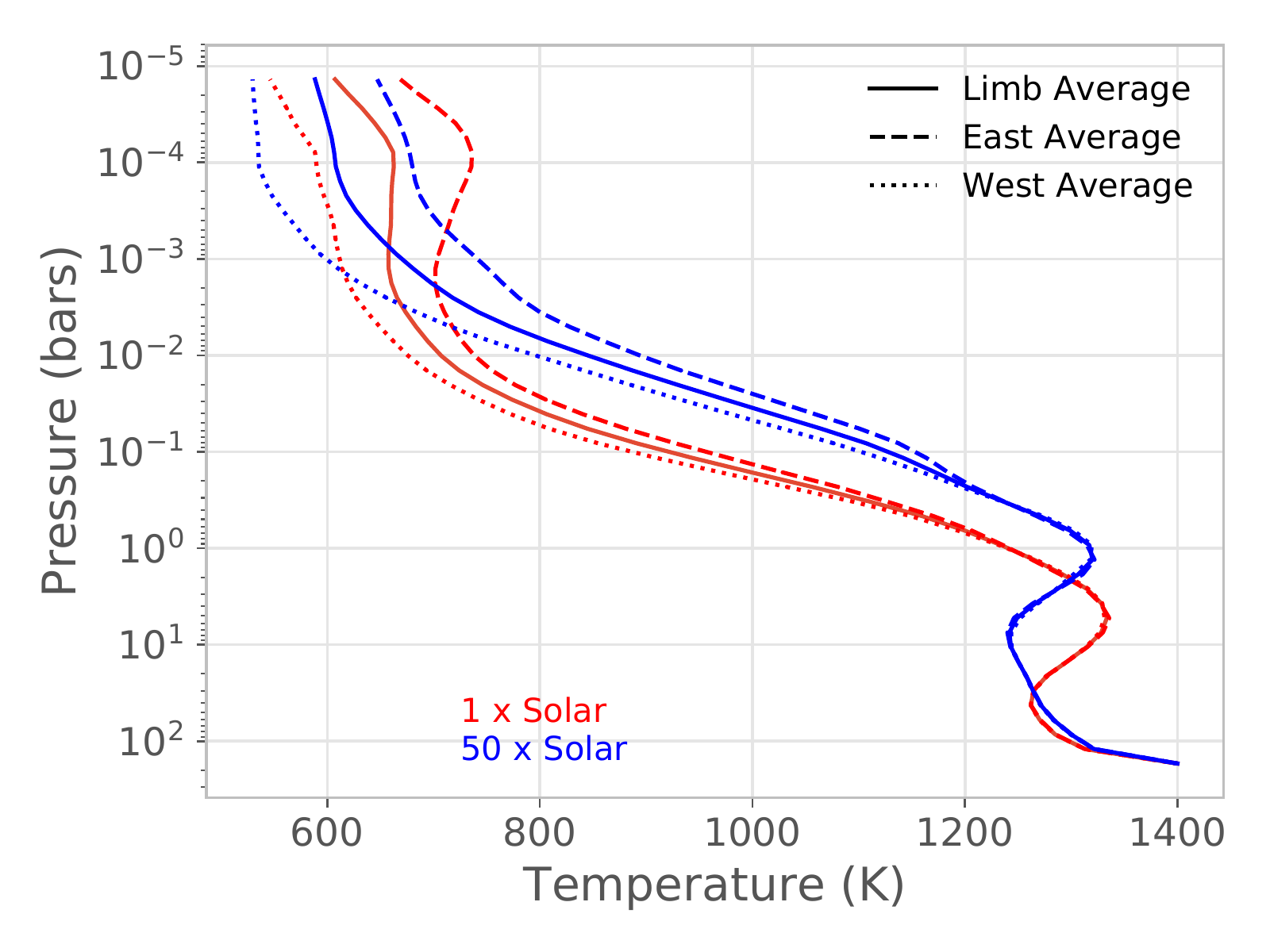}
	\includegraphics[width=0.48\linewidth]{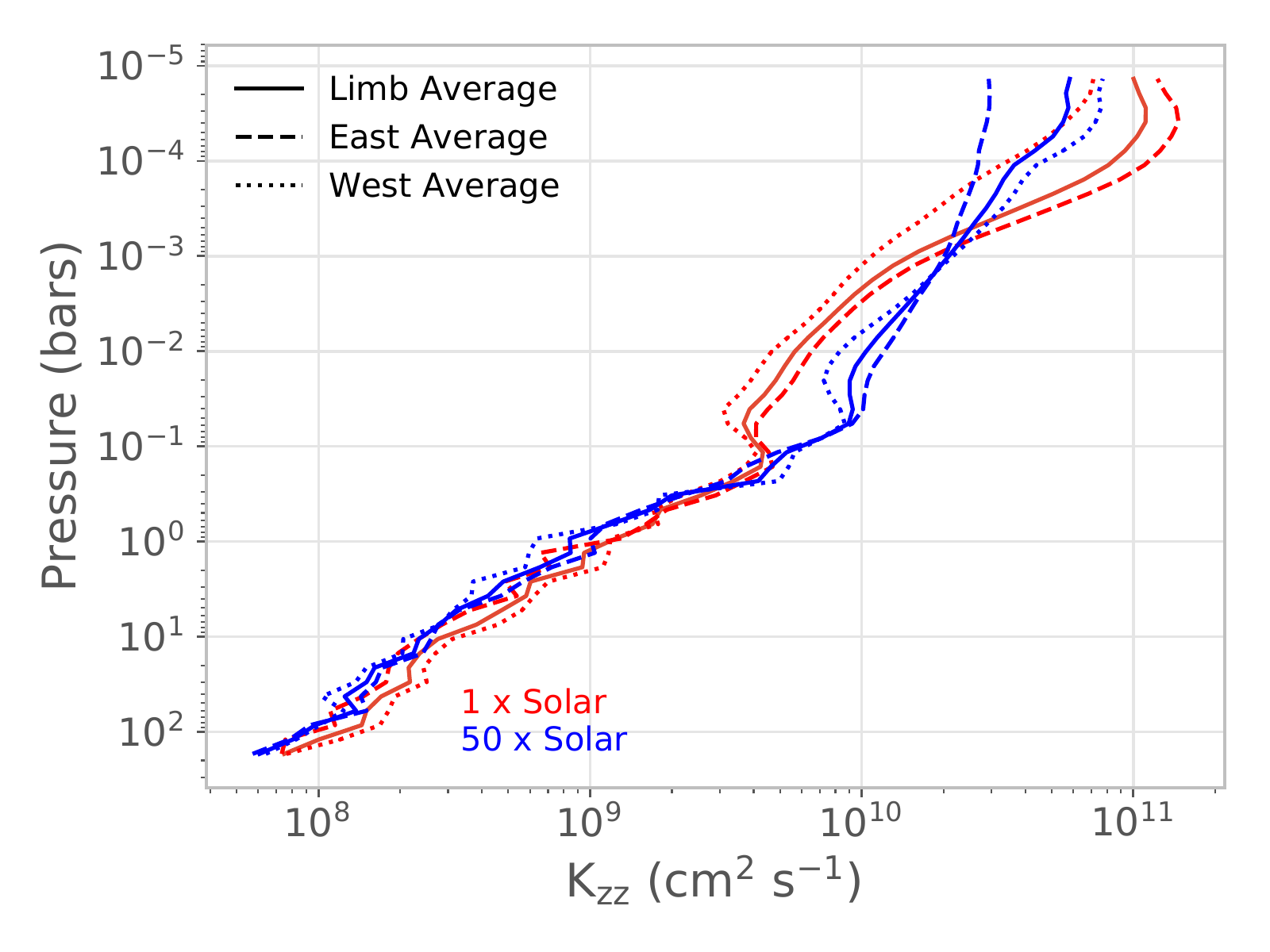}
	\caption{Temperature (left) and vertical mixing parameter $K_{zz}$ (right) profiles as a function of pressure at the orbital phase of the transit (since HAT-P-11b has a significant eccentricity).  These profiles are obtained from a SPARC GCM model for HAT-P-11b and are used as inputs in our microphysical cloud models. Transmission spectroscopy probes the atmosphere at pressures roughly between $10^{-1}-10^{-4}$ bars.}
	\label{fig:GCM_Kzz}
\end{figure*}

\begin{figure*}
	\centering
	\includegraphics[width=0.8\linewidth]{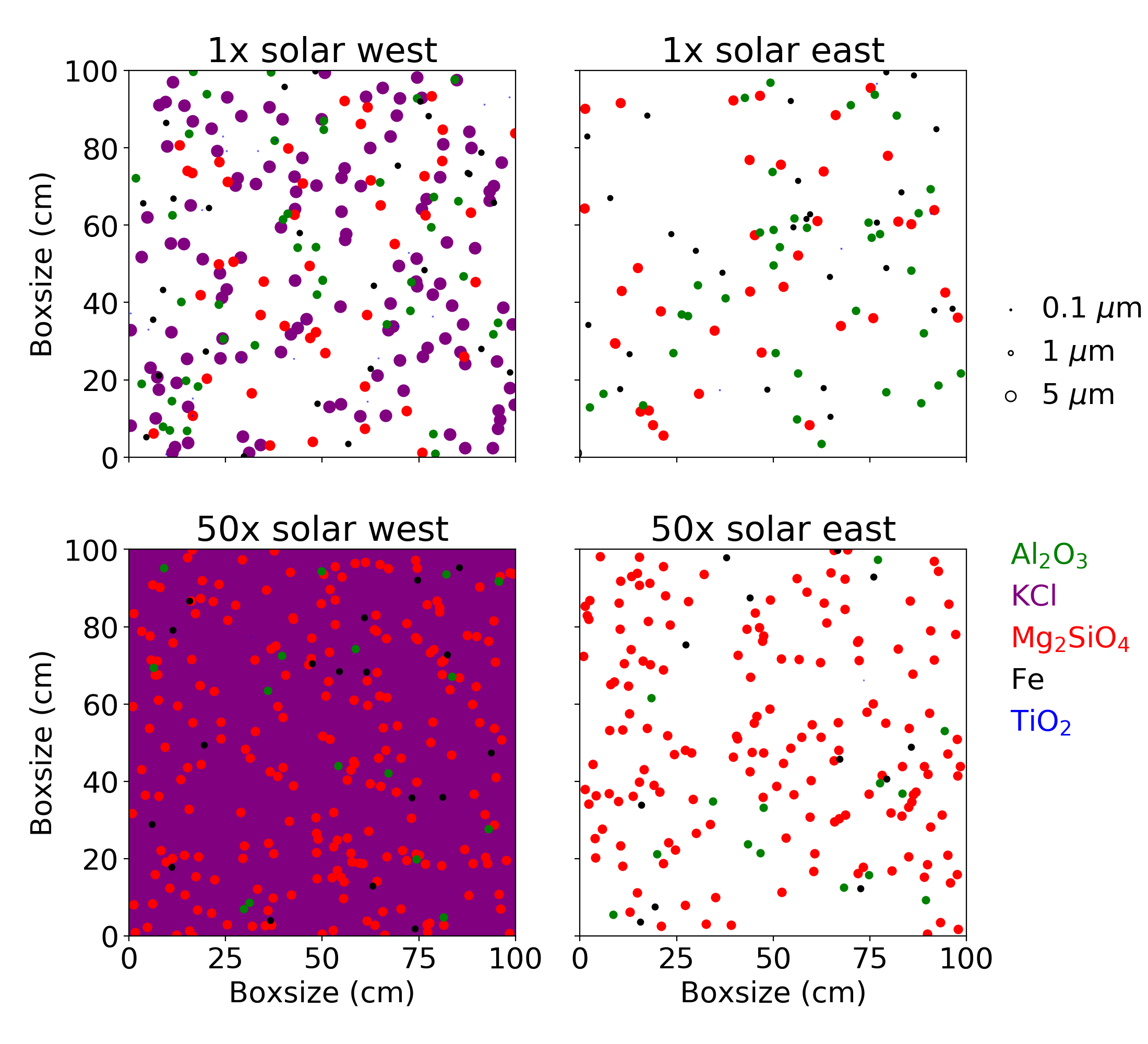}
	\caption{Plot windows showing 2D slices of the atmospheric condensate compositions for a 1$\times$ solar and 50$\times$ solar metallicity atmosphere.  The slices sample the atmosphere on the east and west limbs at $\tau \sim 1$ and show the number of condensate particles contained in a 100 cm $\times$ 100 cm $\times$ 100 cm volume.  Condensates on the two limbs have distinct compositions and increasing the metallicity has a significant effect on condensate number density, especially on the west limb.  These plots serve as a visual guide and indicate that the scattering cross-section at the wavelengths of interest is mostly dominated by KCl particles.  Mg$_2$SiO$_4$ and Al$_2$O$_3$ particles also make significant contributions to cloud opacity, especially in the 1$\times$ solar metallicity case.}
	\label{fig:2d_slice}
\end{figure*}

We model vertical mixing as a diffusive process with an effective eddy diffusion coefficient $K_{zz}$.  Deviations from this diffusive approximation are almost guaranteed for tidally locked planets, which are expected to also have vigorous horizontal transport between the day and night sides \citep[e.g.][]{Zhang2018b, Zhang2018c}.  However, it is non-trivial to accurately capture this horizontal transport, and we therefore neglect it for the moment in order to explore the effects of vertical mixing, which is key for cloud formation.  This mixing is typically parameterized as a constant value with or without an inverse dependence on square root of pressure \citep[e.g.][]{Parmentier2013}.  We depart from this formalism and instead use the temperature, pressure, and $K_{zz}$ profiles from the GCM, which should be more representative of the relevant conditions in HAT-P-11b's atmosphere.  We use the GCM results to calculate 1D pressure-temperature profiles that are spatially averaged over the east and west limbs of the planet.  We estimate the corresponding pressure/height dependent $K_{zz}$ values for these locations using mixing length theory:

\begin{equation}
    K_{zz} = w(z)L(z) = \cfrac{\omega H^2}{P} 
\end{equation}
where $w(z)$ is the vertical velocity in m/s and $L(z)$ is a characteristic length scale, in this case the atmospheric pressure scale height. This commonly adopted method \citep[e.g.][]{Moses2011} gives us a height dependent $K_{zz}$ value which we then use in our microphysical cloud models.  We show the resulting $K_{zz}$ and temperature profiles as a function of pressure for the limb average, eastern limb average, and western limb average in Figure~\ref{fig:GCM_Kzz}.  As shown in previous GCM studies exploring the effect of atmospheric metallicity \citep[e.g.][]{Lewis2010, Kataria2014}, the higher metallicity profile of HAT-11b has a higher photosphere due to the higher opacity, which produces a $K_{zz}$ profile that rises more rapidly with height than the lower metallicity model.

\begin{figure*}[t!]
	\centering
	\includegraphics[width=0.9\linewidth]{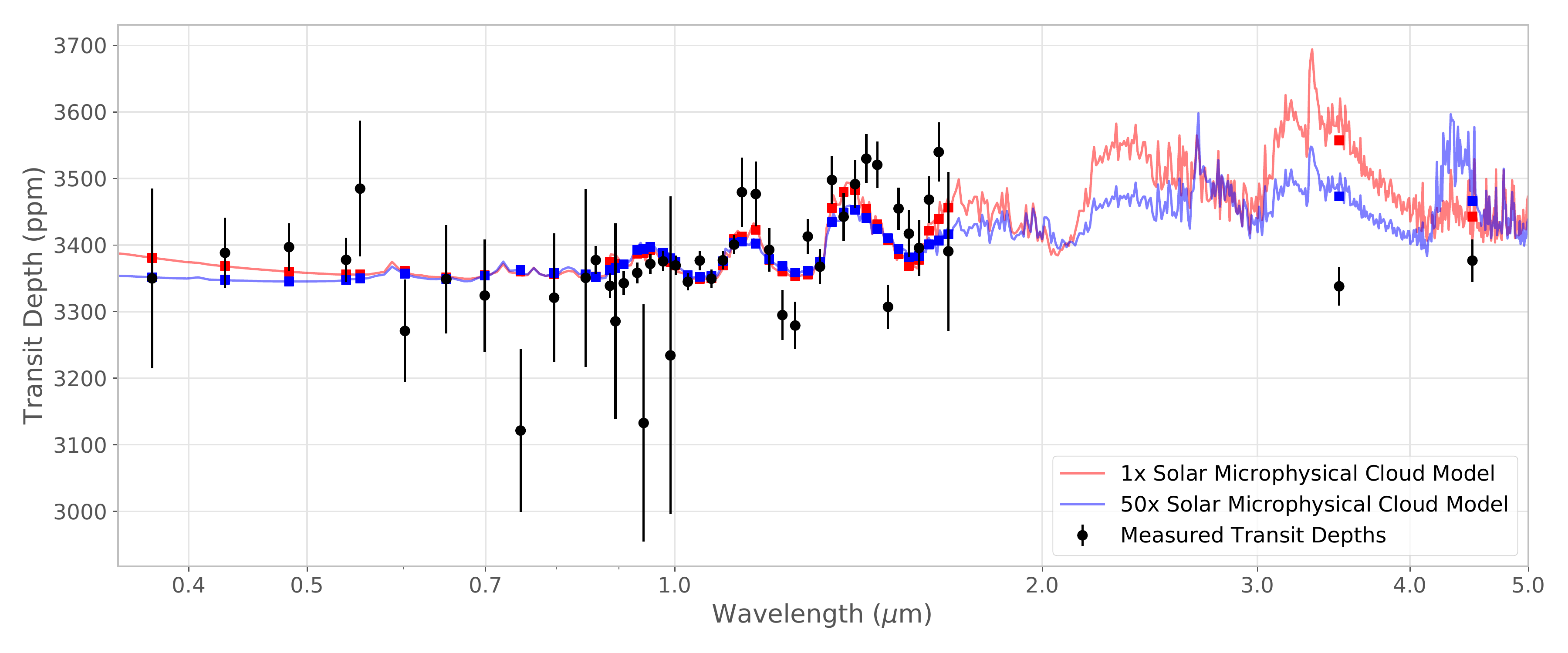}
	\caption{Measured transmission spectrum of HAT-P-11b versus transmission spectra generated by averaging CARMA models for the east and west limbs.  These model spectra fit the measured spectrum quite well without any fine-tuning or parameter fitting.  The \emph{HST} data display a slight preference for the $1\times$ solar metallicity model.  However, both the $1\times$ and $50\times$ solar metallicity models are unable to reproduce the \textit{Spitzer} transit depths.}
	\label{fig:retrieval_model_comparison}
\end{figure*}

\subsection{Microphysical Cloud Model}
\label{sec:MCM}
We use the Community Aerosol and Radiation Model for Atmospheres (CARMA) to determine which species are expected to condense in HAT-P-11b's atmosphere and the corresponding particle size distribution and abundance.  CARMA is a bin-scheme cloud microphysics model that considers microphysical processes such as nucleation, evaporation, condensation, sedimentation, and diffusion.  The strength of bin-scheme microphysics is that it uses discrete bins for particle sizes and makes no prior assumption regarding the size distribution, instead allowing the different bins to `interact' (i.e. exchange mass) via the aforementioned microphysical processes.  For a thorough exposition of the model, we direct the reader to \cite{Gao2018} and \cite{Powell2018}.

We include the following condensible species in our model: Cr, KCl, Al$_2$O$_3$, Mg$_2$SiO$_4$, Fe, and TiO$_2$.  We also consider condensation of metal sulphides but find it to be unimportant.  Na$_2$S, MnS, and ZnS have high nucleation energy barriers that inhibit the formation of these cloud species (Gao et al. 2019, submitted). Another reason ZnS clouds can be neglected is the low abundance of Zn.  We assume that KCl, Cr, TiO$_2$ and Al$_2$O$_3$ can nucleate homogeneously, meaning that they can condense into stable clusters directly from the gas phase and subsequently grow to larger sizes.  In contrast, heterogeneous nucleation requires a foreign surface or `seed' onto which vapor can condense.  Though the majority of Al$_2$O$_3$ condensates likely form via heterogeneous surface reactions \citep[e.g.][]{Helling2018}, assuming homogeneous nucleation is unlikely to greatly affect our results, as Al$_2$O$_3$ condenses at much higher temperatures ($\sim$2000 K) than considered here.  Al$_2$O$_3$ is present in small concentrations at the high altitudes that we probe (Figure~\ref{fig:2d_slice}), but its distribution in this region is primarily controlled by transport processes rather than condensation and nucleation \citep{Gao2018a}.  We assume that Fe and Mg$_2$SiO$_4$ nucleate heterogeneously on TiO$_2$ particles, similar to the treatment of \citet{Helling2018} and related works.  Although Fe can nucleate homogeneously as well, we do not consider it as this process may not be efficient \citep{Lee2018}.

We model the east and west limbs separately, as well as a limb averaged profile, ($T$ and $K_{zz}$) for both solar and 50$\times$ solar metallicity atmospheres.  We neglect the effect of radiative feedback from condensation and cloud formation on the atmosphere's T-P profile.  The resulting particle sizes and number densities of the dominant condensate species are shown in Figure~\ref{fig:2d_slice} as a 2D visualisation of a slice of the atmosphere at a pressure of $\sim$2 mbar ($\tau \sim 1$ for transmission spectroscopy) with a path length of 100 cm through the atmosphere.  In addition, the area covered by the different condensate species is proportional to the geometric cross-section due to each species, thereby visually indicating which species dominate the cloud opacity.

It is immediately evident that for both metallicity cases, the east and west limb-averaged profiles display distinct cloud properties and are dominated by different condensate species.  This is primarily due to the temperature difference between the two limbs, which can be as large as 100-200 K (see Figure~\ref{fig:GCM_Kzz}).  Most notably, the west limb is cool enough for KCl to condense and contribute dominantly to the opacity whereas the east limb is completely devoid of condensed KCl.  The lower temperature of the west limb also causes more nucleation sites to form, additionally increasing the cloud opacity in this region.  The east limb has a significantly lower condensate number density ($< 100 \; \mathrm{m}^{-3}$) and consists of species that have cloud bases deep in the atmosphere but are carried to pressures probed by transmission spectroscopy by strong vertical mixing (Figure~\ref{fig:GCM_Kzz}).  These differences result in distinct predictions for the mid-IR spectra of the two limbs, and suggest that cloud models utilizing the limb averaged pressure-temperature profile may not produce accurate predictions \citep[e.g.][]{Kempton2017}.  Using the average of the transmission spectra rather than the average of the pressure-temperature profile for the two limbs should allow a better comparison of the models with the data.  We therefore compare our retrieval results with model transmission spectra generated by averaging the spectra from the east and west limbs.

Increasing the metallicity from 1$\times$ solar to 50$\times$ solar increases the abundance of condensates by $1-2$ orders of magnitude.  Although the rate of homogeneous nucleation increases when the metallicity increases, the particle sizes tend to be somewhat smaller because there is less gas (per nucleated site) to provide additional condensible material for the growing particle.  KCl overwhelms the absorption cross-section on the west limb while the east limb is much clearer.

Figure~\ref{fig:retrieval_model_comparison} shows transmission spectra generated using CARMA models.  The models provide a good match to the observed absorption features at 0.95, 1.15, and 1.4 $\mu$m while maintaining a relatively flat optical spectrum without any fine-tuning or fitting.  We find that the 1$\times$ solar metallicity atmosphere is a slightly better match for the observed amplitude of the molecular absorption bands and optical scattering between $0.3 - 1.7$ $\mu$m than the $50\times$ solar metallicity model (reduced $\chi^2$ of 1.8 and 2.1 respectively).  However, both of these models predict strong methane absorption in the 3.6 $\mu$m \textit{Spitzer} band, making them a relatively poor match to the observed transit depth in this band.

\section{Atmospheric Retrieval: PLATON}
\label{sec:retrieval}
We use a simple and highly customisable atmospheric retrieval model, \texttt{PLATON}\footnote{Planetary Transmission Atmosphere Tool for Observer Noobs: https://github.com/ideasrule/platon} \citep{Zhang2019} to constrain HAT-P-11b's atmospheric properties using its transmission spectrum. \texttt{PLATON} is based on \texttt{ExoTransmit} \citep{M-RKempton2017} and uses a fast Python based algorithm to compute forward models for planetary atmospheres, which are then compared with the data in a retrieval framework.  \texttt{PLATON} includes opacities for 30 different molecular and atomic species \citep{Zhang2019b}, the majority of which are calculated using line lists from ExoMol \citep{Tennyson2018} and HITRAN \citep{Gordon2017}.  We use nested sampling for our retrievals to accurately capture the posteriors of atmospheric model parameters that may display multi-modality.  More importantly, using nested sampling allows us to compare the Bayesian evidence for different retrievals and rigorously quantify the significance of molecular absorption detection.

We fit for HAT-P-11b's atmospheric properties assuming an isothermal atmosphere in chemical equilibrium.   We allow the planet radius $R_p$, temperature $T$, atmospheric metallicity log $(Z)$, and the carbon-to-oxygen ratio C/O to vary as free parameters in our fit.  We also include scattering from high-altitude clouds, which we discuss in the following section.  All of these parameters have flat priors.  For $R_p$ and $T$ we choose physically motivated lower and upper bounds, while our prior range for metallicity and C/O ratio is dictated by limitations in our model's pre-computed equilibrium chemistry grid (see Table~\ref{table:PLATON_best_fit}).  Our grid limits us to log $(Z) \geq -1$, but we linearly (in $Z$) extrapolate abundances of atoms and molecules containing elements heavier than hydrogen and helium to lower metallicities (down to log $(Z) = -2$) to resolve the posterior distribution on the lower metallicity end.  We verify that linear extrapolation in $Z$ captures the atmospheric composition reasonably well by comparing transmission spectra obtained for atmospheric metallicities between $0.1 \times - 1 \times$ solar from extrapolation and from the pre-computed abundance grid.  We include the stellar radius \citep[$0.683 \pm 0.009 \mathrm{R}_{\odot}$;][]{Deming2011} and planetary mass \citep[$23.4 \pm 1.5 \mathrm{M}_{\oplus}$;][]{Yee2018} as free parameters in our model with Gaussian priors set to the published values.  This ensures that we correctly account for the effects of these uncertainties in our model fits.  We also include an additional parameter (``Error Multiple'' $\sigma_{\mathrm{mult}}$, same for all instruments) that multiplies the errors on the data with a constant factor to account for the errors' under- or over-estimation.

\begin{figure*}
	\centering
	\includegraphics[width=0.48\linewidth]{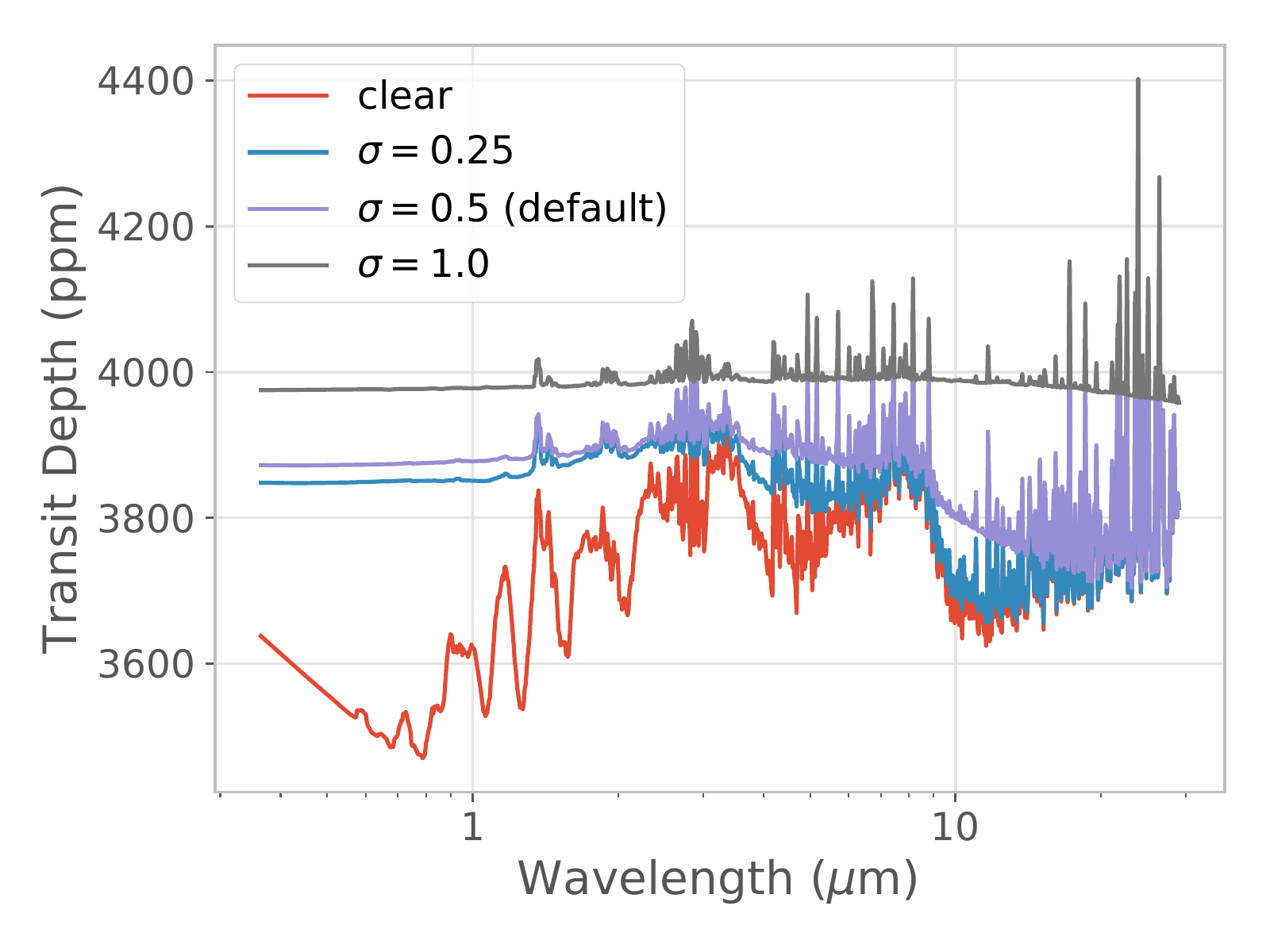}
	\includegraphics[width=0.48\textwidth]{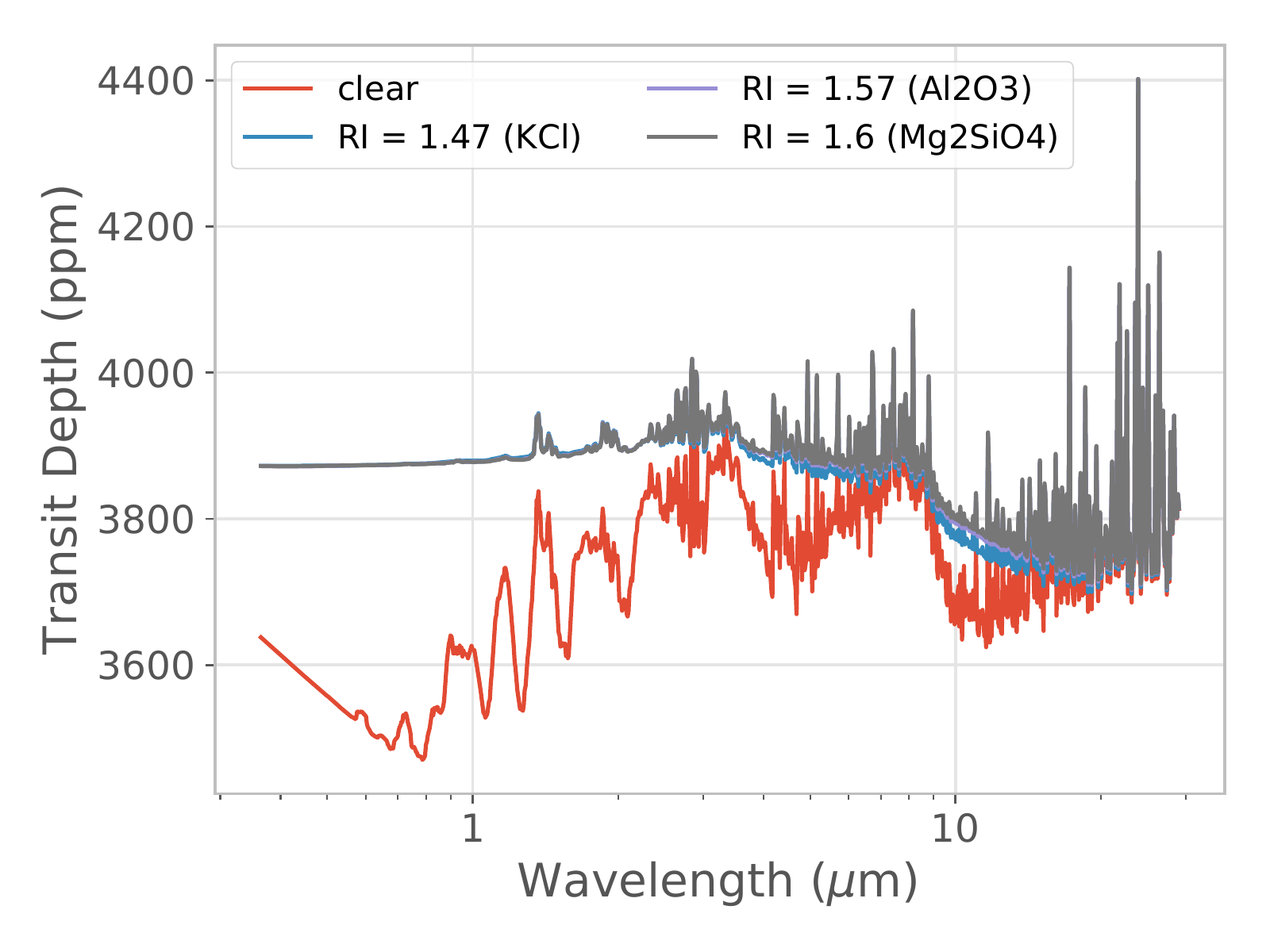}
	\caption{Effect of varying $\sigma$ and refractive index in our Mie scattering model. We assume a particle size $a =1$ $\mu$m, refractive index of 1.5, fractional scale height $f = 1$, particle size distribution width $\sigma = 0.5$, and number density at the base of the atmosphere $n_o = 10^4 \mathrm{cm}^{-3}$ unless specified otherwise.}
	\label{fig:mie_parameterisation_fixed}
\end{figure*}

\subsection{Scattering from Clouds}
\label{sec:clouds_platon}
We model scattering particles with five parameters: a cloud-top pressure ($P_{\mathrm{cloud}}$) below which the atmosphere is opaque at all wavelengths (top of a grey cloud), particle number density $n_0$ at $P_{\mathrm{cloud}}$, a lognormal distribution of particle sizes centered on an effective particle size $a$ with distribution width $\sigma$, and the scale height for particle number density as a fraction $f$ of the gas scale height $H_{gas}$.  This allows for a deep grey cloud that begins to thin as the pressure decreases.  Alternatively, it can be interpreted as a haze layer lying on top of a grey cloud.  The particle size distribution $p(r)$ and number density as a function of height $n(z)$ are given by:

\begin{equation}
p(r) = \frac{1}{\sqrt{2 \pi} \sigma  r} \exp\Big[-\frac{(\ln{r} - \ln{a})^2}{2\sigma^2}\Big],
\label{eq:lognormal_dist}
\end{equation}

\begin{equation}
n (z) = n_0 \; \exp[-z/ (f H_{gas})],
\end{equation}
The extinction cross-section, $\sigma_{ext}$, from condensate particles is then given as:

\begin{equation}
\sigma_{ext} (\lambda, z) = n_0 \; e^{-z/f H_{\mathrm{gas}}} \int p(r) Q_{ext}(\lambda,r) \pi r^2 dr
\end{equation}

We calculate $Q_{ext}$, which depends on the refractive index, using the Mie scattering formalism.  The effective particle size $a$, number density $n_0$, and relative scale height $f$ play a decisive role in shaping the planetary transmission spectrum.  The effective particle size $a$ determines the wavelength where the Rayleigh slope begins ($\lambda \sim 2 \pi a$).  The number density $n_0$ and fractional scale height $f$ set the overall scale of the opacity contribution from scattering (relative to molecular absorption opacity) and are partially degenerate with each other.  We find that $f$ is almost entirely unconstrained by our data and allowing it to vary in our retrievals does not have any significant effect on the posteriors for the other parameters in our model.  We therefore turn to our microphysical cloud models for HAT-P-11b, which indicate the effective particle size is roughly constant in the pressure range 0.1 mbar - 100 mbar and that the effective number density falls off with the pressure scale height $H_{\mathrm{gas}}$.  We fix $f = 1$ unless otherwise specified in order to reduce the number of free parameters and to allow for a more direct comparison with predictions from our microphysical models.

We also keep the value of the refractive index fixed to a single,  wavelength-independent value in our fits.  Our microphysical cloud models predict that condensate clouds in HAT-P-11b's atmosphere will include multiple distinct species.  However, the refractive indices for all these species apart from Fe have a very weak dependence on wavelength and negligible imaginary parts in the $0.1 - 5$ $\mu$m region spanned by our data (e.g. see \citealp{Kitzmann2018}).  Adopting a wavelength independent real value for the refractive index also speeds up our model computations enormously, which is a necessary requirement for retrieval codes.  Figure~\ref{fig:mie_parameterisation_fixed} shows that the shape of the predicted transmission spectrum is relatively insensitive to the exact value we assume for the refractive index in our wavelength range of interest.  We set this parameter equal to 1.5, as this is fairly representative of the dominant cloud species (KCl) predicted by our forward models.

Although the particle size distribution can take an arbitrary functional form, the distribution of large particles that are abundant enough to contribute most significantly to scattering may be captured by a lognormal distribution.  We keep the width of the lognormal distribution fixed in our fits.  Varying this parameter mimics the effect of increasing particle size as a broader distribution shifts the effective size of the particles to larger values and large particles tend to dominate the cloud opacity \citep[e.g.][]{Wakeford2015}.  Therefore, variations in the distribution width are degenerate with changes in particle size distributions.  Increasing the distribution width makes the spectrum flatter in a given wavelength range, as does increasing the effective particle size (see Figure~\ref{fig:mie_parameterisation_fixed}).  We fix $\sigma = 0.5$, which agrees well with typical values for aerosols in the Earth's atmosphere \citep[e.g.][]{Pinnick1978, Ackerman2001, Elias2009, Shen2015} and produces a scattering behaviour that is roughly compatible with that produced by the CARMA model with its non-parameterized particle size distribution.

\subsection{Retrieval Results}
\label{sec:retrieval_results}
\subsubsection{HST WFC3}
\begin{figure}[b]
	\centering
	\includegraphics[width=\linewidth]{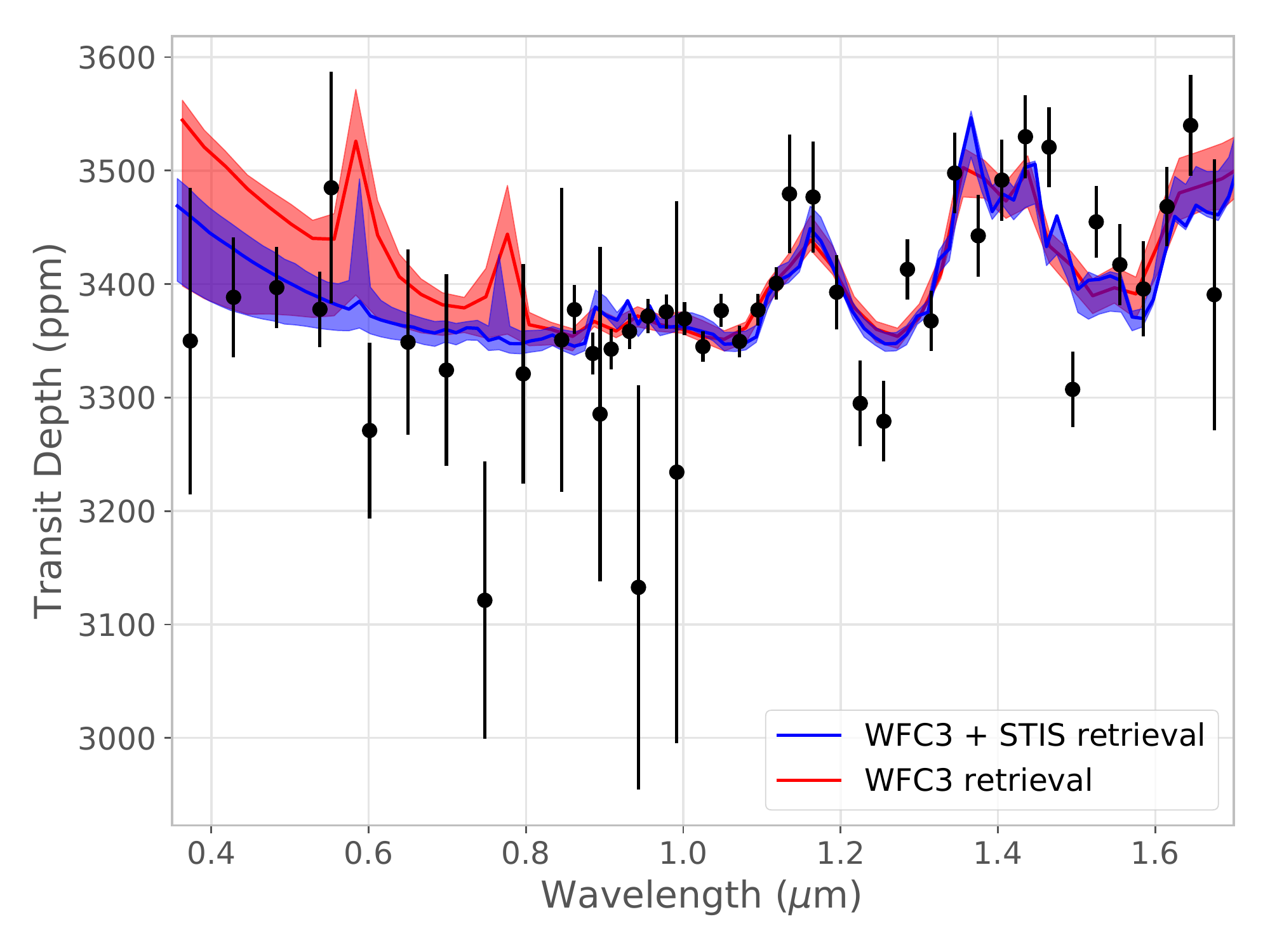}
	\caption{Transmission spectrum in the \textit{HST} WFC3 and STIS bandpasses (black filled circles) with best-fit Mie scattering model spectra from \texttt{PLATON} overplotted along with the $1\sigma$ contours.}
	\label{fig:HST_spectrum_fit}
\end{figure}

We begin by fitting the molecular absorption features in the WFC3 G102 and G141 bandpasses, as these features provide the strongest constraints on the planet's atmospheric composition.  Because these data span a relatively limited wavelength range, a simplified cloud model with a single opaque cloud deck is adequate.  Nonetheless, we `fit' for Mie scattering parameters for later comparison of best-fit models with models that match the entire \emph{HST} transmission spectrum.  We fit for temperature, atmospheric metallicity, and C/O ratio as well, assuming chemical equilibrium.  The resulting best-fit model is shown in Figure~\ref{fig:HST_spectrum_fit} and the corresponding constraints on the model parameters are given in Table \ref{table:PLATON_best_fit}.  The steep rise in transit depth longward of 1.5 $\mu$m hints at the presence of methane in the atmosphere.  We verify this by confirming that this upward rise disappears if methane is removed from our atmospheric models.

\begin{figure*}
	\centering
	\includegraphics[width=0.9\linewidth]{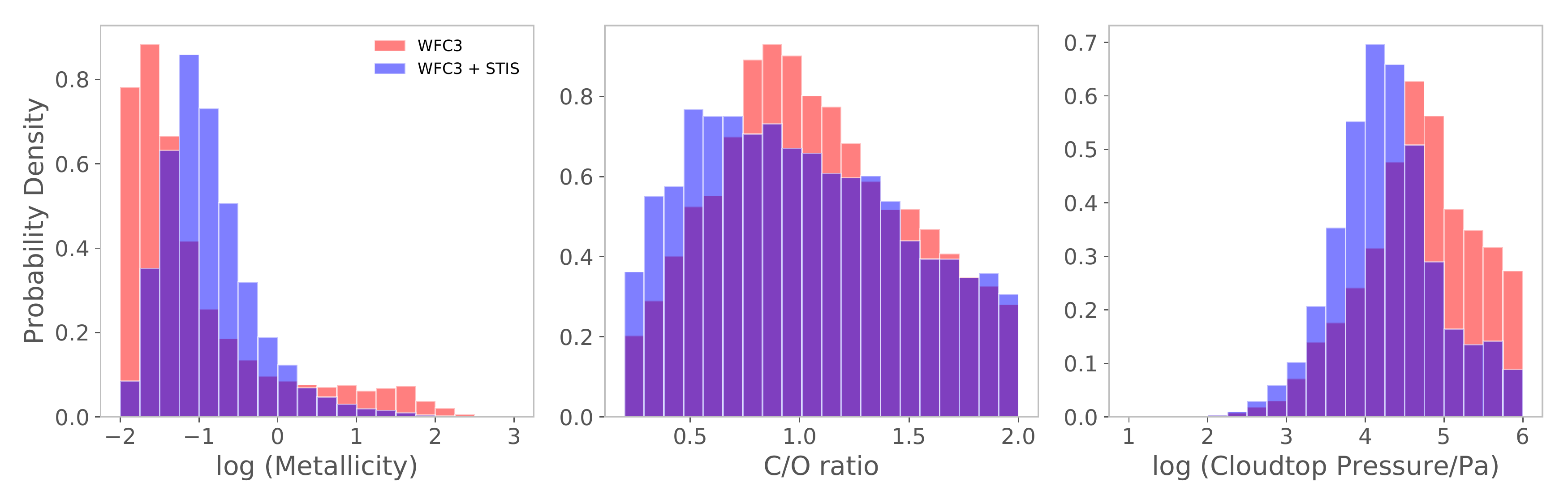}
	\caption{Marginalized posterior probability distributions for the metallicity, C/O ratio, and cloud-top pressure from a fit to the WFC3 data alone and a fit to the WFC3 + STIS dataset.}
	\label{fig:wfc3_hst_1d_hist}
\end{figure*}

We find that HAT-P-11b's atmospheric parameters, in particular its metallicity, are poorly constrained in these fits (see Figure~\ref{fig:wfc3_hst_1d_hist} and Table~~\ref{table:PLATON_best_fit}).  The limited wavelength range of the WFC3 data limits our ability to uniquely infer the metallicity and cloud top pressure.  As for the C/O ratio, the presence of absorption features due to water does not automatically imply a C/O ratio $<0.9$ for planets with equilibrium temperatures $\lesssim$800-1000 K as it does for hot Jupiters\footnote{The exact transition temperature depends on other properties such as atmospheric metallicity and surface gravity.} \citep{Madhusudhan2012, Kreidberg2015a, Heng2018}.  Below $\sim$800 K, methane is the thermodynamically favored carbon-bearing species in hot Neptunes, except at very high atmospheric metallicities \citep{Moses2013}.  Adding more carbon relative to oxygen does not therefore increase the abundance of CO at the expense of water.  Our models indicate that increasing the C/O ratio (even to values greater than one) at temperatures below 800 K has a negligible effect on the water abundance and the methane abundance simply increases linearly with C/O.

The results from this retrieval differ significantly from those presented in \citetalias{Fraine2014} primarily for three reasons.  Firstly, we include WFC3 G102 data here that have small uncertainties and consequently a strong influence on the retrieved posteriors.  The addition of WFC3 G102 data shifts the peak of the metallicity posteriors to lower values.  When we utilize only the WFC3 G141 data (or WFC3 G141 + \emph{Spitzer} data with an offset for the \emph{Spitzer} data), our retrieved results agree with \citetalias{Fraine2014}'s.  Secondly, we apply a wavelength dependent stellar activity correction that changes the spectrum in such a way that a low metallicity - deep cloud solution fits the data.  To test whether this shift to low metallicity is due to our stellar activity correction, we combined the WFC3 G141 spectrum from \citetalias{Fraine2014} and our WFC3 G102 spectrum and performed retrieval analysis on the corrected and uncorrected version of the combined spectrum.  We found that applying the stellar activity correction shifts the posteriors to low metallicity.  Thirdly, we choose a different prior for atmospheric metallicity and extend it to $0.01 \times$ solar so as to resolve the posterior for the retrieved metallicity.  \citetalias{Fraine2014} only explored atmospheric metallicities $\geq 1 \times$ solar in their retrievals and we find that restricting our prior space to match theirs results in significantly better agreement.  Additionally, our models do not favor atmospheric metallicities $\gtrsim 100 \times$ solar primarily because our spectrum, unlike the one published in \citetalias{Fraine2014}, favors the presence of methane in the atmosphere (see \S~\ref{sec:mol_detection_sig} for more details).

\begin{table*}
\small
	\centering
    	\caption{Median Parameters and 68\% Confidence Intervals (CI) from \texttt{PLATON} Retrieval} 
    	\label{table:PLATON_best_fit}
    	\renewcommand{\arraystretch}{1.2}
    	\begin{center}
        	\begin{tabular}{lccc>{\bf}c>{\bf}ccc}
        		\hline \hline
        		Parameter & Prior & \multicolumn{2}{c}{HST WFC3} & \multicolumn{2}{c}{{\bf HST WFC3 + STIS}\footnote{We regard this to be the most reliable retrieval. See \S~\ref{sec:retrieval_results} and \S~\ref{sec:discussion}}} & \multicolumn{2}{c}{HST + Spitzer} \\ 
        		 & & Median & 68\% CI & Median & 68\% CI & Median & 68\% CI  \\ \hline
        		Isothermal Temperature (K) & [500, 1200] & 941 & [726, 1114] & 740 & [635, 876] & 736 & [540, 1026]   \\
        		log (Metallicity/Z$_{\odot}$) & [-2, 3] & -1.39 & [-1.79, -0.16] &  -0.98 & [-1.40, -0.36] & 2.04 & [0.12, 2.75]  \\
        		C/O & [0.2, 2] & 1.03 & [0.62, 1.56]  & 0.97 & [0.51, 1.56] & 0.63 & [0.30, 1.49] \\
        		log (Cloudtop Pressure/Pa) & [1, 6] & 4.71 & [3.96, 5.46] & 4.25 & [3.67, 4.88] &  2.94 & [2.02, 4.77] \\
                log (Particle Size/m) & [-8, -5] &  -6.69 & [-7.61, -5.62] &  -6.67 & [-7.59, -5.65] &  -6.60  & [-7.58, -5.60] \\
                log (Number Density/m$^{-3}$) & [-10, 15] &  -1.70 & [-7.11, 3.92] & -1.70  & [-7.09, 3.67] & -0.59 & [-6.65, 5.42]  \\
                Error Multiple ($\sigma_{\mathrm{mult}}$) & [0, 4]  &  1.46 & [1.29, 1.69]  & 1.32  & [1.19, 1.48] &  1.67 & [1.51, 1.87]  \\
        		\hline
        	\end{tabular}
    	\end{center}
\end{table*}

\subsubsection{HST WFC3 + STIS}
\begin{figure*}
    \centering
    \includegraphics[width=0.8\linewidth]{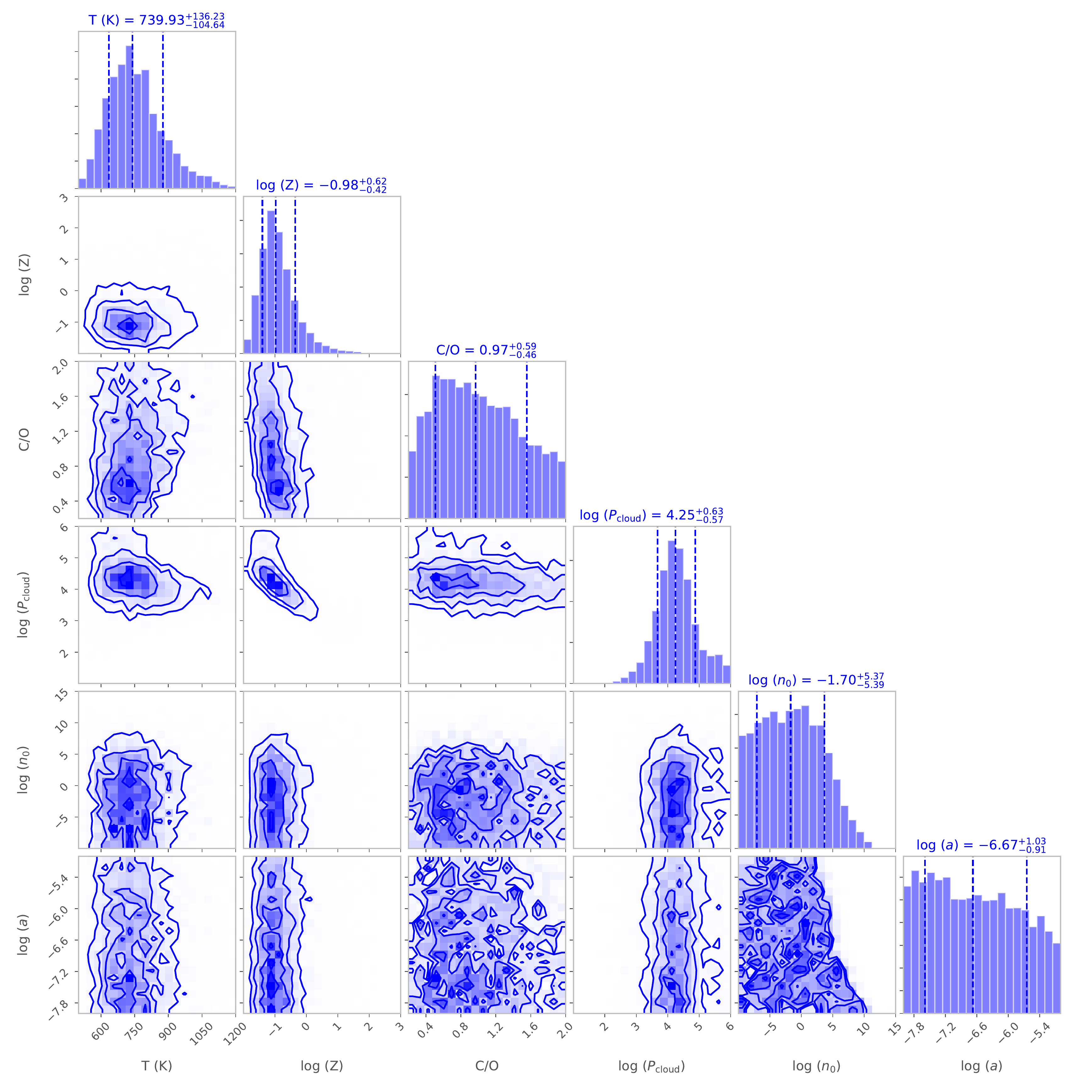}
    \caption{Posterior probability distributions for fits of the \emph{HST} dataset.  Median parameter values and 68\% confidence intervals for the marginalized 1D posterior probability distributions are indicated with vertical dashed lines.}
    \label{fig:hst_posterior}
\end{figure*}

Next, we see how the inclusion of STIS data alters the posteriors for these parameters.  Because our data now span a much larger wavelength range, we must include wavelength-dependent scattering in our model (\S~\ref{sec:clouds_platon}).  The best-fit model is shown in Figure~\ref{fig:HST_spectrum_fit}, parameter constraints are tabulated in Table~\ref{table:PLATON_best_fit}, and the full posteriors for key atmospheric parameters are shown in Figure~\ref{fig:hst_posterior}.  The data place relatively tight constraints on the cloud-top pressure, indicating that we are probing down to $\sim 100$ mbar.  This is in rough agreement with the inferred (grey) cloud top pressures of $10 - 50$ mbar for CARMA models.  The constraints on atmospheric metallicity are significantly tighter than those provided by WFC3 data alone, with a $2 \sigma$ confidence interval of $0.02 - 4.6 \; \times$ solar.  The posterior for atmospheric metallicity has a skewed shape with a long tail towards high metallicities.  We find that the $3 \sigma$ upper limit for metallicity is $86 \; \times$ solar, indicating that enhanced metallicities are still consistent with our data.  Unlike \citetalias{Fraine2014}, our fits prefer lower atmospheric metallicities.  Nonetheless, for metallicities greater than the lower prior bound in \citetalias{Fraine2014} ($1 \times$ solar), our metallicity posteriors are in qualitative agreement with the ones published in \citetalias{Fraine2014}.  The addition of the STIS data to WFC3 data limits the degeneracy between cloudtop pressure and atmospheric metallicity (see Figure~\ref{fig:hst_posterior}) encountered by \citetalias{Fraine2014}, resulting in correspondingly narrower constraints on these properties \citep{Benneke2012a}.

We show the marginalized posterior probability distributions for metallicity, C/O, and cloudtop pressure in Figure~\ref{fig:wfc3_hst_1d_hist}.  The \emph{HST} STIS data provide additional constraints on atmospheric properties by disfavoring models with very low metallicity (log $(Z) \lesssim -1.5$, and correspondingly high cloud top pressure $P_{\mathrm{cloud}}$) and high metallicity (log $(Z) \gtrsim 1$).  This is apparent in Figure~\ref{fig:HST_spectrum_fit} where we see that the STIS data narrow the range of model transmission spectra that agree within $\pm 1 \sigma$.

\begin{figure*}
    \centering
    \includegraphics[width=0.8\linewidth]{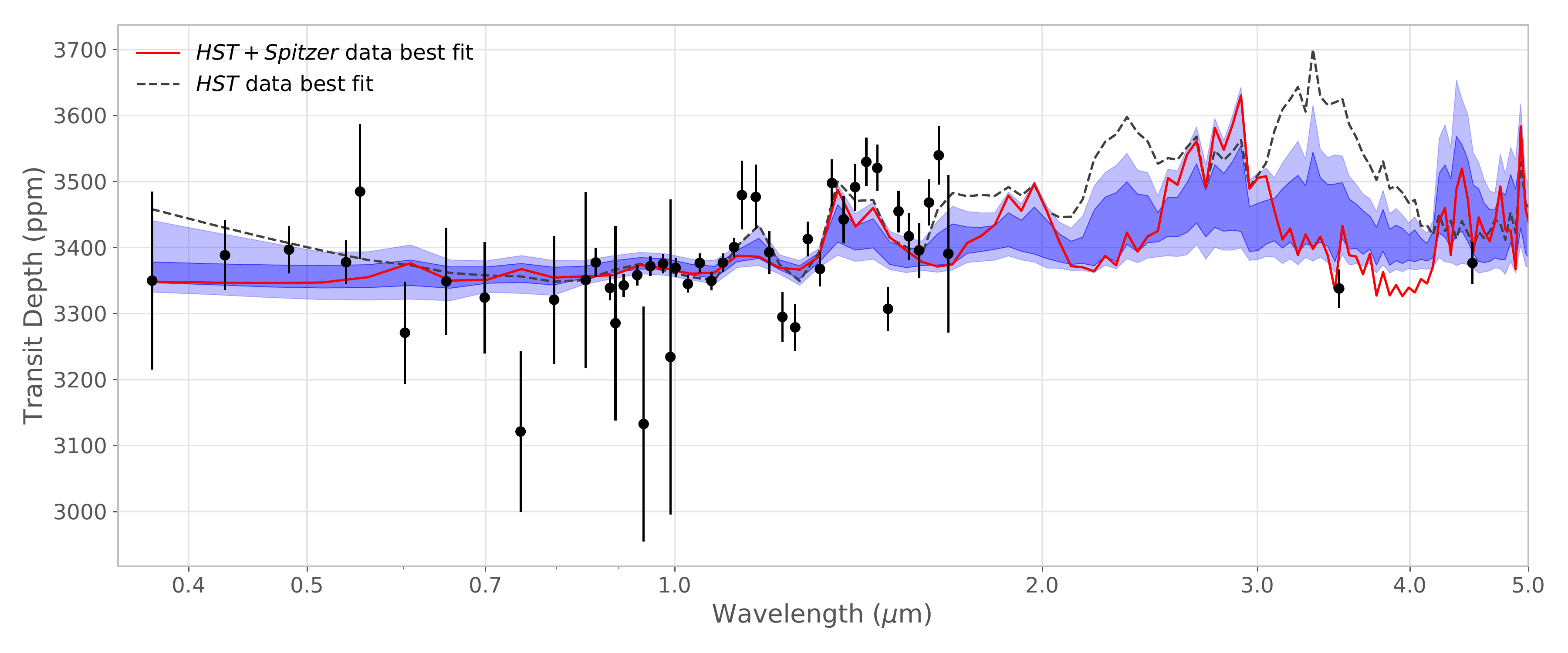}
    \caption{Transmission spectrum for HAT-P-11b including both \emph{HST} and \emph{Spitzer} data (black filled circles) along with the best-fit model from \texttt{PLATON} and corresponding $1\sigma$ and $2\sigma$ contours (dark blue and light blue, respectively).  The best-fit model for \emph{HST} data is also shown for comparison, which predicts a much larger transit depth at 3.6 $\mu$m.  The inclusion of the \textit{Spitzer} transit depths shifts the models toward solutions with high atmospheric metallicity, which suppresses the depth of the absorption features in the WFC3 bands and decreases the overall quality of the fit in this region.}
    \label{fig:full_transmission_spectrum}
\end{figure*}

\subsubsection{HST + Spitzer data}
We carry out a final set of fits including both the \emph{HST} STIS + WFC3 and \emph{Spitzer} transit depths.  The full transmission spectrum with the best-fit model from \texttt{PLATON} is shown in Figure~\ref{fig:full_transmission_spectrum} and the median and confidence intervals for retrieved parameters are given in Table~\ref{table:PLATON_best_fit}.  Our 3.6 $\mu$m \textit{Spitzer} transit depth is low relative to the \textit{HST} data and discrepant with the depth predicted by the best-fit model to the \emph{HST} data.  We are unable to find a single model that can simultaneously match the observed strength of the WFC3 absorption features while fitting the noticeably shallower \emph{Spitzer} transit depths.

The inclusion of \textit{Spitzer} data worsens the constraints on most atmospheric parameters (Table~\ref{table:PLATON_best_fit}).  The acceptable temperature and cloudtop pressure ranges now span the entire prior range.  The constraints on metallicity from this fit are inconsistent with results from the \emph{HST}-only fits.  The preferred metallicity rises to a few $100 \times$ solar, which allows the models to fit the flat baseline of the data by reducing the scale height while still maintaining some molecular absorption and reducing the relative abundance of methane in the atmosphere.  We find that the particle size and number density are relatively unconstrained in both the \emph{HST}-only and \emph{HST} + \emph{Spitzer} fits.  The upper limit on the number density varies as a function of particle size (as expected) and is marginally higher for the \emph{HST} + \emph{Spitzer} fit.  The error multiple ($\sigma_{\mathrm{mult}}$) parameter, which is a measure of how underestimated the errors in the data are, jumps to $\sim 1.7$, i.e. $> 20$\% larger than the value obtained with \textit{HST} data alone.  In addition, the reduced $\chi^{2}$ value (calculated using the errors on the transit depth measurements) increases from 1.9 for the \emph{HST}-only fit to 2.8 for the full dataset fit.  We therefore conclude that our models are unable to provide a satisfactory fit to the full dataset. Including an offset of $\sim$ 100--150 ppm could reconcile the \emph{Spitzer} depths with the models that fit the \emph{HST} data.  Fitting for this offset in a retrieval framework also yields similar estimates for its magnitude.  However, as emphasized in \S~\ref{sec:stellar_activity} and \ref{sec:wlc_fits}, such a large stellar activity correction is incommensurate with the observed stellar variability.

\subsubsection{Retrievals without methane and/or water opacity}
\label{sec:mol_detection_sig}
We quantify the significance of observed molecular absorption features by using the evidence obtained from nested sampling to compute Bayes factor for model comparisons.  To test for the presence of a certain molecule (and the associated confidence/significance), we remove opacity contributions from the molecule and refit the transmission spectrum while keeping the priors unchanged.  The ratio of the Bayesian evidence for fits with and without the molecular opacity yields the Bayes factor and allows us to quantify the data's preference for one model over the other \citep[e.g.][]{Benneke2013}.  There is significant overlap between methane and water features in the near-infrared region (0.8 -- 1.7 $\mu$m), and we therefore perform three additional retrievals for the \emph{HST} data along with the nominal case described above. In these three retrievals, we remove both water and methane opacity, just water opacity, and just methane opacity. 

\begin{table}[]
    \centering
     \caption{\emph{HST} Retrievals Evidence}
    \begin{tabular}{cccc}
    \hline \hline
    Model & log (Evidence) & Bayes & $\sigma$  \\ 
     &  & factor &   \\ \hline
    Nominal & 368.9 $\pm$ 0.1 & -- & \\
    Without CH$_4$ and H$_2$O & 361.0 $\pm$ 0.1 & 1:2812 & 4.4  \\
    Without CH$_4$ & 364.8 $\pm$ 0.1 & 1:64 & 3.4  \\
    Without  H$_2$O & 366.1 $\pm$ 0.1 & 1:17 & 2.9 \\
    \hline 
    \end{tabular}
    \label{tab:hst_retrieval_evidence}
\end{table}

The evidence, Bayes factor (relative to the nominal model that includes both methane and water opacity), and equivalent $\sigma$ significance for each of the three cases are shown in Table~\ref{tab:hst_retrieval_evidence}.  The combined significance for the presence of water and methane is 4.4 $\sigma$.  The Bayes factor for the two molecules individually is lower than the reported combined significance.  The detection significance for each molecule is sensitive to relatively subtle features of the spectrum and may change due to small differences in the shape of the absorption features.  Notably, the inclusion of \emph{HST} STIS data makes the case for the presence of water and/or methane stronger.  With WFC3 data alone, a similar comparison gives lower values for the Bayes factor for all three retrievals.  This is primarily because the relatively flat optical spectrum excludes very low atmospheric metallicity models (log $Z \lesssim - 1.5$), which possess somewhat higher evidence values (in \emph{HST} WFC3 only retrievals) and therefore weaken the case for the presence of these molecules.

\begin{figure*}
	\centering
    \subfigure[]{\includegraphics[width=0.48\linewidth]{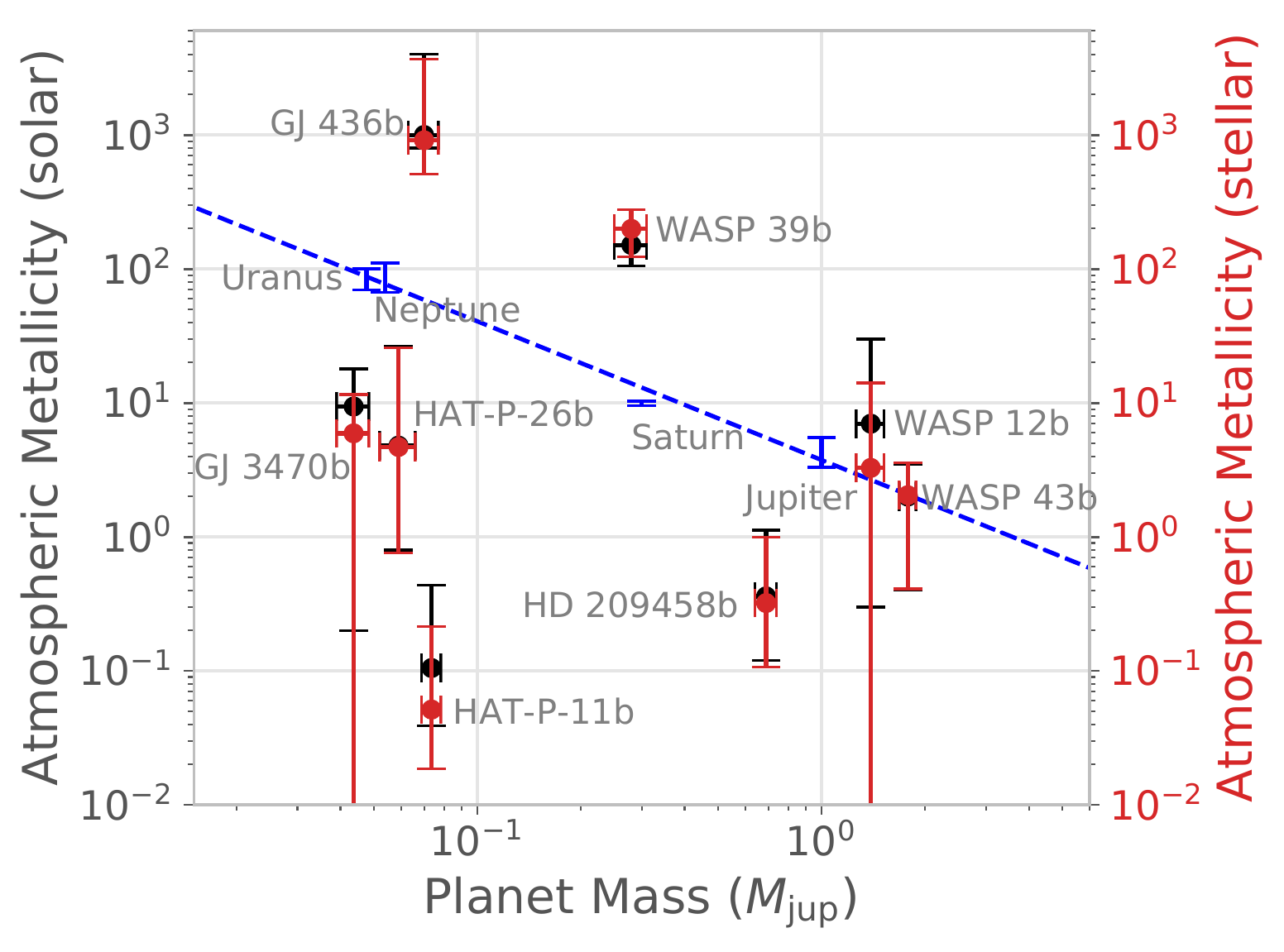}}
	\subfigure[]{\includegraphics[width=0.48\linewidth]{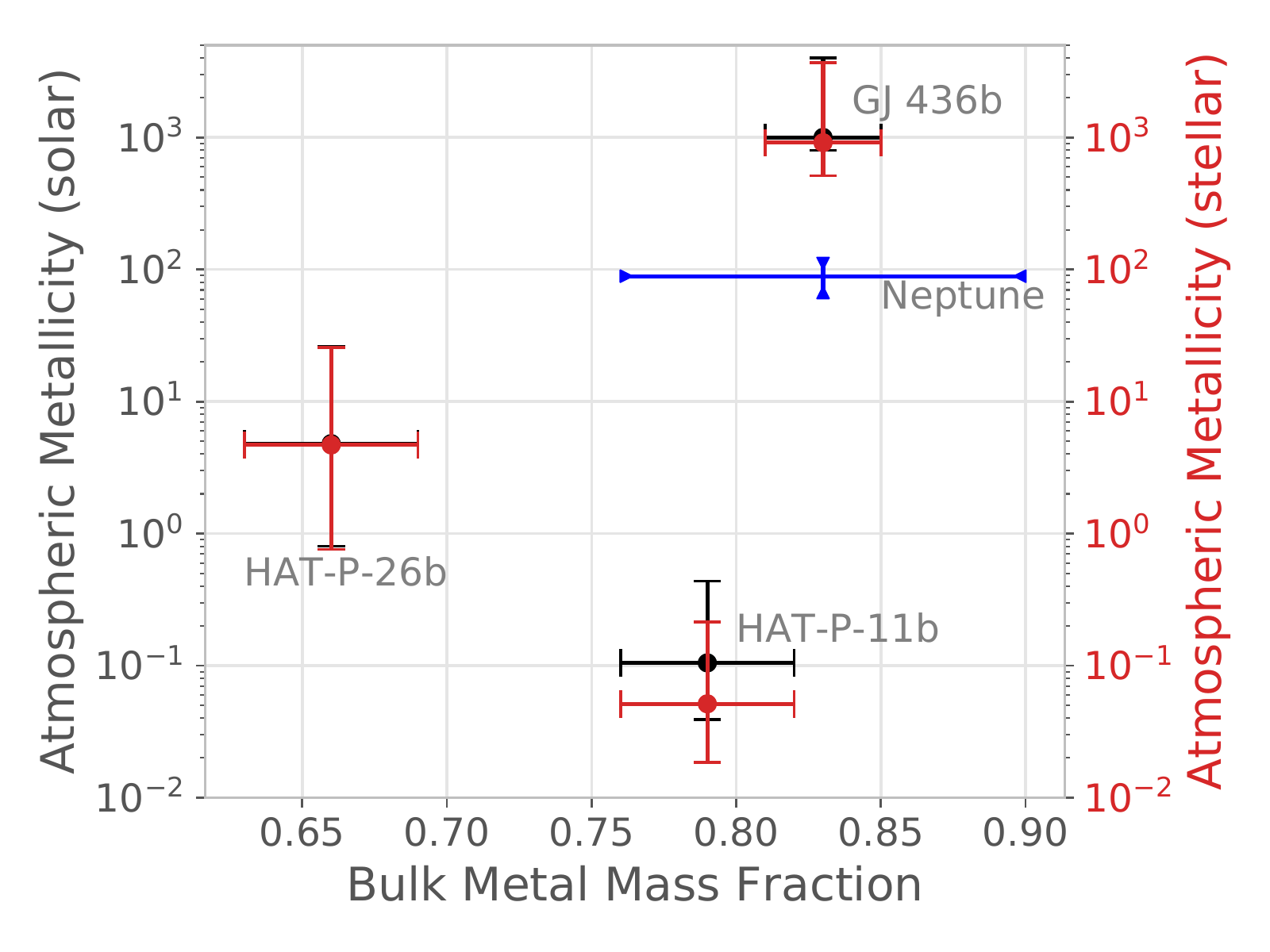}}
	\caption{(a) Atmospheric metallicity versus planet mass for planets observed by \emph{HST} and \emph{Spitzer} \citep{Kreidberg2014, Kreidberg2015a, Wakeford2017b, Morley2017, Brogi2017, Wakeford2018, Benneke2019}. 
	(b) Atmospheric metallicity versus bulk metallicity \cite[obtained from][]{Thorngren2018} for Neptune-class planets.  For Neptune, we plot lower and upper limits rather than 1$\sigma$ error bars \citep{Helled2018}.  GJ 3470b is not included on this plot because the assumptions used to derive bulk metallicity constraints in the \cite{Thorngren2016} models may not be appropriate for planets with such low masses.
	}
	\label{fig:atm_met_trends}
\end{figure*}

This exercise also allows us to investigate whether the disagreement between inferences made from \emph{HST} and \emph{Spitzer} data arises simply due to the absence of methane from the atmosphere.  Vertical mixing and quenching could lower the methane abundance by orders of magnitude relative to the equilibrium values \citep{Moses2011, Moses2013}.  However, quantifying this effect for HAT-P-11b requires a more careful analysis as its temperature-pressure profile overlaps with the equal abundance curve of CH$_4$-CO.  This picture is further complicated by the planet's orbital eccentricity (see \citealp{Visscher2012}).  We test whether our fit to the \emph{HST} data without CH$_4$ opacity fits the \emph{Spitzer} data any better.  We find that removing methane's opacity requires a larger abundance of water to match the strength of the spectral features in the WFC3 bandpass.  This pushes the best-fit models to higher metallicities (lower abundances/metallicities are ruled out by the STIS data).  The best-fit models thus obtained match the 3.6 $\mu$m depth quite well but the higher atmospheric metallicities imply the presence of a substantial amount of CO and CO2 as well, which increases the 4.5 $\mu$m model depth and make it as discrepant with the data as the 3.6 $\mu$m depth is in our nominal model, which includes methane opacity.

\section{Discussion and Conclusions}
\label{sec:discussion}

Our picture of HAT-P-11b's atmosphere is primarily driven by the \emph{HST} observations, which provide a self-consistent, spectrally resolved picture of the planet's atmosphere over nine separate transit observations.  The fact that we see clear evidence for molecular absorption across multiple visits and multiple bands leads us to conclude that any plausible model for this planet's atmosphere must be able to reproduce the observed shape of these absorption (water + methane) bands.  These models all overestimate the observed transit depth in the 3.6 $\mu$m \emph{Spitzer} band; this may indicate that methane is under-abundant in HAT-P-11b's atmosphere as compared to the predictions of our equilibrium chemistry models.  However, comparison of Bayesian evidence for \emph{HST} retrievals suggests that methane is indeed present.  We are unable to resolve these apparent contradictions with the current dataset, but future spectroscopic observations of this planet with the \emph{James Webb Space Telescope} (\emph{JWST}) should provide a much clearer picture of its transmission spectrum in the mid-infrared wavelengths probed by the \emph{Spitzer} photometry.

\begin{figure*}
	\centering
    \subfigure[1 $\times$ solar CARMA models]{\includegraphics[width=0.48\linewidth]{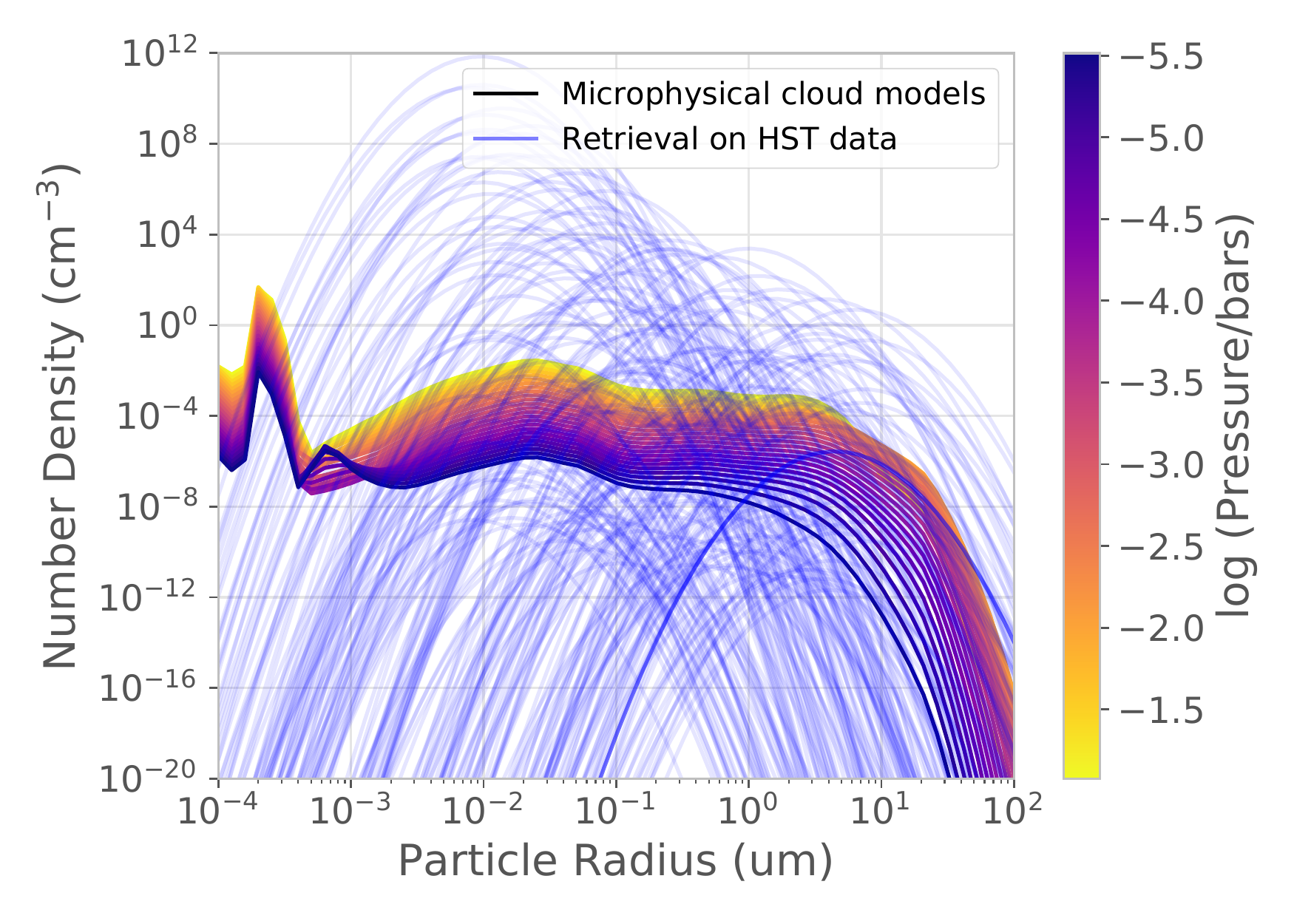}}
	\subfigure[50 $\times$ solar CARMA models]{\includegraphics[width=0.48\linewidth]{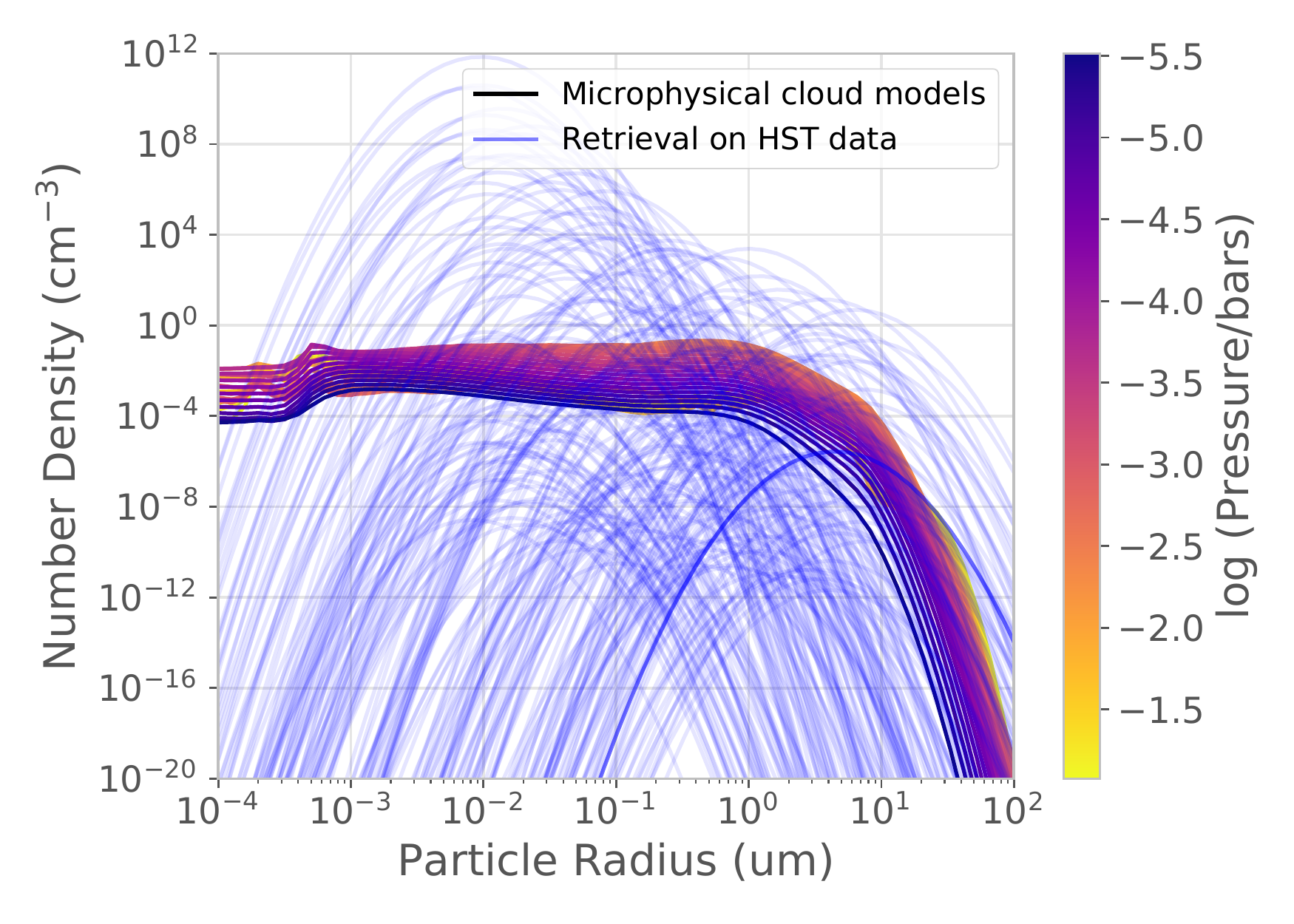}}
	\caption{Particle number density as a function of radius from our microphysical cloud models at different pressures/heights in the atmosphere.  We overplot a sample of lognormal particle size distributions at 10 mbar from our retrievals for comparison.  The best-fit size distribution is highlighted with a dark blue line.  All profiles correspond to models with high likelihoods.
	}
	\label{fig:global_part_size_dist}
\end{figure*}

If we focus our attention for now on the \emph{HST}-only fits, our updated results point to a significantly lower value for the planet's atmospheric metallicity than that reported by \citetalias{Fraine2014}.  This runs counter to the trend observed in the solar system (Figure~\ref{fig:atm_met_trends} (a)):  Uranus and Neptune have atmospheric C/H ratios between $70-100\times$ that of the Sun, while Jupiter's C/H ratio is just a few times solar (\citealp{Wong2004, Fletcher2009, Karkoschka2011, Sromovsky2011}, see also e.g. \citealp{Kreidberg2014}).  Although there are relatively few published constraints on the atmospheric metallicities of Neptune-mass planets around other stars, GJ 436b appears to have an atmospheric metallicity of at least $200\times$ solar \citep{Madhusudhan2011, Moses2013, Morley2017}.  However, HAT-P-26b \citep{Wakeford2017b} provides a counter-example of an extrasolar Neptune with a relativley low atmospheric metallicity ($4.8_{-4.0}^{+21.5} \times$ solar).  Our new observations suggest that HAT-P-11b is more similar to HAT-P-26b than it is to either Neptune or GJ 436b.  The low atmospheric metallicity of HAT-P-11b is all the more striking because it orbits a metal rich star ([Fe/H] = +0.3).  The composition of the planet's atmosphere therefore verges on being almost identical to that of the primordial gas that formed the star.  This diversity in atmospheric composition of Neptune-mass planets suggests that they may not be a homogeneous planet population.

Comparison of atmospheric metallicity with bulk metallicities (mass fraction) calculated by \cite{Thorngren2018} indicates that Neptune class planets may possess low metallicity envelopes despite having a high bulk metal fraction (Figure~\ref{fig:atm_met_trends} (b)).  This implies that most of the solids, which have the potential to enrich the envelope, ought to have finished accreting before the initiation of substantial gas accretion from the disk.  It also requires mixing in the interior to not be strong enough to significantly enrich the envelope.  We expect that the sample of Neptune-mass planets with well-measured atmospheric metallicities will be significantly expanded by \emph{JWST}, providing a much clearer view of the statistical properties of this population of planets.

In addition to providing improved constraints on HAT-P-11b's atmospheric metallicity, our updated transmission spectrum provides us with an opportunity to explore the properties of the scattering particles in this planet's atmosphere.  We find that transmission spectra for our microphysical cloud models agree quite well with the observed \emph{HST} spectrum (Figure~\ref{fig:retrieval_model_comparison}).  In Figure~\ref{fig:global_part_size_dist}, we compare our retrieved cloud properties to those predicted by the models.  The data do not put narrow constraints on these retrieved cloud properties and there is a degeneracy between mean particle size and number density (as evident in Figure~~\ref{fig:hst_posterior}).  Regardless, the upper limit on mean particle size and its corresponding number density is roughly commensurate with predictions from the microphysical cloud models.  Improved constraints provided by new data in the future should enable us to compare the predictions of the forward model and the retrieved parameters more rigorously.  Moreover, the good agreement between the CARMA models and the retrieved models from PLATON (which uses local condensation from GG-chem\footnote{GG-chem is an open source thermo-chemical equilibrium code that calculates abundances of different molecular and atomic species given gas elemental composition, temperature, and pressure \citep[][\texttt{https://github.com/pw31/GGchem}]{Woitke2018}.} to deplete the gas phase) is reassuring because it is usually unclear if the amount of retrieved cloud opacity is realistic or not compared to the gas phase chemistry and condensation.

In the future, more accurate microphysical cloud models will be crucial for improving our understanding of the properties of these atmospheres.  Better a priori predictions for cloud formation could allow future \emph{JWST} observers to identify and prioritize observations of planets with relatively cloud-free terminators, while model-based constraints on cloud properties would help to limit degeneracies between cloud properties and atmospheric metallicity for planets with cloudy atmospheres.  Our observations of HAT-P-11b serve as a useful illustration of both the limitations of our current understanding of cloud formation in these atmospheres, and also the power of spectrally resolved data with broad wavelength coverage to provide useful constraints on atmospheric composition despite our limited understanding of relevant cloud formation processes.

\section*{Acknowledgements}
This work is based on observations from the Hubble Space Telescope, operated by AURA, Inc. on behalf of NASA/ESA.  This work also includes observations made with the Spitzer Space Telescope, operated by the Jet Propulsion Laboratory, California Institute of Technology under a contract with NASA.  Support for this work was provided by NASA through Space Telescope Science Institute grants GO-14260 and GO-14767.  P. Gao and I. Wong acknowledge the generous support of the Heising-Simons Foundation via the 51 Pegasi b fellowships in Planetary Astronomy.  J. K. Barstow acknowledges funding support from the Royal Astronomical Society Research Fellowship.  G. W. Henry acknowledges support from NASA, NSF, Tennessee State University, and the State of Tennessee through its Centers of Excellence program.  We would like to thank D. Deming and J. Fraine for exchanges regarding the WFC3 G141 dataset.  Y. C. also thanks C. V. Morley, S. H\"{o}rst, Y. Kawashima, Y. Wu, E. Lee, and Y. Yung for inspiring conversations.

\bibliographystyle{apj}
\bibliography{manuscript_hat11}

\appendix

\begin{figure*}[h]
	\centering
	\includegraphics[width=0.6\linewidth]{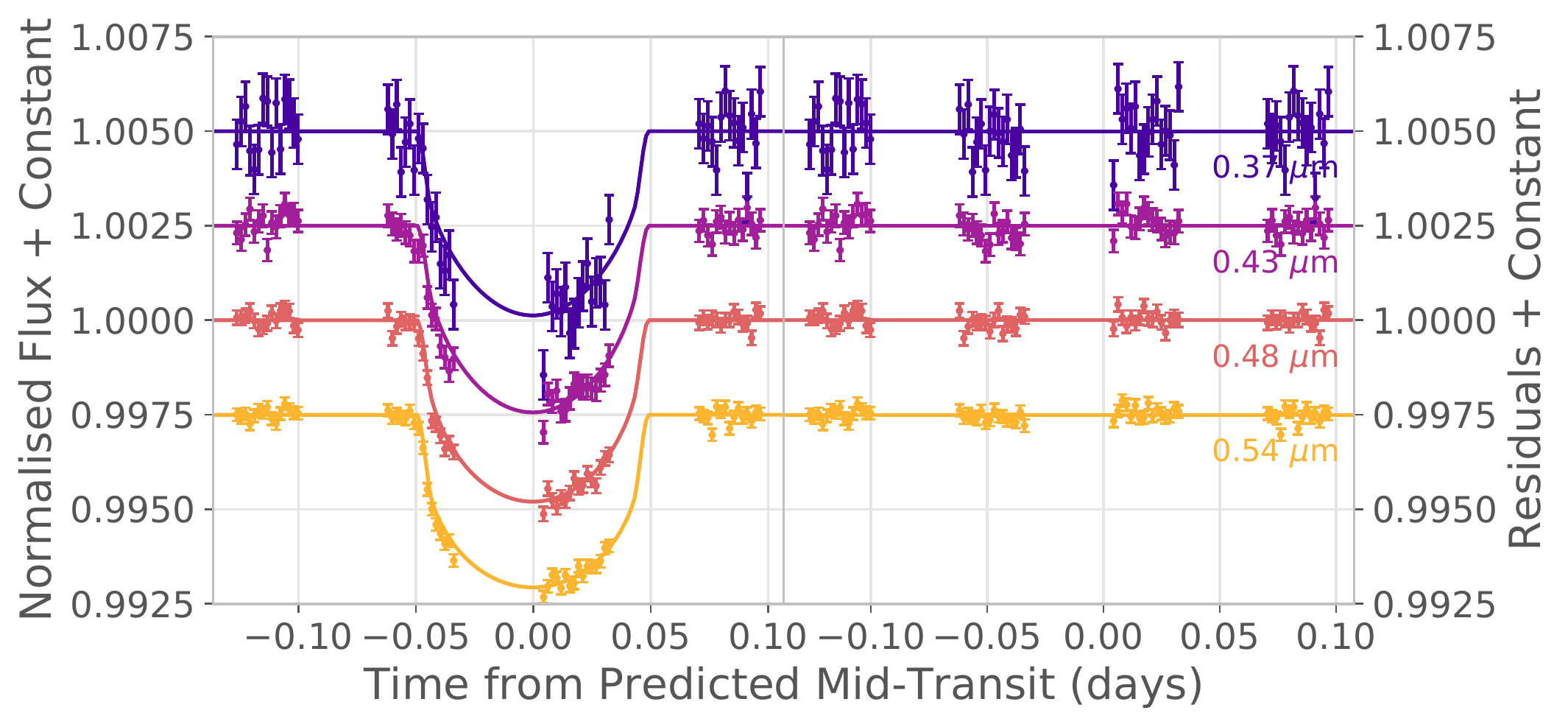}
	\includegraphics[width=0.6\linewidth]{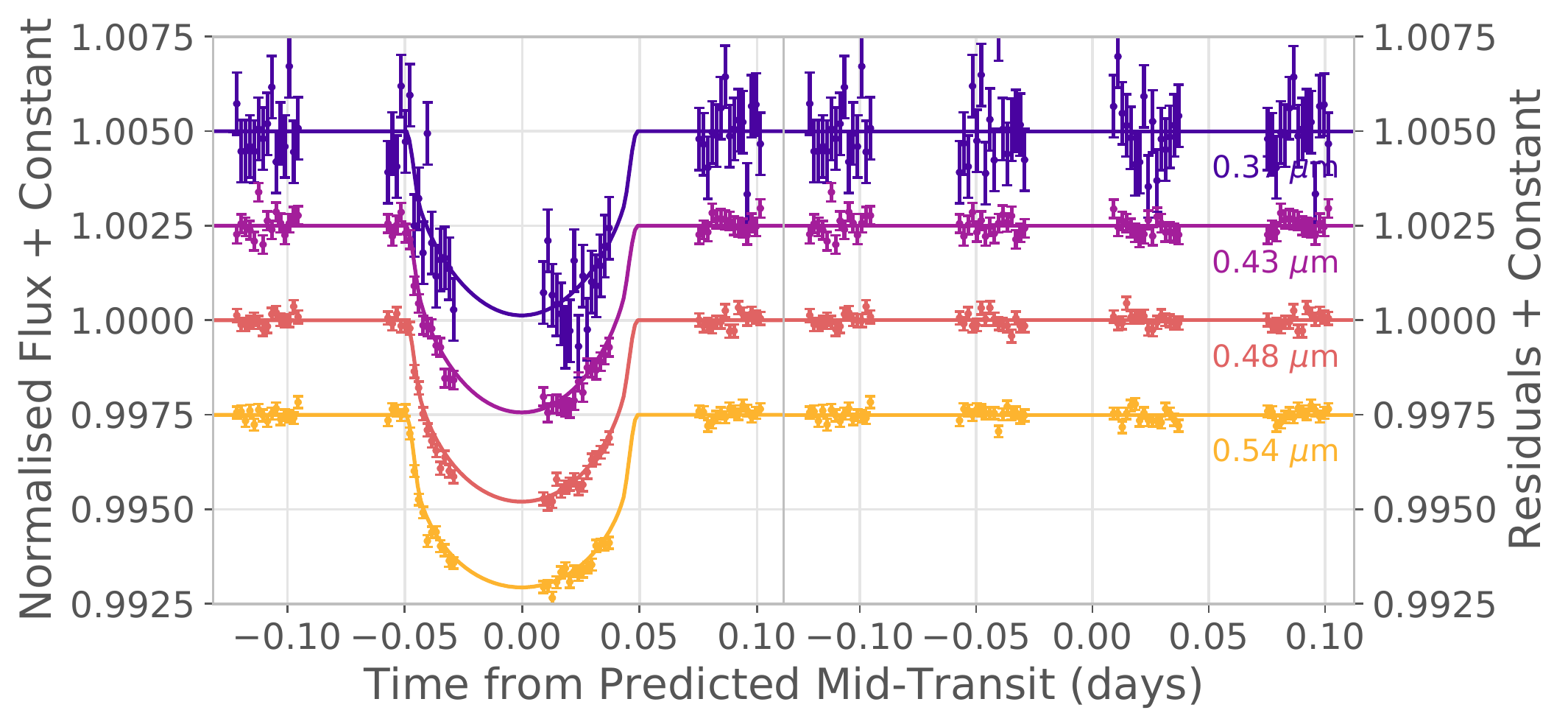}
	\caption{\textit{HST} STIS G430L wavelength dependent light curves for visit 1 and 2.}
	\label{fig:G430_WaveLC_fits}
\end{figure*}

\begin{figure*}
	\centering
	\includegraphics[width=0.6\linewidth]{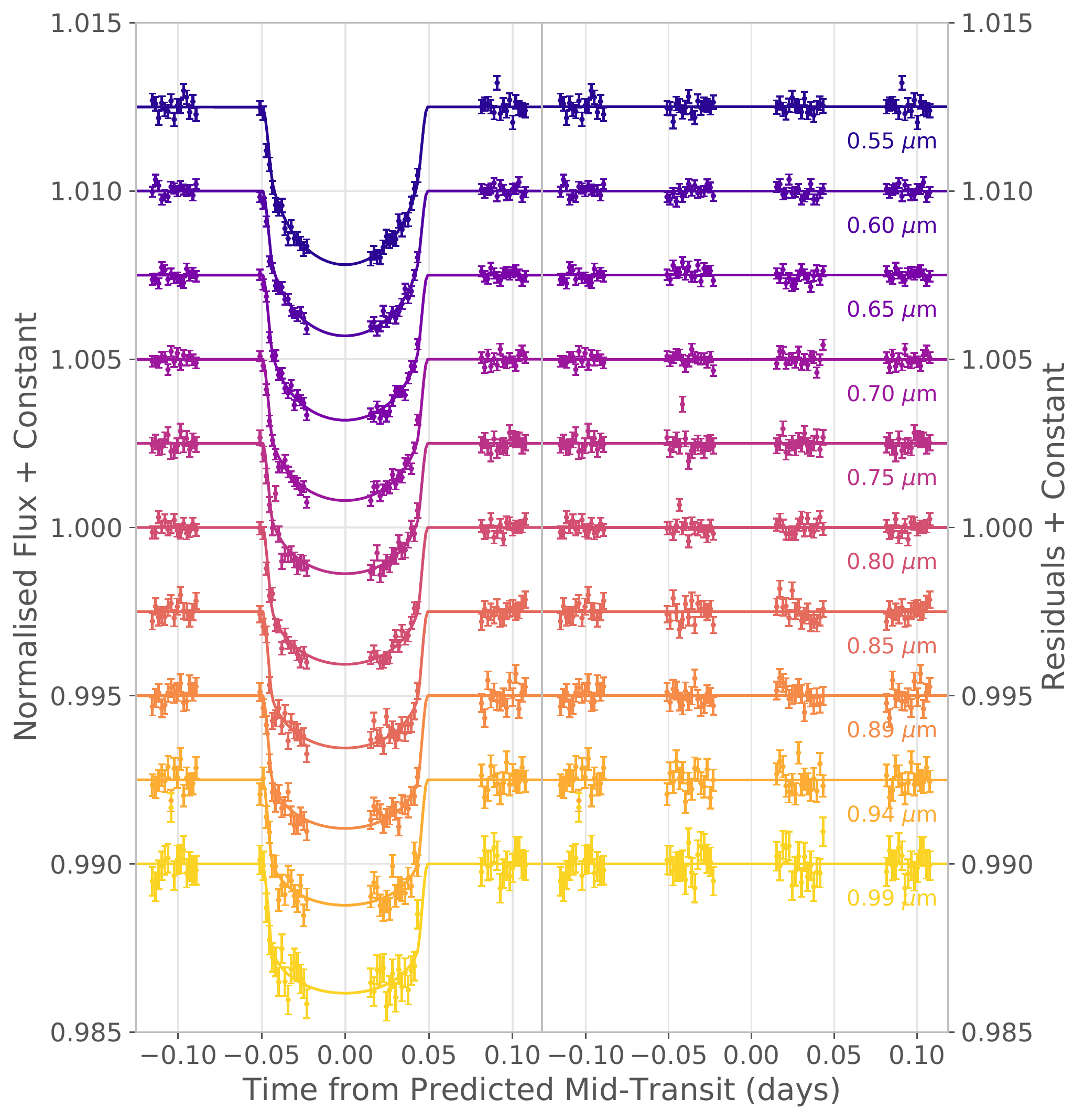}
	\caption{\textit{HST} STIS G750L wavelength dependent light-curves.}
	\label{fig:G750_WaveLC_fits}
\end{figure*}

\begin{figure*}
	\centering
	\includegraphics[width=0.45\linewidth]{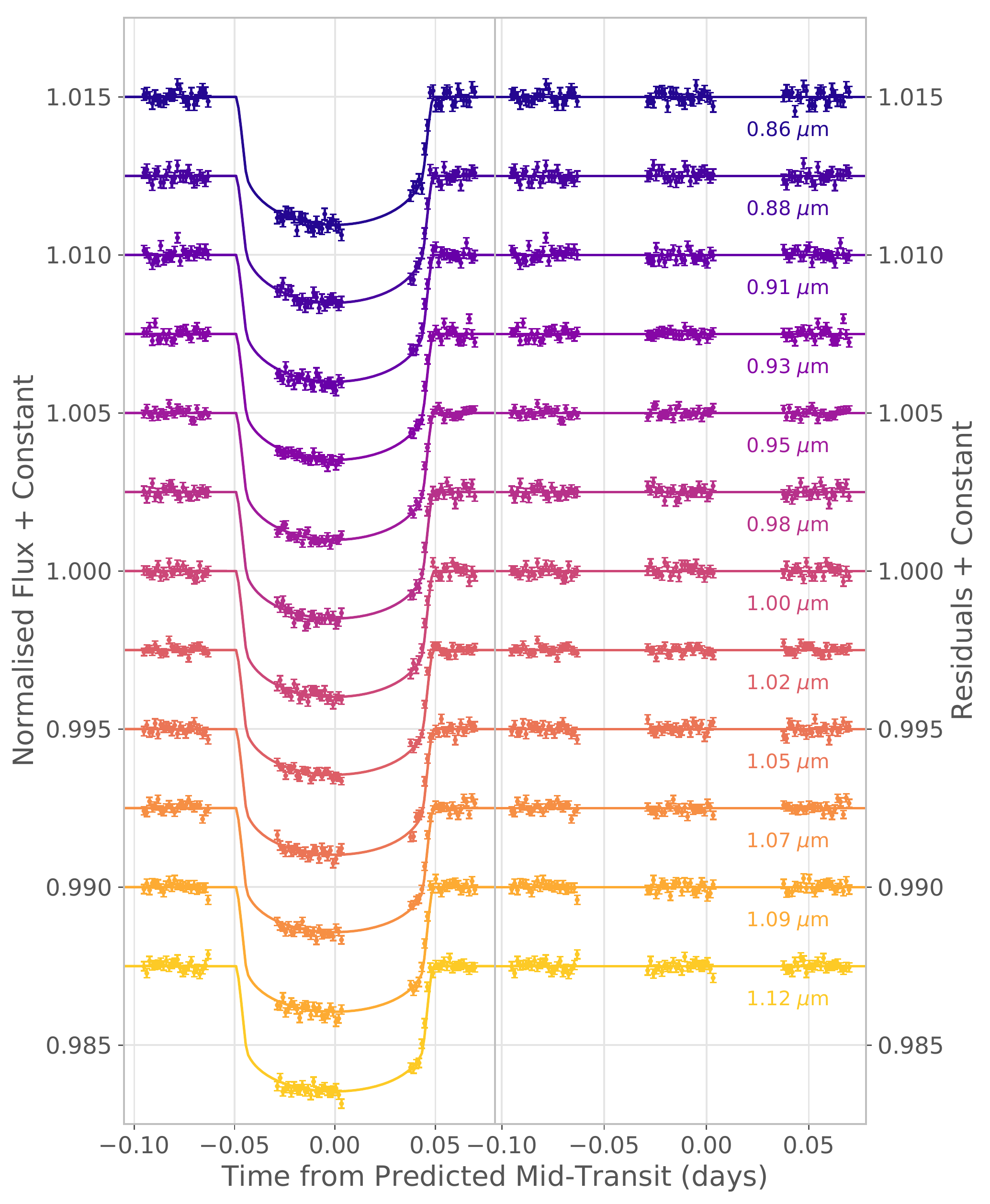}
	\includegraphics[width=0.45\linewidth]{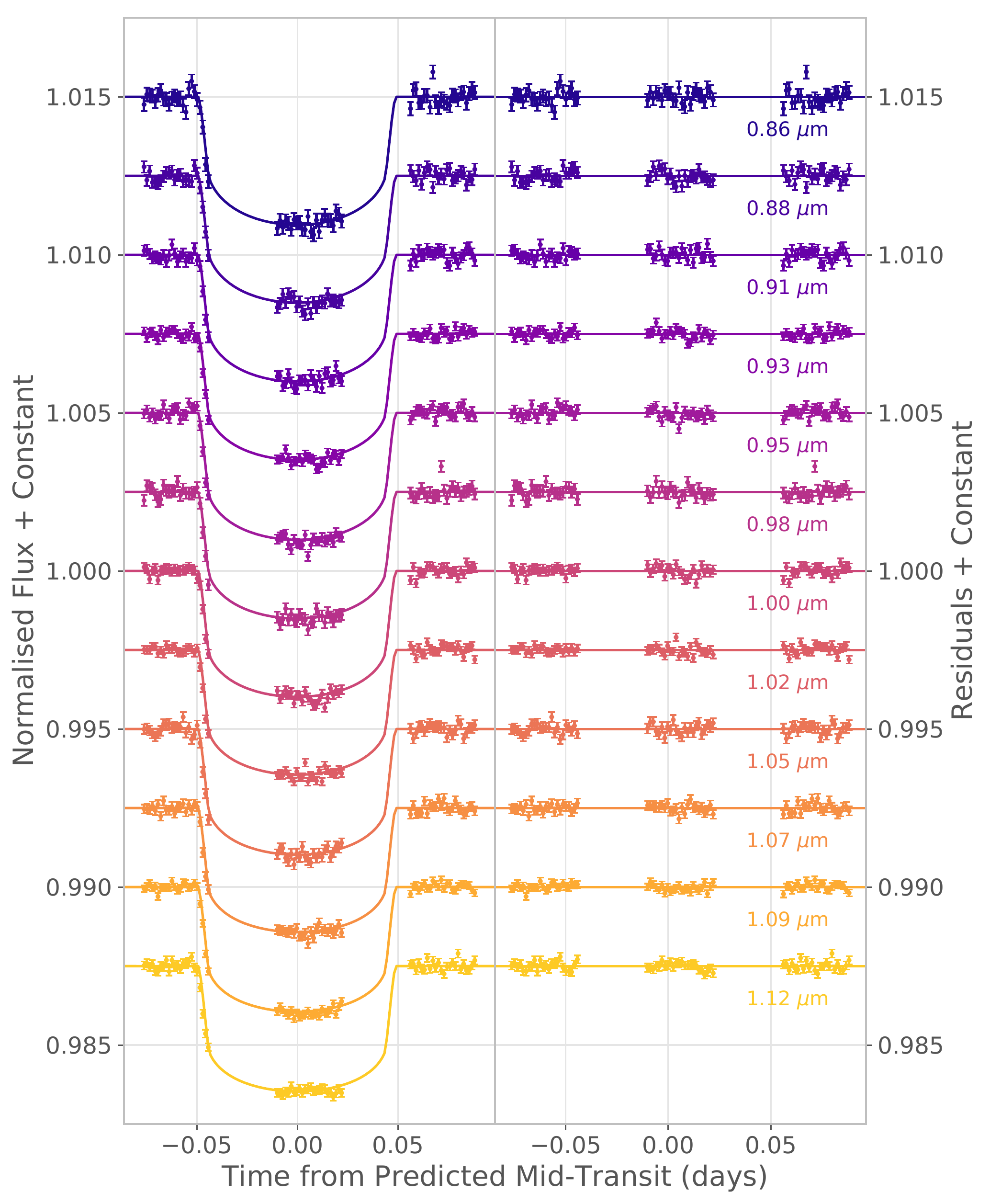}
	\caption{\textit{HST} WFC3 G102 wavelength dependent light curves for visit 1 and 2.}
	\label{fig:G102_1_WaveLC_fits}
\end{figure*}

\begin{figure*}
	\centering
	\includegraphics[width=0.45\linewidth]{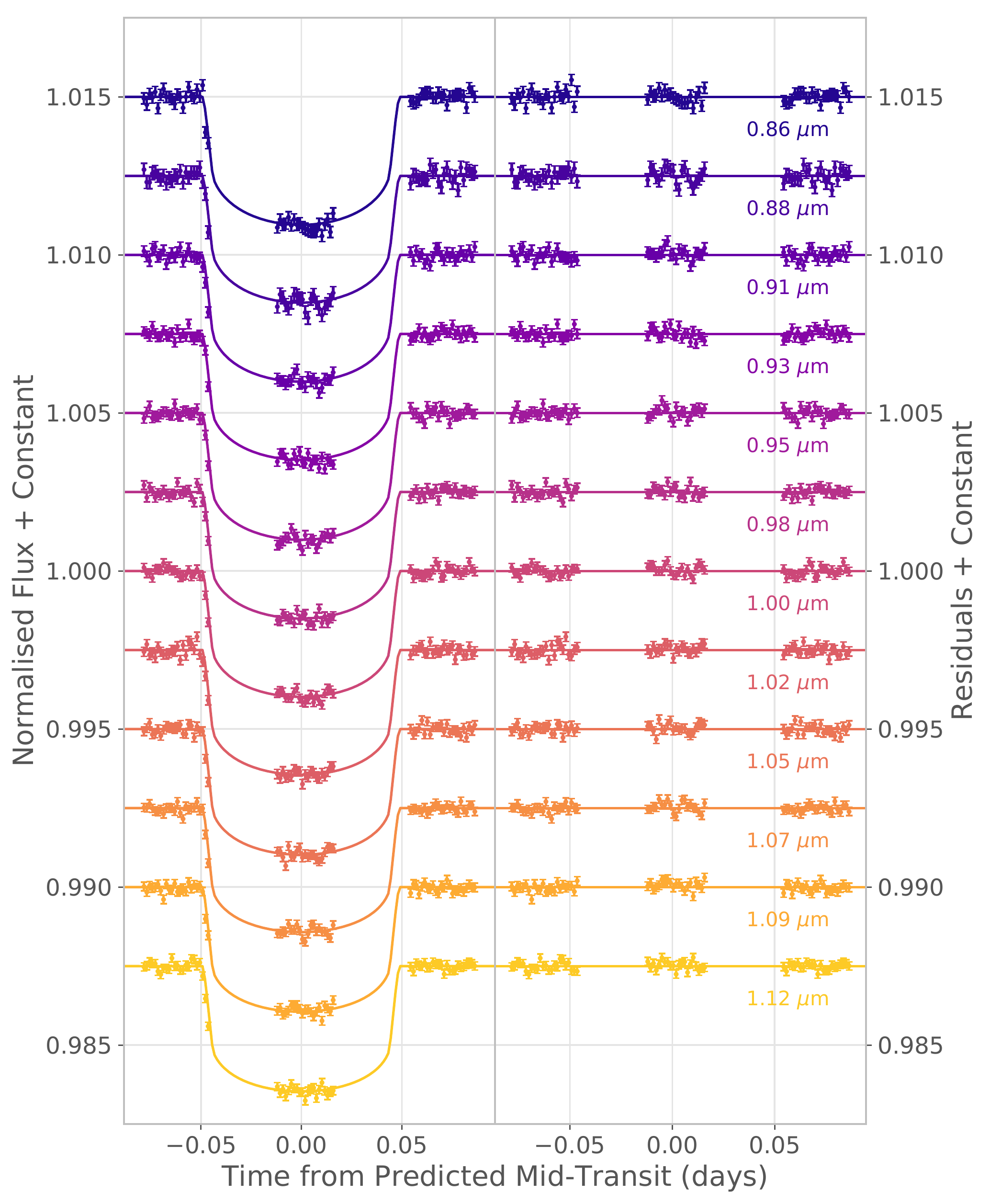}
	\includegraphics[width=0.45\linewidth]{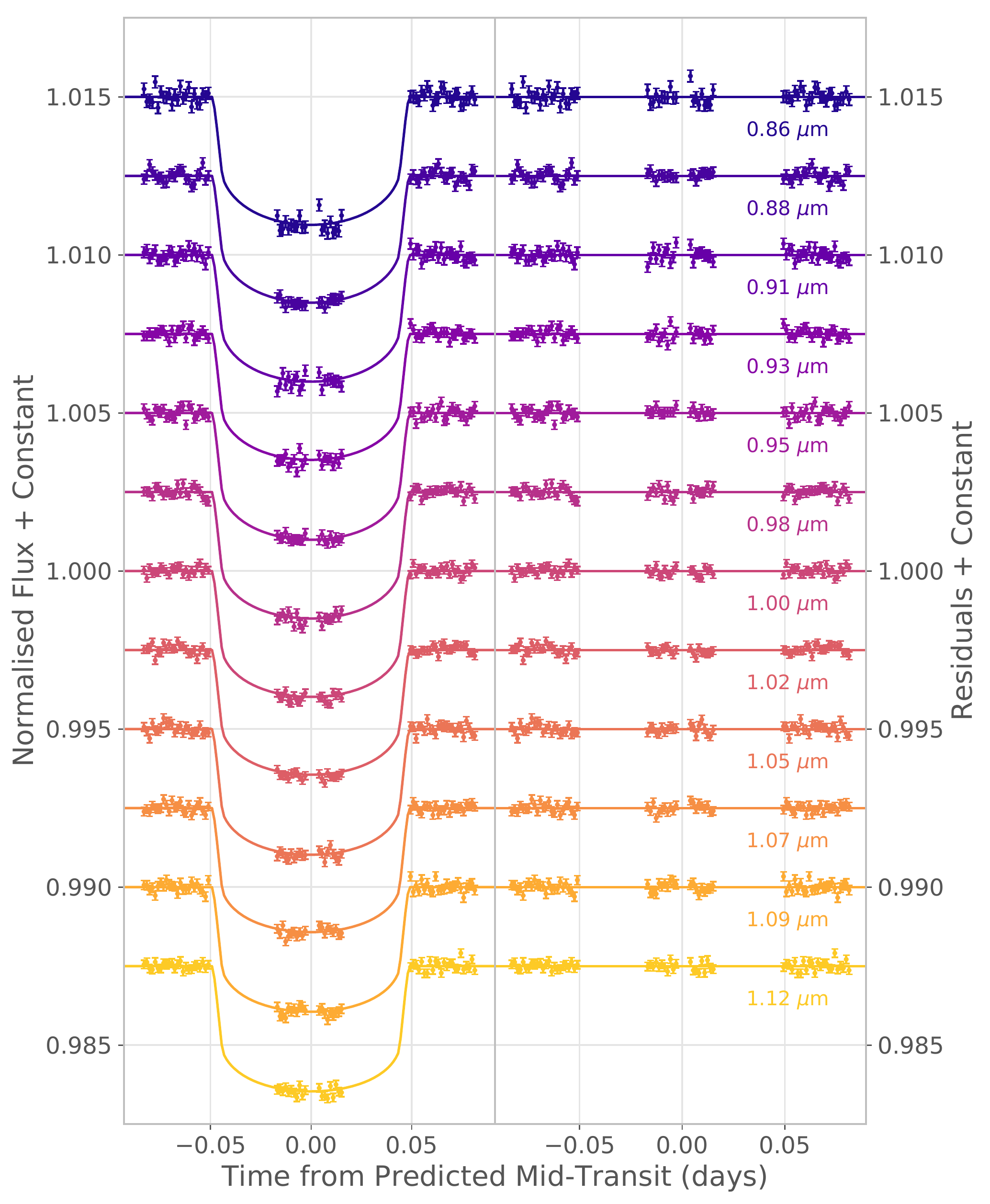}
	\caption{\textit{HST} WFC3 G102 wavelength dependent light curves for visit 3 and 4.}
	\label{fig:G102_2_WaveLC_fits}
\end{figure*}

\begin{figure*}
	\centering
	\includegraphics[width=0.45\linewidth]{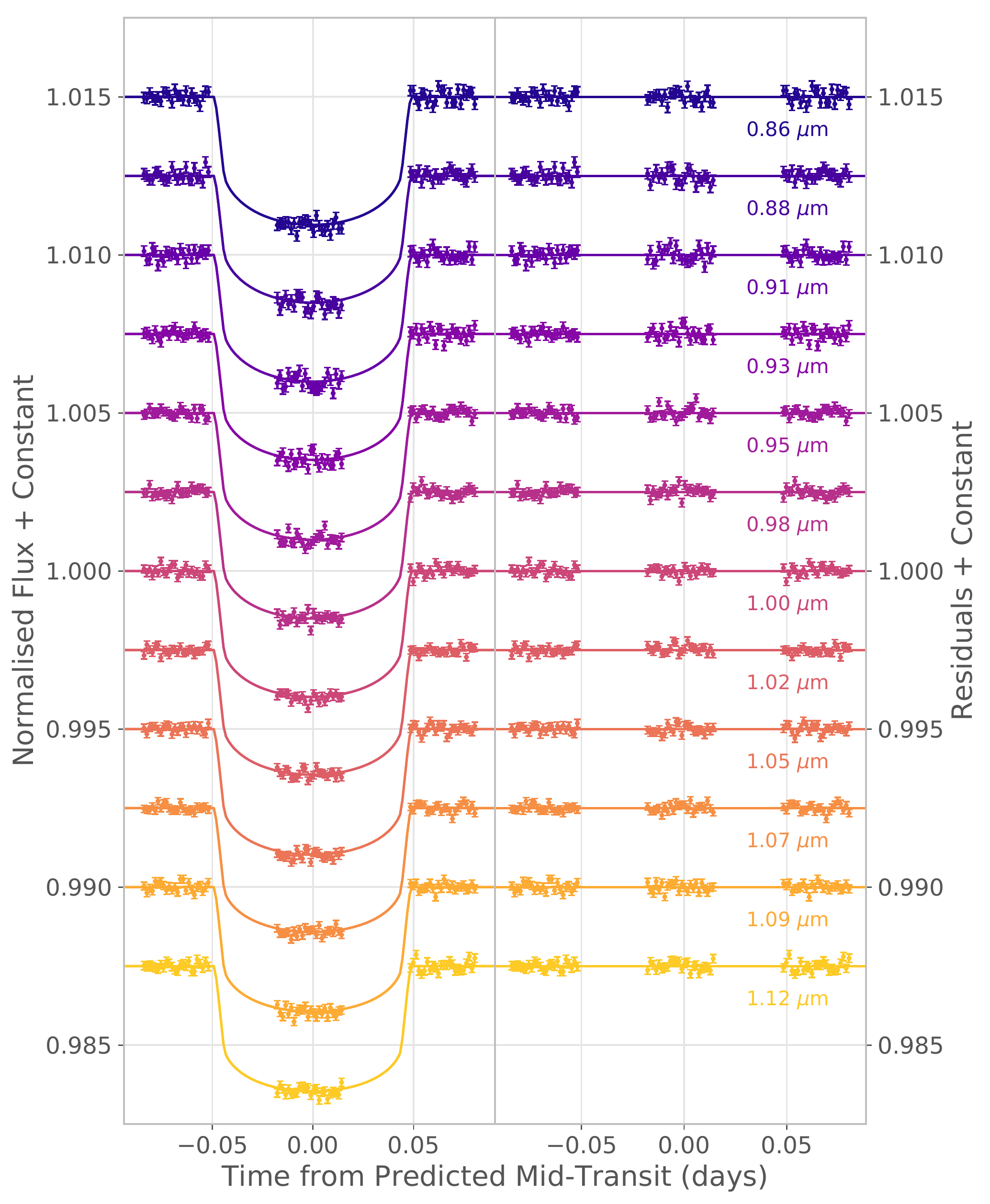}
	\caption{\textit{HST} WFC3 G102 wavelength dependent light curves for visit 5.}
	\label{fig:G102_3_WaveLC_fits}
\end{figure*}

\begin{figure*}
	\centering
	\includegraphics[width=0.55\linewidth]{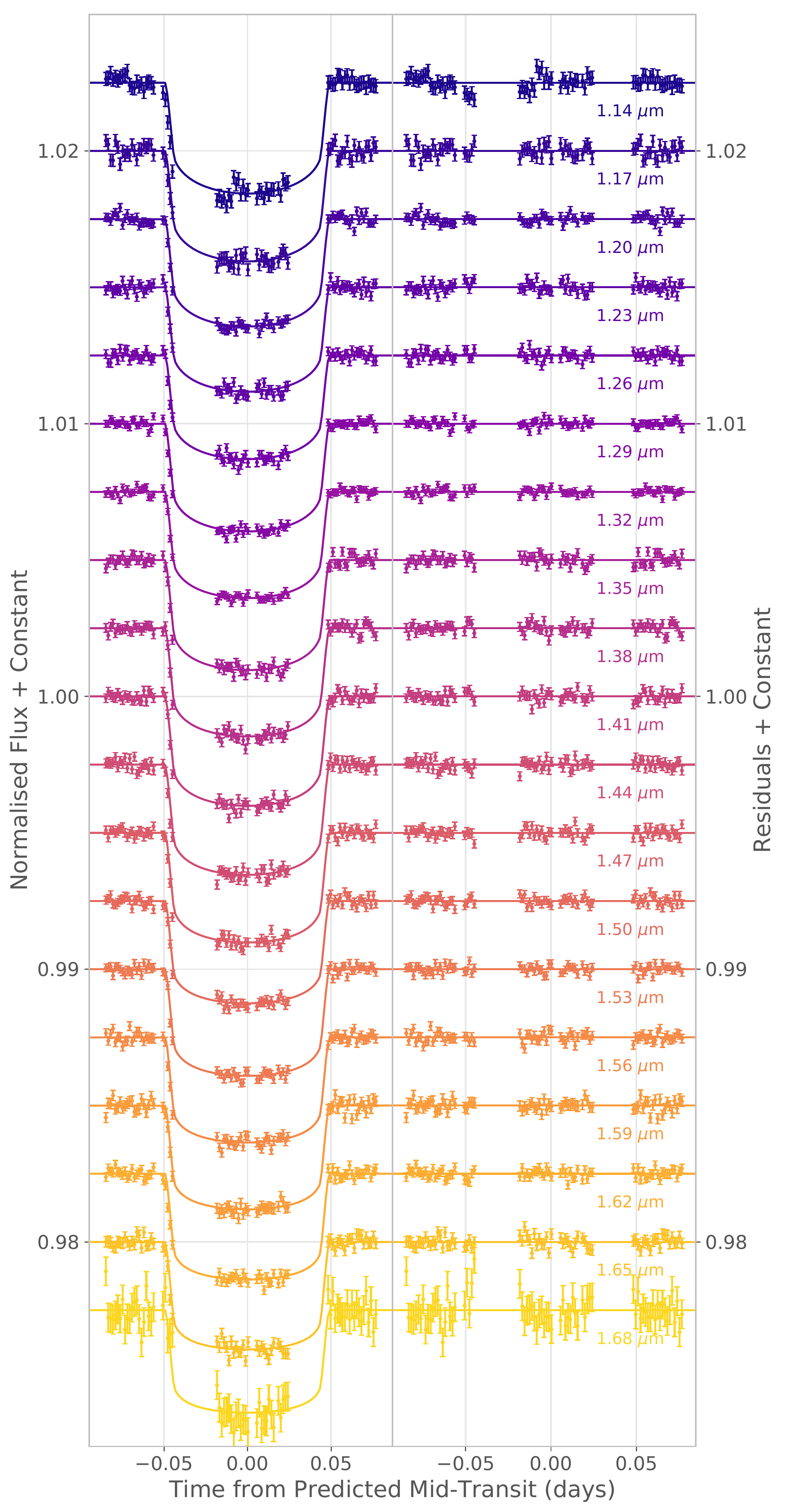}
	\caption{\textit{HST} WFC3 G141 wavelength dependent light curves.}
	\label{fig:G141_WaveLC_fits}
\end{figure*}

\end{document}